%% file: PARADIGM_satellite_destruction_and_quenching.tex
\documentclass[fleqn,usenatbib]{mnras}

\usepackage{newtxtext,newtxmath}

\usepackage[T1]{fontenc}

\DeclareRobustCommand{\VAN}[3]{#2}
\let\VANthebibliography\thebibliography
\def\thebibliography{\DeclareRobustCommand{\VAN}[3]{##3}\VANthebibliography}

\usepackage{graphicx}	
\usepackage{amsmath}	
\usepackage[dvipsnames]{xcolor}

\newcommand{\MSUN}{{\rm M}_{\sun}}
\newcommand{\MHOST}{M_{\rm 200c}}
\newcommand{\RHOST}{r_{\rm 200c}}

\newcommand{\MSTAR}{M_{\rm *}}

\newcommand{\targetMerger}{$z \approx 2$}
\newcommand{\vgn}{VG}
\newcommand{\tng}{TNG}
\newcommand{\XXC}{\textbf{SM2}}
\newcommand{\XVC}{\textbf{SM1}}
\newcommand{\XC}{\textbf{FM}}
\newcommand{\XCX}{\textbf{LM1}}
\newcommand{\XCXX}{\textbf{LM2}}
\newcommand{\EF}{\textbf{EF}}
\newcommand{\LF}{\textbf{LF}}
\newcommand{\XCEC}{\textbf{FM-EC}}

\title[MW satellite survival and quenching]{The PARADIGM project II: The lifetimes and quenching of satellites in Milky Way-mass haloes}

\author[G. D. Joshi et al.]{
Gandhali D. Joshi,$^{1}$\thanks{E-mail: gandhali.d.joshi@durham.ac.uk}
Andrew Pontzen,$^{1}$
Oscar Agertz,$^{2}$
Justin Read$^{3}$
and Martin P. Rey$^{4}$
\\
$^{1}$Institute for Computational Cosmology, Durham University, Lower Mountjoy, South Rd, Durham DH1 3LE, UK \\
$^{2}$Lund Observatory, Division of Astrophysics, Department of Physics, Lund University, Box 43, SE-221 00 Lund, Sweden \\
$^{3}$Department of Physics, University of Surrey, Guildford GU2 7XH, UK \\
$^{4}$University of Bath, Department of Physics, Claverton Down, Bath, BA2 7AY, UK 
}

\date{Accepted XXX. Received YYY; in original form ZZZ}

\pubyear{\the\year{}}

\begin{document}
\label{firstpage}
\pagerange{\pageref{firstpage}--\pageref{lastpage}}
\maketitle

\begin{abstract}
The abundance and star-formation histories of satellites of Milky Way (MW)-like galaxies are linked to their hosts’ assembly histories. To explore this connection, we use the PARADIGM suite of zoom-in hydrodynamical simulations of MW-mass haloes, evolving the same initial conditions spanning various halo assembly histories with the VINTERGATAN and IllustrisTNG models. Our VINTERGATAN simulations overpredict the number of satellites compared to observations (and to IllustrisTNG) due to a higher $M_{*}$ at fixed $M_{\rm tot}$. Despite this difference, the two models show good qualitative agreement for both satellite disruption fractions and timescales, and quenching. The number of satellites rises rapidly until $z=1$ and then remains nearly constant. The fraction of satellites from each epoch that are disrupted by $z=0$ decreases steadily from nearly 100\% to 0\% during $4>z>0.1$. These fractions are higher for VINTERGATAN than IllustrisTNG, except for massive satellites ($M_{*}>10^{7}\MSUN$) at $z>0.5$. This difference is largely due to varying distributions of pericentric distance, orbital period and number of orbits, in turn determined by which sub(haloes) are populated with galaxies by the two models. The time between accretion and disruption also remains approximately constant over $2>z>0.3$ at $6-8$~Gyr. For surviving satellites at $z=0$, both models recover the observed trend of $M_{*}>10^{7}\MSUN$ satellites quenching more recently ($<8$~Gyr ago) and within $1.5\,\RHOST$ of the host, while lower mass satellites quench earlier and often outside the host. Our results provide constraints on satellite accretion, quenching and disruption timescales, while highlighting the convergent trends from two very different galaxy formation models.
\end{abstract}

\begin{keywords}
galaxies: formation -- galaxies: evolution -- galaxies: dwarf -- galaxies: abundances -- galaxies: interactions
\end{keywords}



\section{Introduction}
The abundance and properties of satellite galaxies are closely tied to their host halo and can therefore be a valuable probe of the its evolution and that of its central galaxy. Once accreted onto their host halo, satellites are subject to several environmental processes that have a direct impact on their mass content, morphology and star-formation rates (SFRs) and indeed whether they are star-forming or quenched. Milky Way (MW)-mass hosts are a particularly interesting regime in which to study the fate of low-mass, dwarf satellite galaxies of stellar mass $10^{4}<M_{*}/\MSUN<10^{9}$ \citep[e.g.][]{Collins2022}. Dwarf satellites are more susceptible to environmental processes such as tidal stripping and shocking \citep[e.g.][]{Read2006,Mayer2006,Sales2010,Fattahi2018} and ram-pressure stripping \citep[e.g.][]{Mayer2006,McConnachie2007,Nichols2011,Kenney2014,Kazantzidis2017,Boselli2021,Yang2022} due to their lower masses and shallower potentials compared to $L^{*}$ galaxies, and therefore provide an important testbed for the modelling of the interaction between intrinsic processes and environment.

Over the last few decades, we have been able to build up a near complete accounting of satellites around our own MW and the nearby Andromeda galaxy (M31) and more generally, the Local Group (LG) \citep[e.g.][]{McConnachie2012,Martin2013,Martin2016,McConnachie2018,Gozman2024}. Extending beyond the LG, we have also observed satellites in other galaxies through targeted observations \citep[e.g.][]{Merritt2014,Carlin2016,Smercina2018,Crnojevic2019,Bennet2020,MutluPakdil2024,Muller2025a,Muller2025b}. Broader surveys such as ELVES \citep{Carlsten2022}, SAGA \citep{Geha2017,Mao2021,Mao2024}, xSAGA \citep{Wu2022}, UNIONS \citep{Heesters2025} and EUCLID \citep{Marleau2025} are now also providing a more complete census of satellites in the Local Universe. Despite these advances, we are as yet unable to observe such low-mass satellites beyond a few 10s of Mpcs (extending to 100s of Mpcs in the case of UNIONS and EUCLID) and therefore infer their evolution directly. Understanding the fate of satellites from early cosmic epochs and their impact on the central galaxy and its stellar halo requires access to the complete evolutionary history of all satellites that were accreted onto the host halo and that eventually merged with it. For this, we turn to simulations. 

In recent years, cosmological and zoom-in simulations have allowed us to study dwarf satellites in a range of environments. For satellites of MW-mass hosts in particular, simulations of large volumes (i.e. of the order of (10s of Mpc)$^{3}$) such as IllustrisTNG50 \citep{TNG50Pillepich2019,TNG50Nelson2019}, ROMULUS25 \citep{Tremmel2017Romulus25} and FIREBox \citep{Feldmann2023Firebox} naturally contain several dozens of MW analogues while still being able to resolve classical MW satellites. Additionally, several galaxy formation models specifically focus on this regime with high-resolution zoom-in simulations of MW-, M31- and LG-like hosts, e.g. FIRE \citep{Hopkins2014,Hopkins2017,GarrisonKimmel2019}, APOSTLE \citep{Sawala2016Apostle}, LATTE \citep{Wetzel2016Latte}, AURIGA \citep{Grand2017Auriga}, NIHAO \citep{Wang2015Nihao,Buck2019}, HESTIA \citep{Libeskind2020Hestia}, ARTEMIS \citep{Font2020Artemis,Font2021} and the DC Justice League \citep{Applebaum2021JusticeLeague}. These simulations give us access to the assembly histories of MW-like haloes and their satellites, and crucially, can provide insights into how the currently observable properties of such satellites connect to their accretion histories.

Observations suggest that there is a large diversity in the population of dwarf satellites around MW-mass galaxies, with substantial variations in: luminosity functions (LFs) and radial distributions \citep{McConnachie2012,Geha2017,Smercina2018,Bennet2020,Mao2021,Carlsten2022,Wu2022,Muller2024,Muller2025b}; sizes, structural properties and morphologies \citep{McConnachie2012,Geha2017,Crnojevic2019,Mao2021,Carlsten2022,Gozman2024}; and colours, metallicities and star-formation (SF) status \citep{McConnachie2012,Crnojevic2019,Mao2021,Carlsten2022,Greene2023,Geha2024}. What ultimately drives this diversity in satellite properties, whether it is the varying assembly histories of their hosts, the structure of the host environment and the internal structure of the dwarfs themselves, or the circumgalactic- and interstellar medium (CGM and ISM respectively) of the host galaxy with which the satellites interact, remains an active area of research. 

The abundances, mass functions, and distributions of properties of satellites around MW-like hosts are the result of a complex interplay of environmental influences and internal evolutionary processes within the satellites. Along with the continued accretion of new satellites into the host halo, two processes in particular determine the overall abundance of satellites: a) tidal stripping taken to its extreme results in the complete disruption of satellites, with their DM, stars, and any remaining gaseous material becoming part of the MW halo, and b) dynamical friction causes satellite orbits to decay, eventually leading them to merge with the central. The properties of disrupted satellites and the timescales over which they disrupt therefore have crucial implications for the understanding of the \emph{ex-situ} stellar populations of the MW and the formation of the MW stellar halo. On the other hand, the majority of present-day MW satellites appear to have been quenched several Gyr ago \citep[e.g.][]{Weisz2015}, suggesting either that they have survived over such long time intervals within the MW host or that they were quenched externally and were accreted more recently. Therefore, understanding satellite disruption provides an important context for studying satellite quenching in these systems and vice versa. 

In simulations, several previous studies have explored the disruption of satellites in MW-mass hosts, starting with N-body simulations \citep[e.g.][]{Hayashi2003,Taffoni2003,Taylor2004,BoylanKolchin2007,Choi2009,Read2019,Mazzarini2020} and semi-analytical modelling \citep[e.g.][]{Bullock2001,Bullock2005,Faltenbacher2005,Purcell2007,DOnghia2010,Henriques2010}. Earlier studies based on N-body simulations such as \citet{Bullock2001} and \citet{Taffoni2003} predicted on the order of $\sim400$ subhaloes of masses $\gtrsim 10^{7}\MSUN$ to have been tidally disrupted in MW-mass hosts. They also found that lower mass subhaloes undergo substantial mass loss, which slows down their orbital decay leading to longer disruption timescales compared to more massive subhaloes. \citet{BoylanKolchin2007}, using N-body simulations including a stellar bulge and central BH highlighted the importance of a central BH in efficiently disrupting satellites, although the effect is only seen once the satellite orbit decays enough to bring it within the sphere of influence of the BH. Similarly, semi-analytical efforts such as those of \citet{Bullock2005} show that by incorporating significant mass loss from satellites over up to 10~Gyr, their models are better able to match the observed properties of central galaxies such as their LFs, colours, intracluster light (ICL) and others. Studies such as \citet{DOnghia2010,Penarrubia2010,Brooks2013,GarrisonKimmel2017} and \citet{Read2019} also emphasize the importance of the MW stellar disc itself, not just the host halo, in disrupting satellites. More recently, cosmological hydrodynamical simulations have been able to reach high enough resolutions to examine the accretion and subsequent fates of dwarf satellites, particularly in the context of the MW stellar halo and stellar streams \citep[e.g.][]{Kelley2019,Grimozzi2024,Riley2024,Shipp2024,Pathak2025}. Using the AURIGA simulations, \citet[][]{Shipp2024} show that the majority of all accreted satellites of MW hosts have been phase-mixed with the central galaxy by $z=0$, with a further $\sim30$ per cent having been disrupted and currently seen as stellar streams. \citet[][]{Grimozzi2024} consider the evolution of the mass-metallicity relations of disrupted satellites compared to the present-day surviving satellites in the ARTEMIS simulations and find disruption timescales of $<1$~Gyr for the satellites accreted very early ($z>3$), which is the majority of their low-mass satellites ($M_{*}<10^{7}\MSUN$), while for more massive satellites which tend to be accreted at later times, the disruption timescales can be as long as 9~Gyr. \citet{Pathak2025}, with the DC Justice League simulations, show that while the majority of accreted ultra-low-mass satellites ($\MSTAR<10^{6}\MSUN$) can survive over 11~Gyr to present day, intermediate- ($\MSTAR=10^{6}-10^{8}\MSUN$) and high-mass ($\MSTAR>10^{8}\MSUN$) satellites have shorter disruption timescales of up to $\sim8$ and $\sim5$~Gyr respectively.

The large diversity in observed properties of satellites around MW-mass hosts and the fate of such accreted satellites is likely in part due to the variations of assembly histories of MW-like haloes, and in the case of hydrodynamical simulations, also due to the differences in the numerical methods and galaxy formation models they employ. To tackle both these issues, in \citet{Joshi2025} we introduced the PARADIGM project, built on the earlier VINTERGATAN-GM suite of zoom-in simulations of MW-like galaxies \citep{Rey2022,Rey2023} with the VINTERGATAN model \citep{Agertz2021}. This initial suite consisted of one fiducial set of initial conditions (ICs) and four `genetically modified' (GM) versions using the code \textsc{GenetIC} \citep{Roth2016,Rey2018,Stopyra2021} to alter the strength of a significant \targetMerger{} merger while maintaining the $z=0$ halo mass and, to the greatest extent possible, the overall environment the halo is embedded in. This approach allowed us to investigate the impact of halo assembly history on e.g. the chemodynamics of the MW stellar halo \citep{Rey2023} and the evolution of satellite abundances around MW-mass hosts \citep{Joshi2024}, in a controlled manner. The PARADIGM project extends this suite by evolving the same set of ICs with the IllustrisTNG model \citep{TNGMethodsPillepich2018,TNGMethodsWeinberger2017}. This approach is in contrast and complementary to efforts such as the AGORA collaboration \citep{Kim2014Agora,RocaFabrega2021Agora}, which compare the behaviour of different simulation codes in evolving a single set of ICs and explore how well these simulations can be made to agree. For example, \citet{RodriguezCardoso2025} examine satellite quenching is the CosmoRun of the AGORA project and show that it is primarily driven by strangulation and gas removal through ram-pressure stripping according to several different models. Instead our project focuses on how the simulation codes respond to controlled modifications to assembly and merger histories. The two models we compare employ complementary strategies for simulating the Universe with contrasting emphases. While the VINTERGATAN model aims to resolve star-formation (SF) and the physics of the ISM down to the coldest and densest gas, the IllustrisTNG model aims to closely replicate the global population of present-day galaxies and structures by recovering several well known scaling relations. By comparing the response of the two models to varying ICs, we can gain insight into whether and how the complex interplay between the different physical processes they model can lead to similar emergent trends and hence, show which trends are to be regarded as robust predictions of $\Lambda$CDM.

In this paper, we explore the evolution of satellite abundances and track satellites selected at multiple epochs to their disruption. We then constrain the disruption timescales of satellites accreted at various epochs and investigate their dependence on several properties of the host and the satellites themselves. We also explore the star-formation status of the surviving satellites at $z=0$ and how their quenching times connect to the disruption timescales, for a comprehensive look at the fate of satellites in MW-mass hosts. The paper is structured as follows. In Section~\ref{sec:methods} we provide details of the simulation setup, a brief summary of the two galaxy formation models, and the methods used to measure disruption timescales and quenching times. In Section~\ref{sec:resultsHostProps} we discuss the properties of the host haloes and central galaxies. Sections~\ref{sec:resultsSatNumbers}-\ref{sec:resultsHaloFormTimes} consider the abundances and disruption of satellites and their connection to halo assembly. Section~\ref{sec:discContributingFactors} explores the impact of various orbital and intrinsic properties of the satellites and the structure of their hosts, as well as the how the two models populate DM haloes with galaxies, on satellite disruption. In Section~\ref{sec:resultsQuenching} we investigate when and where surviving satellites were quenched. Finally, we compare our results to previous work in Section~\ref{sec:discussion} and summarise our conclusions in Section~\ref{sec:summary}.


\section{Simulations and methods} \label{sec:methods}

\subsection{Simulation setup}
We use the PARADIGM suite of simulations which consists of two sets of zoom-in cosmological simulations of MW-mass haloes, evolving the same set of initial conditions (ICs) with two different galaxy formation models, namely VINTERGATAN \citep{Agertz2021} and IllustrisTNG \citep[][]{TNGMethodsPillepich2018,TNGMethodsWeinberger2017}, specifically the TNG50 configuration \citep[][]{TNG50Pillepich2019,TNG50Nelson2019}. The haloes targeted by the zoom-in simulations were initially selected from a uniform resolution DM-only (DMO) simulation of a $(50\,h^{-1}{\rm Mpc})^{3}$ volume resolved by $512^{3}$ particles. MW candidate haloes were chosen to have a mass range of $\MHOST=(7.5\times10^{11}-1.5\times10^{12})\MSUN$ and with no neighbours within $5\,\RHOST$ more massive than them \citep[see][for details]{Rey2022}. Of these, the fiducial galaxy \XC{} was selected to have a merger history similar to that expected of the MW, with its last major merger occurring at $z\sim2$ and having a quiet merger history afterwards. Two other haloes were also selected: \EF{} (early former) which forms early and has its last major merger at $z\sim3$, and \LF{} (late former) which forms later and has several major mergers with the latest occurring at $z\sim0.7$. Particles within $3\,\RHOST$ of each of these haloes at $z=0$ were tracked back to their positions in the ICs and used to define the zoom-in region.

The \XC{} simulation was also used as the reference to generate four `genetically-modified' (GM) ICs using \textsc{genetic} \citep{Roth2016,Rey2018,Stopyra2021}, which allows us to perform targeted alterations to the ICs while retaining, as much as possible, the large-scale structure surrounding the halo. The ICs were modified by altering the overdensity of the region in which the halo is embedded at early times, with the goal of modifying the strength of a significant $z \approx 2$ merger. These are referred to as the \XXC{}, \XVC{}, \XCX{} and \XCXX{} ICs, in increasing order of the mass of the secondary system involved in the $z\approx2$ merger, with the \XC{} ratio being intermediate between \XVC{} and \XCX{}. Complete details on the design of all ICs and the initial suite of VINTERGATAN-GM simulations is provided in \citet{Rey2022} and \citet{Rey2023}, while details of the corresponding IllustrisTNG-GM simulations are provided in \citet{Joshi2025}. Note that the VINTERGATAN \EF{} simulation was only run to $z=0.11$ i.e. $t=12.36$~Gyr due to computational time limitations, and hence it is excluded from results at $z<0.2$ unless stated otherwise. Finally, we also generate an additional set of ICs based on the \XC{} ICs, but resulting in an earlier collapsing halo, to study the relative impact of merger history and halo collapse time on our results. Due to computational costs, these ICs were only evolved with the IllustrisTNG model, producing the \tng{} \XCEC{} run. Hence, we discuss it separately in Section~\ref{sec:resultsHaloFormTimes}, where we study the impact of collapse time and merger history on our results.

All other ICs were evolved with both galaxy formation models VINTERGATAN and IllustrisTNG (hereafter referred to as \vgn{} and \tng{} respectively). In all simulations, we adopt cosmological parameters for a flat $\Lambda$CDM cosmology with $\Omega_{\rm m,0}=0.3139$, $h=0.6727$, $n_{s}=0.9645$ and $\sigma_{8}=0.8440$ \citep{Planck2016}. In both sets of simulations, DM is modelled by particles of mass $2\times 10^{5}\MSUN$. We provide a brief description of both models here and refer the reader to \citet{Rey2023} and \citet{Joshi2025}, as well as the papers from each of the collaborations that developed them for further details.

\paragraph*{\vgn{} simulations}: The \vgn{} simulations were performed with \textsc{ramses} \citep{Teyssier2002}, an adaptive mesh refinement (AMR) code that uses a particle-mesh algorithm to solve Poisson's equation and an HLCC Riemann solver \citep{Toro1994} for fluid dynamics, with an ideal equation of state for gas with $\gamma=5/3$. The AMR strategy allows us to reach a minimum effective gas cell radius of 11~pc (and median 88~pc) within the central galaxy. The \vgn{} galaxy formation model \citep{Agertz2021} includes prescriptions for resolved SF in cold, dense gas, feedback from supernovae and (SNeIa and SNeII), stellar winds, metal enrichment, gas cooling through emission from metal-lines, and heating and photo-ionization by a UV background \citep{Haardt1996} that is uniform but evolves with cosmic time. The ISM model resolves gas to temperatures as low as 10~K and SF occurs in cells with density $>100\,m_{\rm p}\,{\rm cm}^{-3}$ where $m_{\rm p}$ is the mass of a proton and temperature $<100$~K through a Poisson process following the Schmidt law \citep{Schmidt1959} and assuming a Chabrier \citep{Chabrier2003} initial mass function (IMF). The model accounts for the injection of mass, momentum and metals from SNeIa, SNeII and stellar winds from O, B and AGB stars. Supernova feedback is modelled following the \citet{Kim2015} prescription, in the form of energy injection when the cooling radius is large and momentum injection otherwise. Feedback from active galactic nuclei (AGN) is not included in this model. Finally, galaxies are identified in the simulations using the Amiga Halo Finder \citep[\textsc{ahf};][]{Gill2004,Knollmann2009}, which identifies peaks in the matter density distribution using an AMR strategy and defines bound haloes through a reiterative unbinding process.

\paragraph*{\tng{} simulations}: The \tng{} simulations were generated with \textsc{arepo} \citep{Springel2010}, a moving mesh code that solves the equations of magneto-hydrodynamics (MHD) using a finite-volume method on an unstructured Voronoi-tesselated moving mesh. The simulations attain a minimum effective gas cell radius of 46~pc (and median 174~pc). The \tng{} galaxy formation model \citep{TNGMethodsPillepich2018,TNGMethodsWeinberger2017} includes prescriptions for SF, feedback from supernovae (SNeIa and SNeII) and AGB stars, metal enrichment, gas cooling through metal-line emission, and gas heating by a uniform, time-dependent UV background \citep{FaucherGiguere2009}. For SF, the ISM is treated as a two-phase model with an effective equation of state (eEOS), and SF occurs in cells with density $>0.1\,m_{p}\,{\rm cm}^{-3}$ and a variable maximum temperature depending on the eEOS down to a minimum of $10^{4}$~K. SF then proceeds following a Kennicutt-Schmidt law \citep{Schmidt1959,Kennicutt1998} and assuming a Chabrier IMF \citep{Chabrier2003}. The \tng{} model also includes the formation of black holes (BHs) and their growth through Bondi accretion. BHs are seeded with an initial mass of $1.2\times10^{6}\MSUN$ in haloes of mass $>7.4\times 10^{10}\MSUN$, which are always positioned at the centres of their host haloes. In our simulations, the central MW always hosts a BH. In most satellite subsamples discussed below, the satellites do not host a BH; however, in a few subsamples, a single satellite is found to host a BH of mass $(1-9)\times10^{6}\MSUN$. BH feedback is modelled as a continuous injection of thermal energy in the high-accretion mode and as a stochastically released momentum kick in the low-accretion mode. Finally, galaxies are identified in the simulations through a two-step process, first with a friends-of-friends \citep[FOF][]{Davis1985} algorithm to determine haloes, followed by using \textsc{subfind} \citep{Springel2001,Dolag2009} to identify bound structures within the haloes with a single unbinding step.

\begin{table*}
    \caption{Properties of the host haloes and central MW galaxies at $z=0$. The properties are as follows: (Columns 2-5) Host virial radius and total, gaseous and stellar mass enclosed within $\RHOST$; (Column 6) halo formation time, measured as the time at which the halo has assembled 50 per cent its present day mass; (Columns 7 \& 8) stellar half-mass radius (3D) and half-light radius in \emph{V}-band (2D, face-projection); (Columns 9-11) total, gaseous and stellar galaxy mass, measured within $3\times R_{*,1/2}$.} \label{tab:centralProps}
    \centering
    \begin{tabular}{l|c|c|c|c|c|c|c|c|c|c}
    \hline
    Simulation & $\RHOST$ & $M_{\rm 200c,tot}$ & $M_{\rm 200c,gas}$ & $M_{\rm 200c,*}$ & $t_{\rm form}$ & $R_{\rm *,1/2}$ & $R_{\rm *,1/2,V-band}$ & $M_{\rm gal,tot}$ & $M_{\rm gal,gas}$ & $M_{\rm gal,*}$ \\
    & [kpc] & [$10^{10}\MSUN$] & [$10^{10}\MSUN$] & [$10^{10}\MSUN$] & [Gyr] & [kpc] & [kpc] & [$10^{10}\MSUN$] & [$10^{10}\MSUN$] & [$10^{10}\MSUN$] \\
    \hline
    \multicolumn{11}{c}{TNG} \\
    \hline
    \input{figures/halo_props_all.tex}
    \end{tabular}
\end{table*}

\subsection{Satellite properties}

\subsubsection{Sample selection} \label{sec:methodsSample}
Throughout the paper, we select samples of satellites at a given epoch and then track these satellites forward to their disruption and backward to their accretion. In a given snapshot, satellites are selected with the following criteria:
\begin{itemize}
    \item $\boldsymbol{0.1\,\RHOST<d<d_{\rm max}}$ where $d$ is the distance from the host centre and $\RHOST$ is the host's virial radius. The inner region is excised to remove any gas/stellar clumps from the central galaxy. For most of the analysis, we focus on satellites found within the host's virial radius i.e. $d_{\rm max}=\RHOST$. However, when comparing to observational results, we instead use $d_{\rm max}=300$~kpc to match the observational constraints. Additionally, we consider satellites out to $d_{\rm max}=3\,\RHOST$ in Section~\ref{sec:smhmRelations}, when comparing how the two models populate DM haloes with galaxies.
    \item $\boldsymbol{{M_{\rm DM}/M_{\rm tot}>0.1}}$ where $M_{\rm DM}$ and $M_{\rm tot}$ are the galaxy's DM and total mass. Satellites are required to have a minimum proportion of DM to further remove clumps that are unlikely to be galaxies, while still retaining galaxies that may have been significantly stripped of their DM but not their stellar matter.
    \item $\boldsymbol{{f_{\rm hi-res}>0.9}}$ where $f_{\rm hi-res}$ is the fraction of DM particles in the galaxy that are of the highest resolution. We remove satellites with significant contamination from low-resolution DM particles which are likely to be affected by boundary effects at the edges of the zoom-region. This affects less than 2 per cent of the (sub)haloes found within $3\,\RHOST$.
    \item $\boldsymbol{{M_{*}/n_{*}>10^{5}\MSUN}}$ where $n_{*}$ and $M_{*}$ are the number and mass of stellar particles in the galaxy. This criterion further removes satellites with significant contamination from low-resolution stellar particles. This only affects the \tng{} simulations, since these simulations are allowed to form stars anywhere within the simulation volume, not just in the high-resolution zoom region. This affects up to 8 per cent of luminous (sub)haloes within $3\,\RHOST$ of the host.
    \item $\boldsymbol{{M_{\rm tot}>10^{7}\MSUN}}$ and $\boldsymbol{{M_{*}>10^{6}\MSUN}}$. These mass limits correspond to a minimum of 35 and 160 stellar particles in the \tng{} and \vgn{} simulations respectively (due to the higher baryon resolution in the \vgn{} simulations) and 55 DM particles, ensuring the satellites are sufficiently well resolved. Note that in practice none of the \vgn{} satellites and less than 5 per cent of the \tng{} have fewer than a 100 DM particles; we set this lower limit to retain any satellites that might have experienced extreme stripping of DM.
\end{itemize}
For the sample selection process, we rely on the properties of the (sub)haloes from the respective halo finders. To evaluate any potential differences between the halo finders, we have also run \textsc{ahf} on the \tng{} \XC{} simulation. In comparing the $z=0$ haloes, we find no significant differences in the halo mass functions (HMFs), stellar mass functions (SMFs) or stellar mass-halo mass (SMHM) relations for satellites selected with the listed criteria. Hence, we do not expect significant biases in the samples selected due to the halo finders.

We use the above criteria to compile a list of satellites at specific epochs selected to be approximately 1~Gyr apart, at $z=4$, 3, 2, 1.5, 1, 0.8, 0.6, 0.5, 0.4, 0.3, 0.2, 0.1 and 0. Additionally, for the five GM runs, we consider two other times namely the beginning and end of the \targetMerger{} merger that is targeted for modification. The beginning of a merger is defined as the time at which the secondary galaxy is last outside the virial radius of the primary halo, while the end of the merger is defined when the secondary galaxy can no longer be identified by the halo finder. These redshifts and pre-/post-merger epochs are referred to as selection epochs through the paper; note that satellites in these samples will have been accreted before this epoch and will have been in the host for a range of periods. We also measure the accretion time for these satellites as described below.

\begin{figure}
    \centering
    \includegraphics[width=\linewidth]{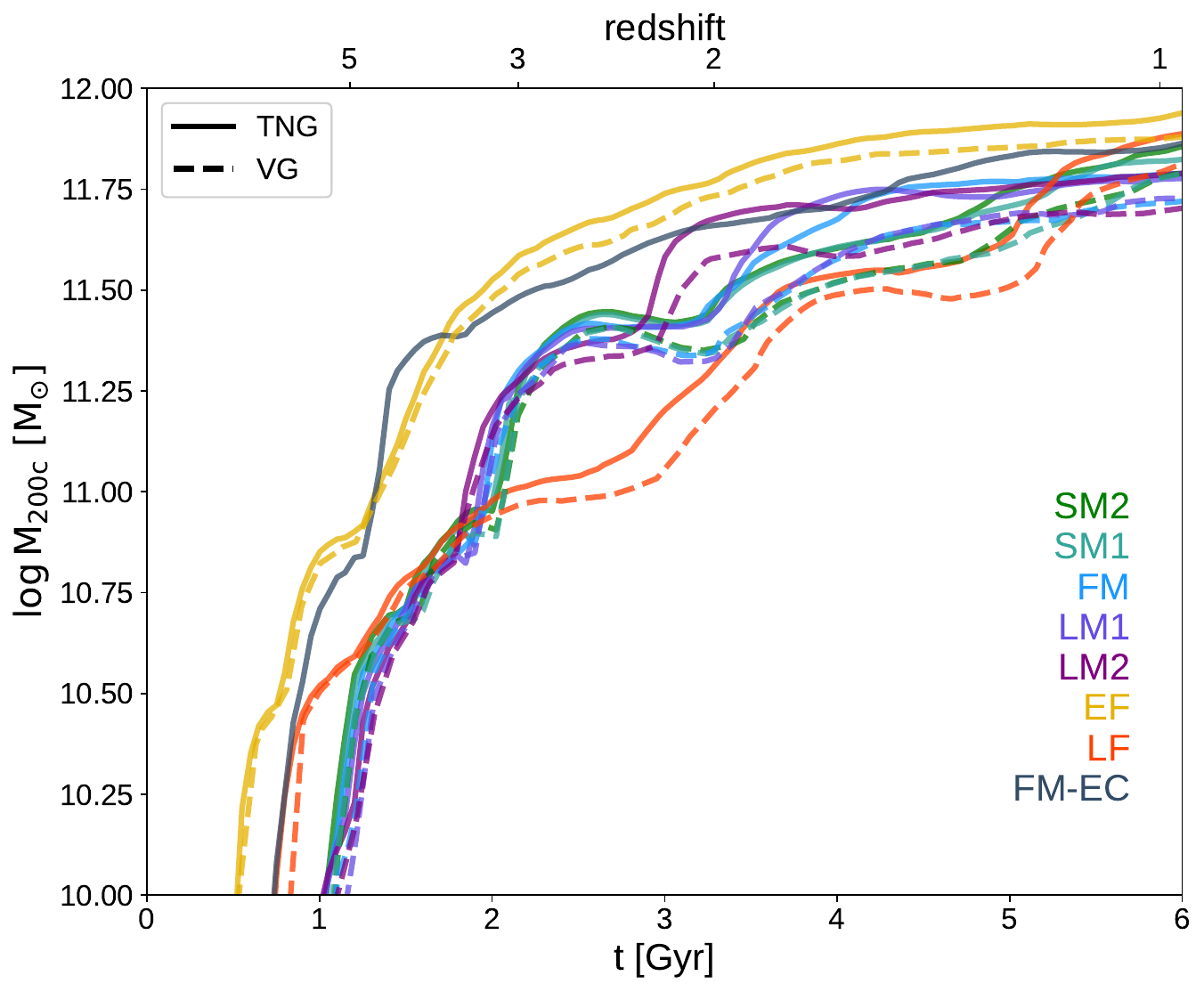}
    \caption{Evolution of the host mass $\MHOST$ until $z=1$ for all haloes in our sample. After this epoch, the masses grow gradually by 0.1-0.25 dex to their final values. The \EF{}, \XC{} and \LF{} haloes have starkly different growth histories. The \EF{} grows rapidly at early times and has a quiet merger history after $z=4$, while \LF{} grows more slowly and through multiple mergers. The \XC{} halo has an intermediate growth history, with a halo formation time intermediate to the former two haloes, significant mergers occurring at $z\approx2$ and a quiet merger history after. The four GM haloes behave similarly to the \XC{} halo until $z\approx3$, when the impact of the GMs are seen on two significant mergers. Finally, the \XCEC{} halo, whose ICs were modified from the \XC{} ICs to have an earlier collapse time has a similar growth history to the \EF{} halo at early times, but slower growth after $z=3$.}
    \label{fig:haloGrowth}
\end{figure}

\begin{figure*}
    \centering
    \includegraphics[width=\linewidth]{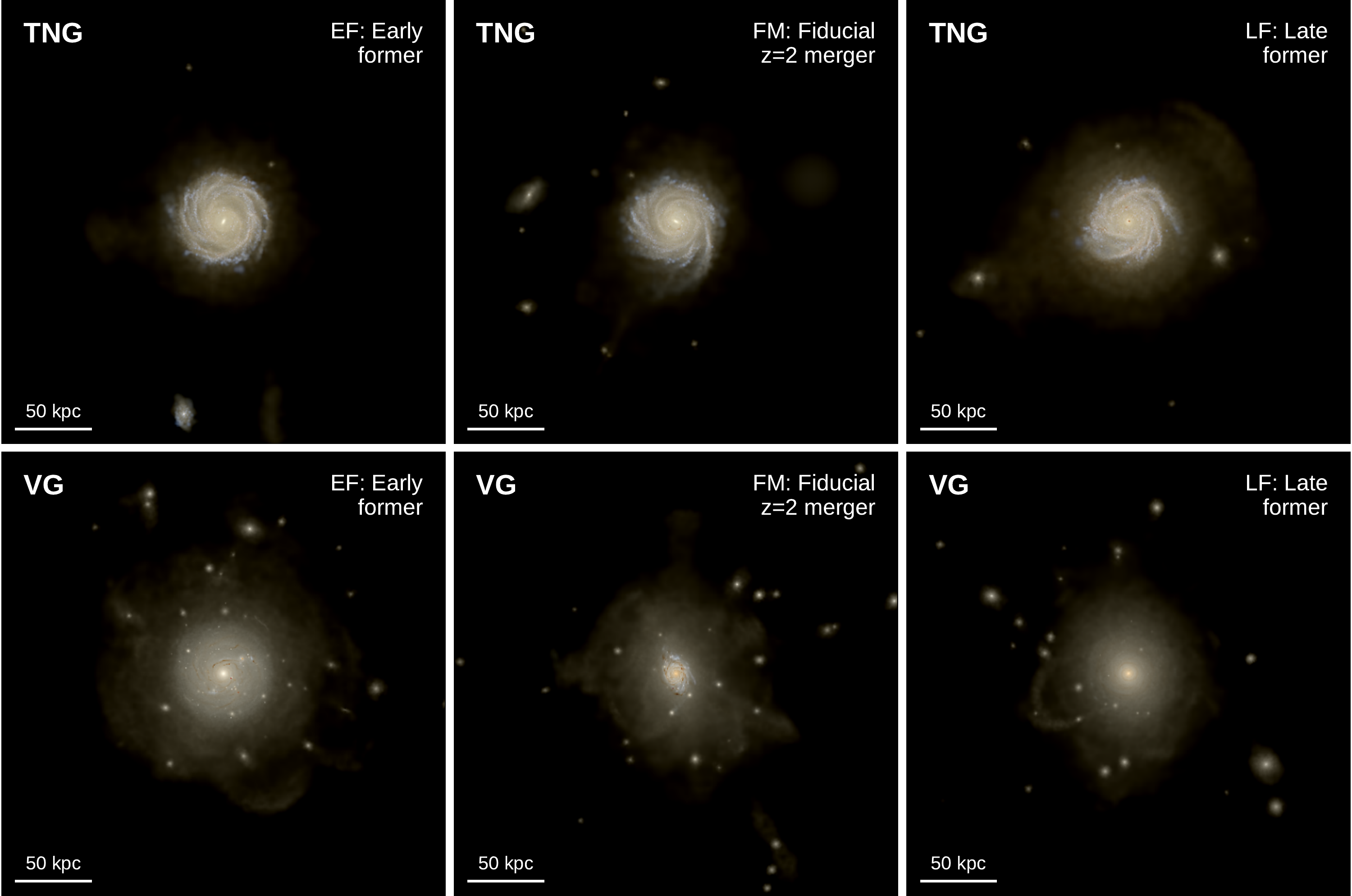}
    \caption{Mock stellar images of the \EF{}, \XC{} and \LF{} haloes at $z=0$ from both sets of simulations in our sample. Each galaxy has been rotated to be face-on. The colours correspond to Johnsons U, V and I bands and encompass $20-32$mag/arcsec$^{2}$ in surface brightness. We use this faint lower limit to highlight the satellites of the haloes. A simple dust correction has been applied following the \citet{Calzetti2000} law. Each image is 300~kpc on a side and uses a projected image of depth 300~kpc. The \tng{} central galaxies are always larger than the \vgn{} ones. Each \vgn{} halo has significantly more satellites than its \tng{} counterpart, and \vgn{} bright satellites appear more concentrated than \tng{} ones.} \label{fig:stellarMaps}
\end{figure*}

\subsubsection{Physical properties}
For the majority of this paper, we rely on the two halo finders to obtain satellite positions, velocities, and halo and stellar masses, unless otherwise specified. Any other properties used are calculated using all particles associated with the satellite according to the halo finder, making extensive use of the \textsc{pynbody} \citep{Pontzen2013} and \textsc{tangos} \citep{Pontzen2018} packages. When comparing to observational results as in Section~\ref{sec:resultsZ0LFs} and \ref{sec:resultsQuenching}, to minimize uncertainties and to be able to make a robust comparison, we calculate a radius $R_{\rm SB=30}$, where the \emph{V}-band surface brightness equals 30~mag/arcsec$^{2}$ in a random projection, for all satellites in our sample at $z=0$. This limit was chosen as a reasonable match to the sensitivity of several recent and upcoming advanced observational facilities. We then measure stellar masses and \emph{V}-band luminosities within this radius.

\subsubsection{Tracking satellites to disruption} \label{sec:methodsTrackForward}
With the satellites selected at each epoch, we determine the time at which the satellite is disrupted using the distributions of their stellar particles in subsequent snapshots. To do so, we first select the 80\% innermost stellar particles of each satellite; these are the only particles that are tracked in subsequent snapshots. The \tng{} set and each of the individual \vgn{} simulations have different numbers of output snapshots and it is therefore important to standardise the cadence of outputs to compare the simulations at various epochs. We therefore choose specific snapshots, at intervals of 125~Myr in time, in which to track the satellites. This allows us to easily compare the two sets of simulations without any dependence on output cadence and also reduces computational costs. At each snapshot, including the starting snapshot (i.e. the selection epoch), we then measure the following properties, including only the tracked particles:
\begin{itemize}
    \item The `centre' of the particles using the shrinking sphere method \citep[][implemented in \textsc{pynbody}]{Power2003}
    \item $R_{50}$, the 3D radius containing 50 per cent of the mass of the stellar particles
\end{itemize}
Galaxies are considered to have been disrupted once $R_{50}/R_{\rm 50,initial}>F_{\rm crit}$, where $R_{\rm 50,initial}$ is the $R_{50}$ at the selection epoch and $F_{\rm crit}$ determines how large a galaxy can expand and still remain an independent structure. We chose $F_{\rm crit}=5$ as an appropriate value based on visual inspection of a subset of the tracked satellites. We also considered our results for $F_{\rm crit}=2$ and $F_{\rm crit}=10$ and found that while the corresponding disruption timescales are shorter or longer respectively, the trends are insensitive to the precise value chosen. Note that this definition does not distinguish between satellites merging with the central and satellites being tidally disrupted and becoming part of the stellar halo. The disruption time is taken as the time interval between the selection epoch and the \emph{first} snapshot where this criterion is met.

\subsubsection{Tracking satellites to accretion} \label{sec:methodsTrackBackward}
In addition to tracking satellites forward to disruption, we similarly follow them backwards from the selection epoch to determine their time of accretion. All stellar particles associated with a satellite at a given selection epoch are tracked into previous snapshots, again with a cadence of 125~Myr, back to the first snapshots when haloes are identified in the simulations ($z\sim19)$, and we keep track of which (sub)halo the majority of tracked particles end up in. Once this history is reconstructed, we remove snapshots in which the satellite particles were tracked to the main galaxy, as well as any snapshots where there are sudden changes in stellar mass according to the halo finders (indicating the satellite particles are temporarily misidentified as part of another galaxy due to halo-finder inaccuracies). With this cleaned assembly history, we can then find the first and last accretion time as the snapshot before the satellite enters the MW host halo's virial radius for the first and last time.  For a small number of satellites ($<$10 per cent at any epoch), we were unable to find an accretion time due to them being found within the host's virial radius at all previous times. These are likely to be unphysical artefacts of the halo finders and we therefore exclude them from our analysis for clarity.

\subsubsection{Star-formation histories of present-day satellites}
The star-formation histories (SFHs) of the present-day satellites are reconstructed using the ages of the stellar particles for each satellite. We also calculate $t_{*,90}$, the time at which the satellite has assembled 90 per cent of its present-day stellar mass, which is used as a proxy for quenching time, as is common practice in observational studies. To compare these quenching times to observational results, we measure the SFHs and quenching times only using stellar particles within $R_{\rm SB=30}$.


\section{The fates of MW satellites} \label{sec:resultsSatAbundance}

\subsection{Host and central MW properties} \label{sec:resultsHostProps}

The host haloes in our samples were selected to be nearly identical in mass and size, but have diverse central galaxies due to their varied assembly and merger histories and large-scale environment. In Table~\ref{tab:centralProps}, we provide key properties of the host haloes and central MW galaxies. At present day, the host masses range from $(9.0-11.5)\times 10^{11} M_{\odot}$ in the \tng{} simulations and $(8.1-9.9)\times 10^{11} M_{\odot}$ in the \vgn{} simulations (note that this approx. 10\% difference in halo mass persists when we use \textsc{ahf} on both simulations, but is reduced considerably to $\sim5$\% when considering DMO versions of the simulations). However, their growth histories at early times differ significantly, as is evident in Fig.~\ref{fig:haloGrowth}. The \EF{}, \XC{} and \LF{} haloes show significantly different rates of growth, with \EF{} growing rapidly at early times and having a quiet merger history after $z=4$ and \LF{} growing slower and through multiple significant mergers until $z=0.6$. The \XC{} halo has a growth intermediate between the two, having significant mergers until shortly after $z\approx2$ and a quiet merger history later. This is reflected in the halo formation times provided in Table~\ref{tab:centralProps}, defined as the time at which the haloes have assembled half their present-day mass. While the \XC{} halo forms at $t=4.0-4.3$~Gyr (corresponding to $z=1.6-1.5$), the \EF{} and \LF{} haloes form approximately 1~Gyr before and after respectively. The four GM haloes \XXC{}, \XVC{}, \XCX{} and \XCXX{} have similar growth histories to the \XC{} halo until $z=3$, when the modifications to the ICs result in varying merger histories, and again after $z\approx 1.5$, as designed. Finally, the \XCEC{} modified halo has a an earlier collapse time (by design) and its growth history largely matches that of the \EF{} halo. 

The different galaxy formation models result in markedly different baryonic content within the MW haloes in the two sets of simulations with the \tng{} haloes containing a factor of 1.8-3.4 more stellar mass and a factor of 1.4-3.3 more gaseous mass from individual simulations. The differences between the central galaxies themselves are even starker, not only in terms of content but also in size and structure. Fig.~\ref{fig:stellarMaps} shows mock stellar images of the \EF{}, \XC{} and \LF{} haloes, showing both the central galaxy and the surrounding satellites in both sets of simulations. Each image is $300 \times 300$~kpc in size and the colours represent luminosities in the Johnsons U, V, and I bands. A simple dust correction is applied using \textsc{pynbody}, following a \citet{Calzetti2000} law. The \tng{} central galaxies are always larger and more massive than the \vgn{} ones, as also evident from Table~\ref{tab:centralProps} and as we have explored in detail in \citet{Joshi2025}. The \vgn{} simulations have significantly more satellites visible in the region (this is also true at larger radii). The brighter \vgn{} satellites are also more concentrated than their \tng{} counterparts, as we discuss in later sections. We explore these differences in satellite abundance and properties and their implications, and the connection to host properties in the following sections.

\subsection{Satellite abundances} \label{sec:resultsSatNumbers}

\subsubsection{Present-day luminosity functions} \label{sec:resultsZ0LFs}

\begin{figure}
    \centering
    \includegraphics[width=\linewidth]{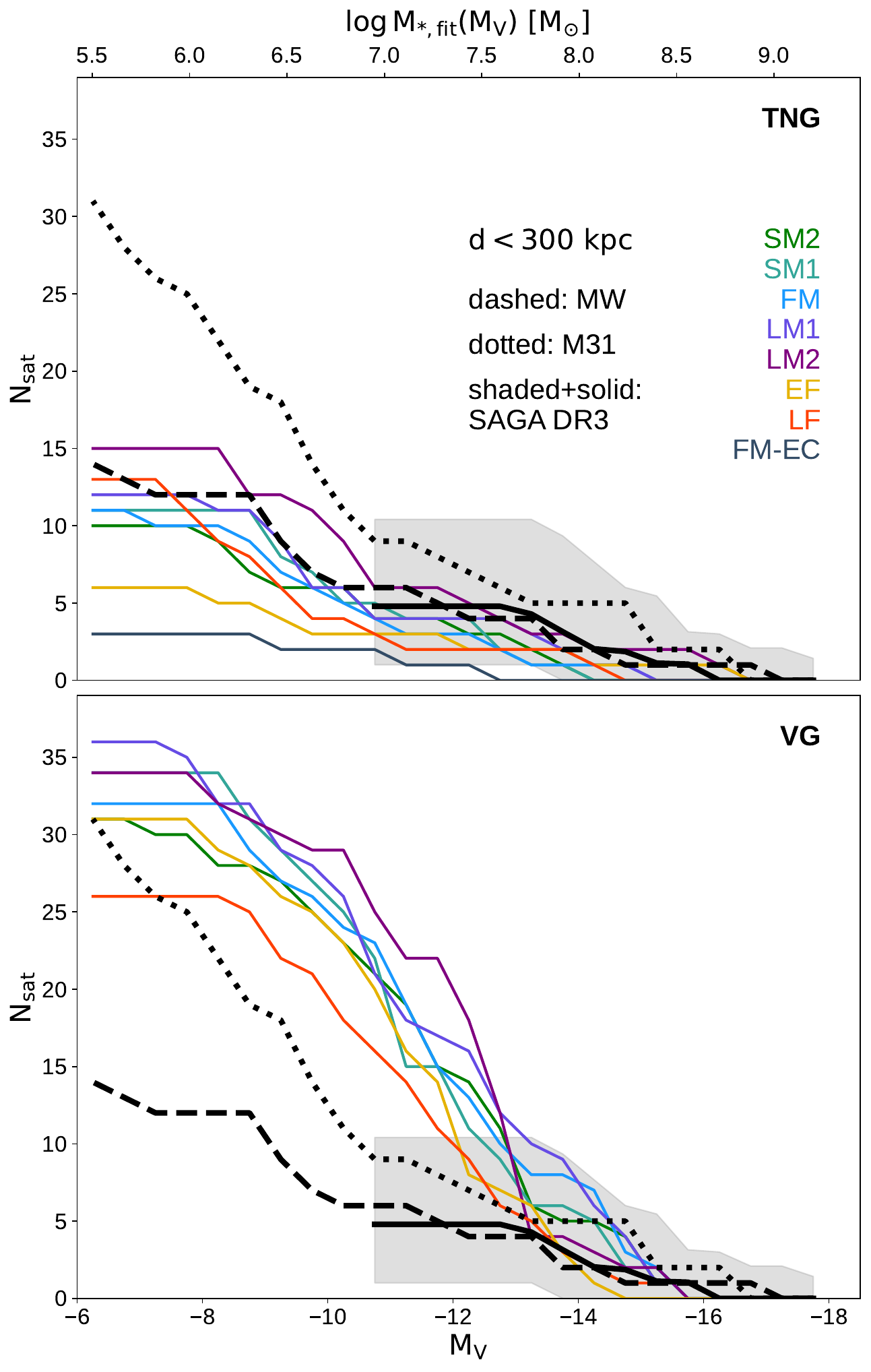}
    \caption{Luminosity functions (LFs) of satellites within 300 kpc of each simulation at $z=0$, indicated by the different colours. \emph{V}-band magnitudes were calculated within a radius $R_{\rm SB=30}$, where the \emph{V}-band surface brightness of the satellite is 30 mag/arcsec$^{-2}$. The corresponding stellar mass (according to a linear fit to our data) is indicated by the top x-axis. The \vgn{} simulations have significantly more satellites at all masses/luminosities compared to the \tng{} simulations. For comparison, the dashed and dotted black curves show the observed MW and M31 LFs from \citet{McConnachie2012}, while the shaded region and solid lines indicate the spread and median LFs of a matched sample from the SAGA Survey DR3 \citep{Mao2024}. The \tng{} LFs are in good agreement with the observed MW LF and within the spread of LFs from the SAGA survey, albeit with marginally fewer bright satellites than average. The \vgn{} simulations overpredict the number of satellites at all but the brightest magnitudes. This is found to be due to the formation of several satellites at early times that produce most if not all of their stellar mass through a single burst of SF (see Section~\ref{sec:resultsZ0LFs} for details).}
    \label{fig:satLFs}
\end{figure}

\begin{figure*}
    \centering
    \includegraphics[width=\linewidth]{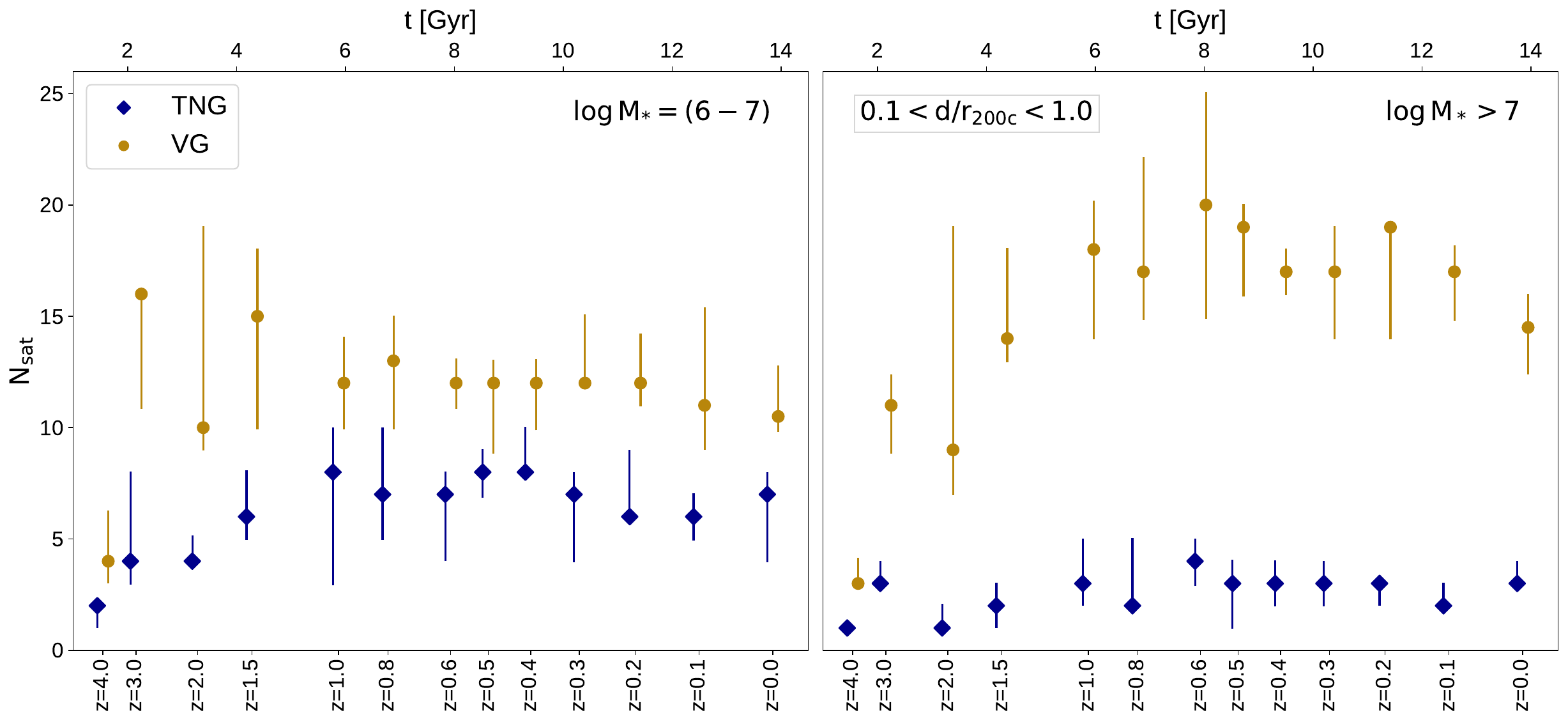}
    \caption{Numbers of satellites within $\RHOST$ as a function of epoch across all assembly histories. The data points show the median value for all simulations generated with the specified code (blue diamonds for \tng{} and gold circles for \vgn{}), while the corresponding errorbars show the 16$^{\rm th}$-$84^{\rm th}$ percentile ranges. The left and right panels show the results for low- and high-mass satellites respectively. Note that the \tng{}/\vgn{} datapoints have been shifted to the left/right by a small amount for clarity. The \vgn{} simulations have significantly higher numbers of satellites at all times compared to the \tng{} ones, and have more massive satellites than lower mass ones, while the opposite is true for the \tng{} satellites. In both sets of simulations however, there is an initial increase in numbers of satellites at early times between $z=4-1.5$, after which the total numbers of satellites remains approximately constant on average.} \label{fig:Nsats1Rvir}
\end{figure*}

We next examine the present-day luminosity functions (LFs) of the MW satellites. Fig.~\ref{fig:satLFs} shows the \emph{V}-band LFs of satellites within 300~kpc of the central MW for each individual simulation in our sample, where absolute magnitude $M_{V}$ is measured within $R_{\rm SB=30}$. The top axis shows the corresponding stellar mass based on an empirical linear fit between $\log{M_{*}}$ and $M_{V}$ for all our satellites at $z=0$. For comparison, we show the observed LFs of the MW and M31 satellites from \citet{McConnachie2012}. Note that we have not applied any completeness corrections to the MW/M31 data as we only consider the classical dwarf regime here. We also show LFs from the SAGA Survey DR3 \citep[][referred to as SAGA-DR3 hereafter]{Mao2024}. To compare with our results, we select hosts in the SAGA-DR3 sample with $M_{\rm halo}=7.5\times10^{11}-1.5\times10^{12}\,\MSUN$ and with at least one confirmed satellite in either their Gold ($M_{*}>10^{7.5}\MSUN$) or Silver ($M_{*}=10^{6.75-7.5}\MSUN$) samples. We further weight the satellites of each host by the incompleteness correction $f_{\rm corr}=N_{\rm sat,corrected}/N_{\rm sat,confirmed}$ for the Gold and Silver samples separately. Finally, we convert the reported \emph{r}-band magnitudes to \emph{V}-band following the prescription from \citet{Jesters2005}. The figure shows the spread of LFs of the SAGA hosts in the shaded region, and the median with the black solid curve.

The \tng{} satellite LFs, especially for the \XC{} and four GM simulations, show good agreement both in shape and normalization with the observed LF for the MW and SAGA-DR3 MW-mass hosts at all luminosities, albeit with slightly fewer bright satellites than the MW or the median for SAGA-DR3. The large-volume IllustrisTNG50 simulation has also been shown to produce satellite stellar mass-functions (SMFs) around MW-mass hosts that are broadly consistent with the observed MW satellite SMFs, albeit with considerable scatter \citep{Engler2021}.

In the \vgn{} case, we find significantly more satellites at all magnitudes compared to observations. We have determined that this is largely due to an early phase of poorly regulated bursty SF. This occurs because in the current \vgn{} simulations, the combination of relatively lower resolution for the dwarfs regime and a high density threshold for SF allows a significant portion of the gas in these objects, which is extremely metal poor, to become eligible for SF simultaneously at $z\sim8$. At later times, SF becomes more regulated when stellar feedback is more effective. These early starbursts occur in most haloes, but their impact is largest for haloes of mass $M_{\rm tot}\lesssim 10^{9}\MSUN$ and negligible at $M_{\rm tot}\gtrsim 10^{10.5}\MSUN$. In the lower mass haloes, these starbursts can form most if not all their present-day stellar mass, resulting in $\sim10^{7}\MSUN$ stellar mass galaxies in $\sim10^{9}\MSUN$ haloes at $z=0$, which should instead host ultra-faint dwarfs. On the other hand, more massive galaxies continue to form stars at later times, such that these early-forming stars are a significantly smaller contribution to their present-day stellar mass. This issue results in a higher SMHM relation for the \vgn{} simulations, as we discuss in detail in Section \ref{sec:smhmRelations}. (Note however that this is not seen in the EDGE simulations of isolated dwarfs which employ a nearly identical model at much higher resolutions and find SMHM relations more consistent with observations \citep{Rey2019,Kim2024,Rey2025}.) We explore this topic in detail, including how to account for this early SF and its relation to satellite abundance and quenching, in an upcoming paper (Rodriguez-Cardoso et al., in prep.). For the current paper, we retain all satellites and examine their fate in the MW-mass hosts since our goal is to understand the trends within strongly differing simulation models at face value, rather than to obtain precise matches to observational constraints. However, we discuss how this issue impacts some of our key results in Appendix \ref{sec:appVGFiltering}.

\subsubsection{Evolution of satellite abundances} \label{sec:resultsNSatEvol}
In Fig.~\ref{fig:Nsats1Rvir}, we show the numbers of satellites at each epoch for the two sets of simulations (blue diamonds for \tng{}, gold circles for \vgn{}). The datapoints shown are the median value for all simulations in the set, while the errorbars indicate the $16^{\rm th}-84^{\rm th}$ percentile ranges. We show the numbers for low- ($10^{6}\geq \MSTAR/\MSUN<10^{7}$) and high-mass ($\MSTAR=>10^{7}\MSUN$) in the left and right panels respectively.

The differences in absolute numbers of satellites discussed above at $z=0$ persist at all cosmic times. Despite this, there are common trends evident between the two sets of simulations. The numbers of satellites initially increase between $4<z<1.5$. After this time, the number of satellites is seen to be approximately constant or showing a mild decrease. Given that the MW host is expected to continue accreting substructure throughout cosmic time, these results indicate a balance between (i) new satellites being brought into the host, and existing satellites either (ii) travelling outside the virial radius or (iii) being disrupted at comparable rates. In the following sections, we explore to what extent satellites are disrupted, over what timescales, and what the impacts of using the two different simulations codes are. 

\begin{figure*}
    \centering
    \includegraphics[width=\linewidth]{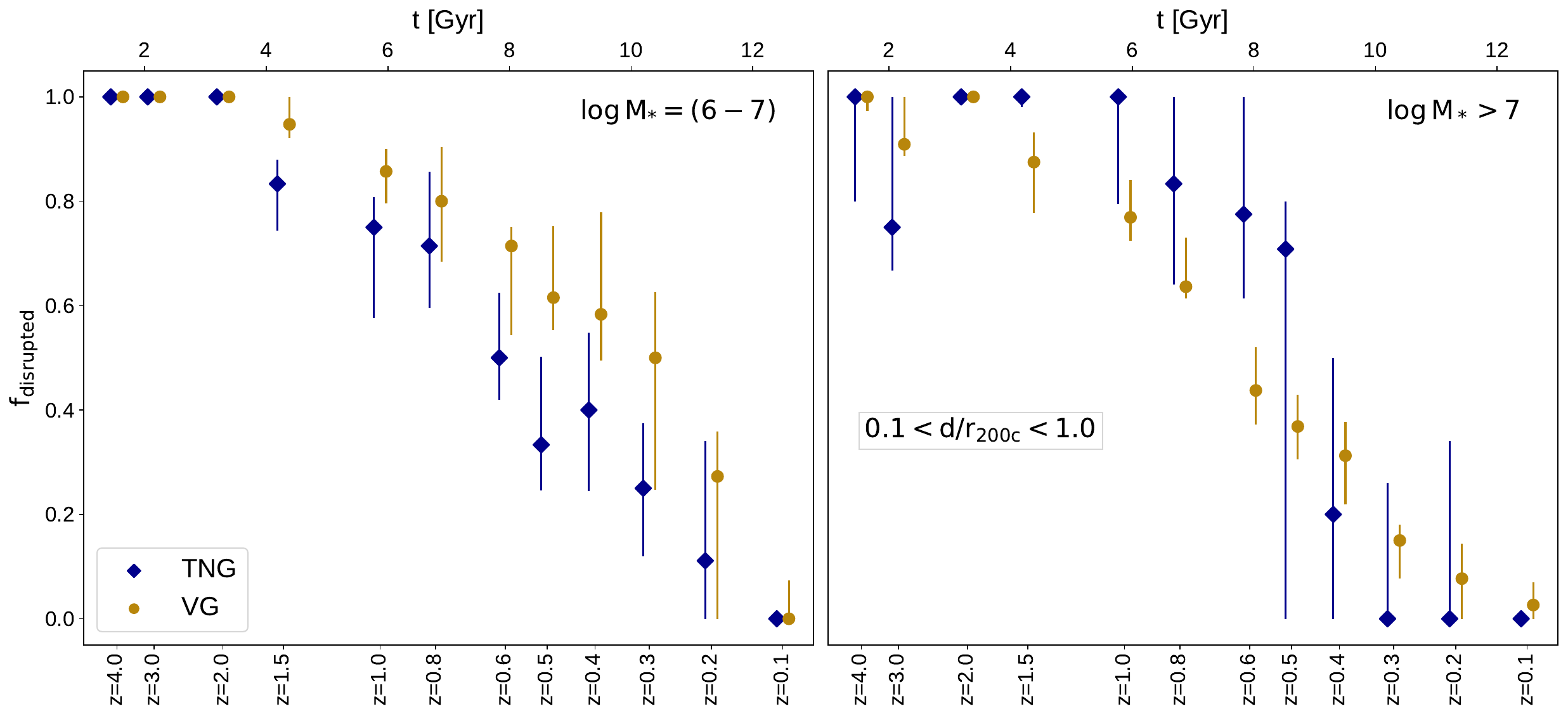}
    \caption{Fraction of satellites at a given epoch that have been disrupted by $z=0$, across all assembly histories. Here we define the time of disruption as the time when $R_{50}/R_{\rm 50,ini}>5$, where $R_{50}$ is the stellar 3D half-mass radius and $R_{\rm 50,ini}$ is this radius at the selection epoch. Colours and symbols are as in Fig.~\ref{fig:Nsats1Rvir}. Over 70 per cent, if not all, satellites from $z\geq1$ ($\gtrsim 8$~Gyr ago) have been disrupted by $z=0$, regardless of mass and in both sets of simulations. After $z=2$, this fraction decreases gradually, to less than 30 per cent by $z=0.2$ i.e. $\sim 2.5$~Gyr ago. High mass satellites are less likely to disrupt compared to low-mass ones and \vgn{} low mass satellites are more likely to be disrupted than \tng{} ones, with the exception of high-mass satellites at $z>0.5$, where both these trends are reversed. The decreasing disruption fractions after $z\sim1$ are to be expected due to the increasingly limited time available, assuming the disruption timescales are comparable.} \label{fig:FDisrupt1Rvir}
\end{figure*}

\begin{figure*}
    \centering
    \includegraphics[width=\linewidth]{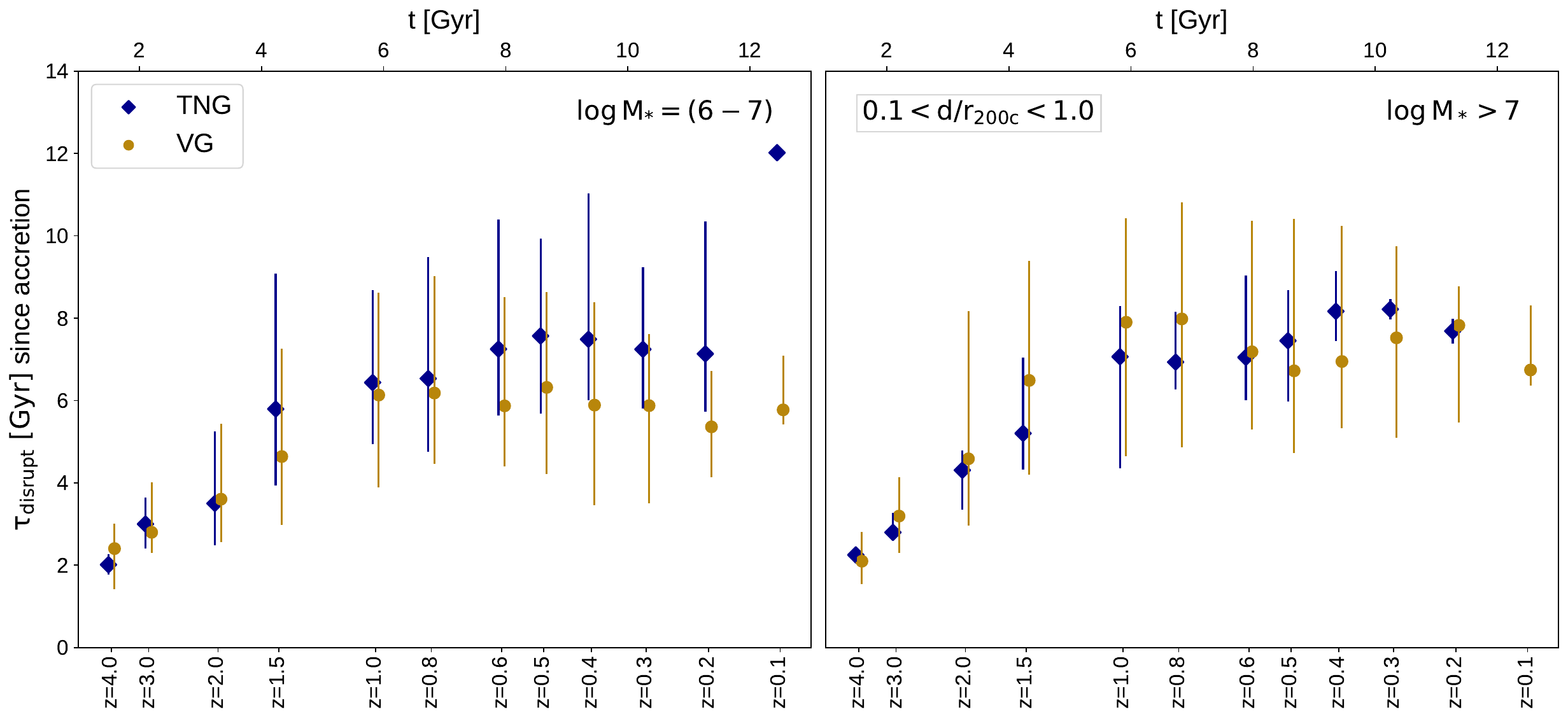}
    \caption{Disruption timescales of satellites since their accretion across all assembly histories. Only satellites that have been disrupted by $z=0$ are included and we combine satellites from all simulations from each set for improved statistics. The datapoints show the median timescale, while errorbars show the $16^{\rm th}-84^{\rm th}$ percentile ranges. Both low- and high-mass satellites have nearly identical distributions and median disruption timescales between the two sets of simulations, which remain nearly constant for the majority of cosmic time between $z=1.5-0.1$ at $6-8$~Gyr, with shorter timescales at earlier times consistent with the shorter dynamical times at these epochs. \tng{} timescales are longer by $\sim1-1.5$~Gyr than \vgn{} for low-mass satellites, while the difference for high-mass satellites is $\pm 1$~Gyr at various epochs.} \label{fig:TDisruptMeanAcc1Rvir}
\end{figure*}

\subsection{Satellite disruption}

\subsubsection{Disruption fractions} \label{sec:resultsDisFracs}

In Fig.~\ref{fig:FDisrupt1Rvir}, we show the fraction of satellites selected at a given epoch which have been disrupted by $z=0$ according to the definition given in Section~\ref{sec:methodsSample}. As with Fig. \ref{fig:Nsats1Rvir}, the datapoints are the median over all assembly histories, and the errorbars indicate the $16^{\rm th}-84^{\rm th}$ percentile ranges. The figure shows that over 70 per cent, if not all, of the satellites in the sample before $z=1$ have been disrupted by $z=0$, in both sets of simulations. At later times, the disruption fractions gradually decrease to less than 30 per cent by $z=0.2$. This decrease is to be expected as there is increasingly limited time available for satellites to be disrupted at later epochs, assuming that the disruption timescale is not too short. We explicitly measure the disruption timescale in the following section, but these results already provide broad limits on the overall disruption timescales for satellites within the virial radius of a MW-mass host of $2.5\ \text{Gyr}\lesssim \tau_{\rm disrupt} \lesssim 8\ \text{Gyr}$. 

The overall trends are remarkably similar between the two simulations, although there are some noticeable differences. Generally, low-mass satellites are more susceptible to disruption than high-mass ones, and \vgn{} satellites are more likely to be disrupted than \tng{} ones by factors of up to 2. Coupled with the higher numbers of \vgn{} satellites (particularly high-mass ones) as seen in the previous section, this accounts for the difference in overall numbers of satellites between the two simulations. The exception to these trends are the high-mass \tng{} satellites at $z>0.5$, which show higher disruption fractions than either their \vgn{} or low-mass \tng{} counterparts at these epochs. However, it is important to keep in mind throughout the paper that the \tng{} high-mass results are particularly affected by low-number statistics. We discuss the reasons for the differences between \tng{} and \vgn{} in detail in Section~\ref{sec:smhmRelations}. Furthermore, we discuss how these results may change if we were to mitigate the impact of the early bursty SF in the \vgn{} model, as well as the results of Sections \ref{sec:resultsNSatEvol} and \ref{sec:resultsDisTimes} in Appendix \ref{sec:appVGFiltering}.

\begin{figure}
    \includegraphics[width=\linewidth]{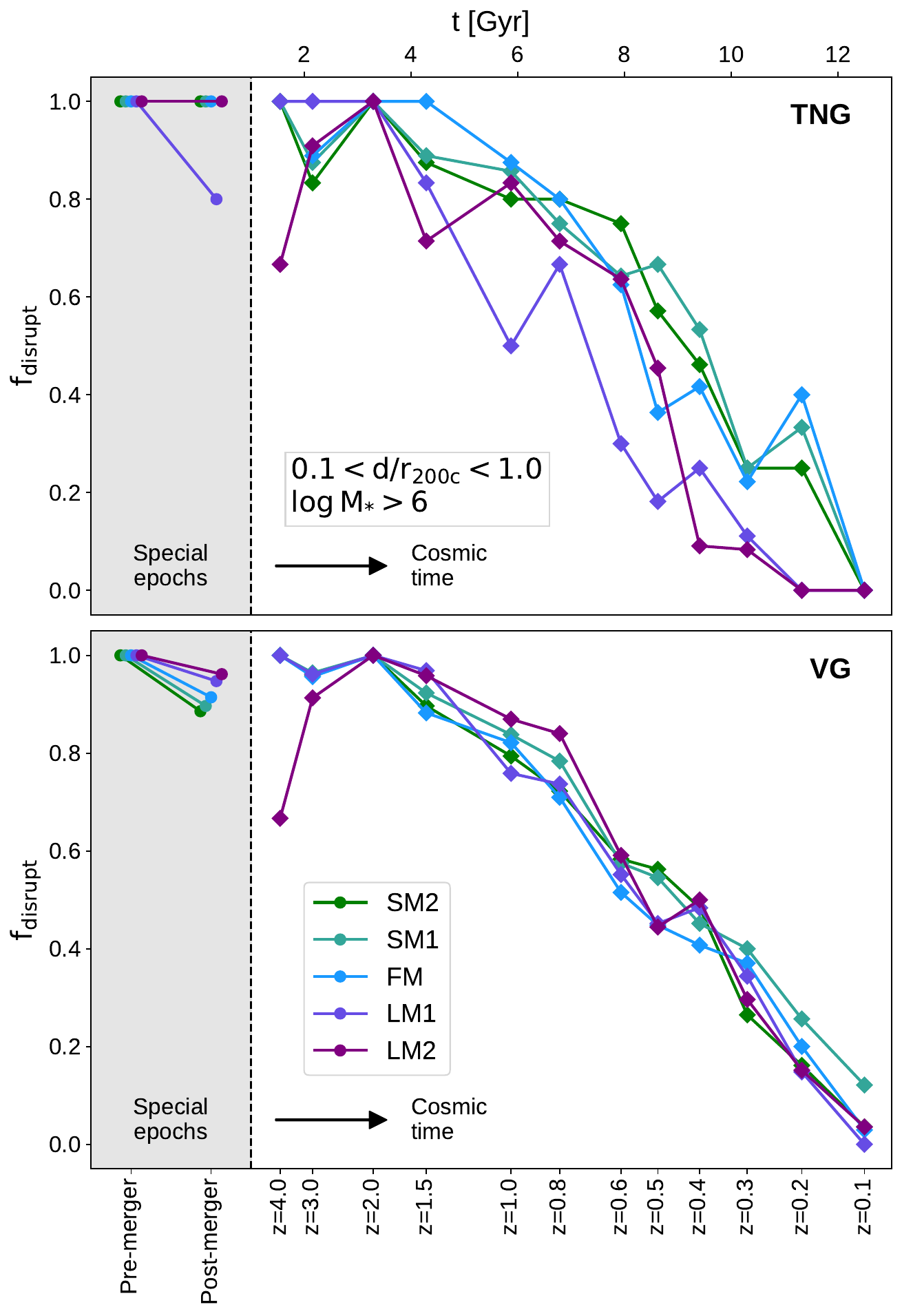}
    \caption{Fraction of all satellites at a given epoch that have been disrupted by $z=0$ for each of the five GM simulations in the \tng{} (top) and \vgn{} (bottom) sets. In addition to the selected redshifts, we also show two significant epochs, before and after the \targetMerger{} merger that is targeted by the GMs, to the left of each panel, marked as the grey region. These datapoints have been shifted by a small amount in the x-direction for clarity. `Pre-merger' is defined as the timestep before the secondary galaxy in the merger enters the virial radius of the primary host. `Post-merger' is defined as the time at which the secondary galaxy coalesces with the primary according to the respective halo finders. There is no significant impact on the disruption fractions from the altered merger histories.} \label{fig:fDisruptMergerHistory}
\end{figure}

\subsubsection{Disruption timescales} \label{sec:resultsDisTimes}

The previous results already provided approximate bounds for satellite disruption timescales. In this section, we directly compare the time intervals over which the satellites are disrupted in order to understand the differences and similarities between the two models. Satellites selected at a given epoch may have been accreted at varying times before this epoch. Hence, to predict how long a satellite can survive inside a MW host before being disrupted, we must account for their accretion times. In Fig.~\ref{fig:TDisruptMeanAcc1Rvir}, we show the disruption timescales since the time of accretion for satellites selected at a given epoch. For each satellite, the overall disruption timescale is defined as the time between accretion and disruption (reminder that the selection epoch is not the same as the accretion time). Only satellites that have been disrupted by $z=0$ are included in the samples. We combine satellites from all merger scenarios for improved statistics, but will return to the effect of merger and assembly history in the following sections. The datapoints and errorbars indicate the median and $16^{\rm th}-84^{\rm th}$ percentile ranges of the stacked distributions. 

The figure shows that the median values, and even ranges of disruption times for low-mass satellites are nearly identical between \tng{} and \vgn{} until $z=0.8$, rising from 1-2~Gyr at early times to $\sim6$~Gyr. At later times, while the disruption timescales for \vgn{} low-mass satellites remain roughly constant at $\sim6$~Gyr, the \tng{} ones are roughly constant at $\sim7$~Gyr. This is consistent with the \tng{} low-mass satellites having slightly lower disruption fractions compared to the \vgn{} ones. High-mass satellites also show the same overall trend, with the timescales being similar until $z=2$, after which the \tng{} satellites have shorter disruption timescales (by $\sim1$~Gyr) until $z=0.6$, and then longer timescales or similar timescales at later times. This is again consistent with the \tng{} disruption fractions being higher than the \vgn{} ones until $z\sim0.5$ and then lower at later times. Note that the lower scatter for the \tng{} high-mass satellites is due to low-number statistics.

Overall, we find that both low- and high-mass satellites show a similar overall trend whereby they have short disruption timescales at early times ($z<2$), in line with the shorter dynamical timescales of the MW halo at these epochs, and therefore high disruption fractions. The disruption timescales gradually increase thereafter to $6-7$~Gyr for low-mass satellites ($7-8$~Gyr for high-mass ones) by $z\sim1$ and then remain constant, and the declining disruption fractions seen in Fig.~\ref{fig:FDisrupt1Rvir} simply reflect the amount of time available for the satellites to be disrupted. Mild differences in the disruption fractions between the \tng{} and \vgn{} models stem from differences in the full distribution of disruption timescales. We discuss the origin of these differences in Section~\ref{sec:discContributingFactors}.

\begin{figure}
    \includegraphics[width=\linewidth]{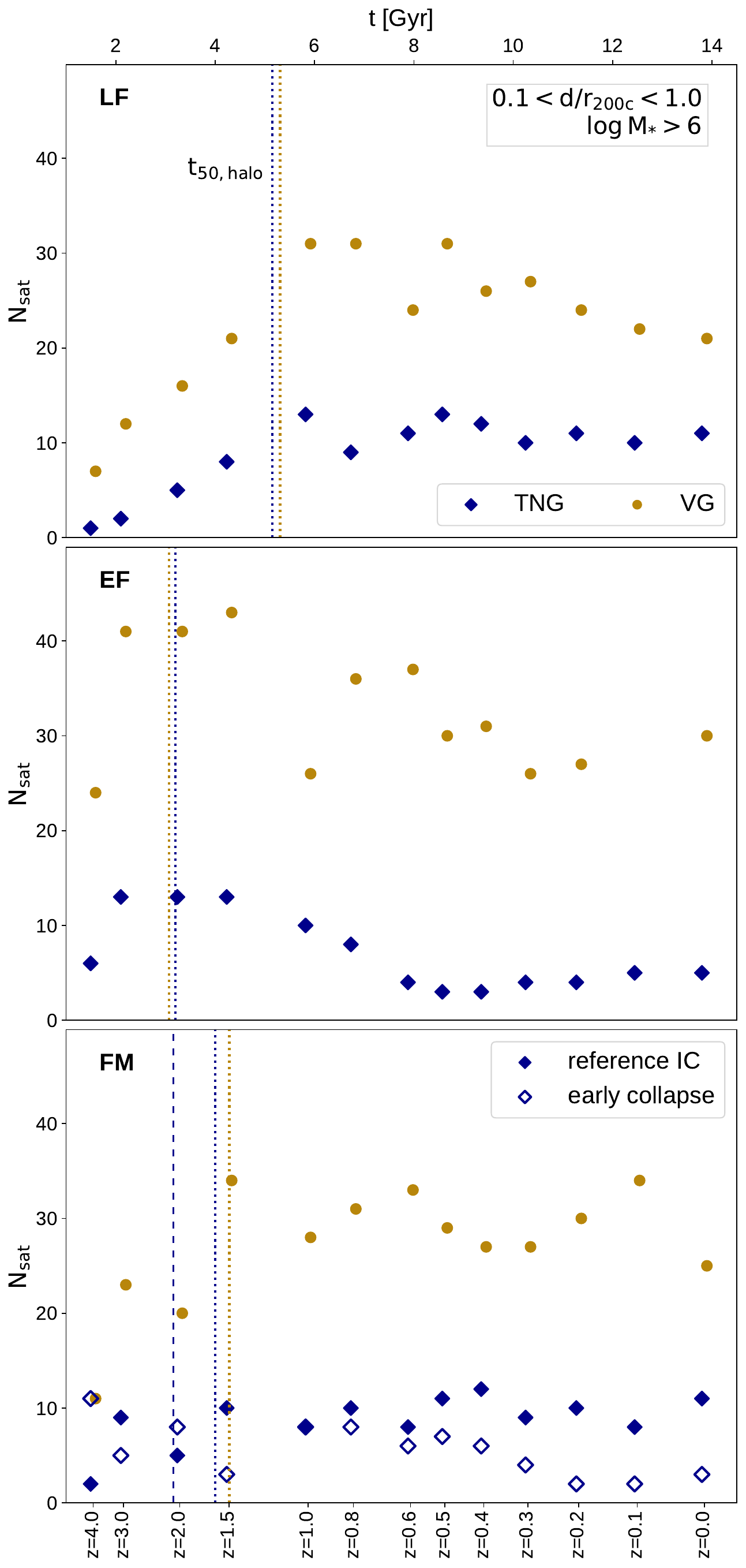}
    \caption{Numbers of all satellites as a function of selection epoch for the \LF{} (top panels), \EF{} (middle panels) and \XC{} (bottom panels) simulations separately. The bottom panels also show results for the \tng{} \XCEC{} simulation (open markers). The formation time of each host halo, $t_{50}$, is indicated by dotted lines (and dashed lines for \XCEC{}). The \tng{} simulations show a dependence of satellite abundance on halo formation time, with the early-forming \EF{} simulation having more satellites at early times and the late-forming \LF{} doing so at later times, and \XC{} having intermediate results. This is confirmed with the \XCEC{} simulation which was modified to have an earlier formation time, but in the same environment as the \XC{} run, and has results nearly identical to the \EF{} simulations. These trends are also present in the case of the \vgn{} satellites albeit with more scatter.} \label{fig:nsatCollapseTime}
\end{figure}

\subsection{Impact of merger and assembly histories} \label{sec:resultsHaloFormTimes}

\subsubsection{Do mergers affect satellite survival?}
It is possible that the merger history of the host halo can impact the results we have discussed above, primarily by determining the mass and other properties of the host halo that can affect satellite disruption. As a secondary effect, the strength and configuration of a merger could impact the orbits of satellites and thus also determine their survival. We have therefore considered the above results separately for the five GM simulations i.e. the fiducial \XC{} and the modified \XXC{}, \XVC{}, \XCX{}, \XCXX{}. As a reminder, the GMs were designed to modify the strength of a significant merger than occurs at $z\approx 2$, such that the mass of the secondary involved in the merger increases in order of \XXC{}, \XVC{}, \XC{}, \XCX{} and \XCXX{}; after this time the galaxy has a relatively quiet merger history. 

Fig.~\ref{fig:fDisruptMergerHistory} shows the disruption fractions for satellites at each epoch, separately for each GM simulation in the \tng{} (top panel) and \vgn{} (bottom panel) sets. We combine low- and high-mass satellites here, as we find no significant difference in the results when considering them separately. In addition to the selected redshifts considered above, we show the results for two important epochs before and after the \targetMerger{} merger targeted by the GMs, to the left of each panel marked as the grey region; `pre-merger' is defined as the timestep before the secondary galaxy enters the virial radius of the primary host and `post-merger' is defined as the time at which it coalesces with the primary central galaxy according to the halo finders. We find no significant trend between merger history and disruption fractions at any time.

The same is true for the impact on disruption timescales. The numbers of satellites do respond to the change in merger history, as we have shown for the \vgn{} simulations in \citet{Joshi2024}, and confirmed for the \tng{} simulations albeit with significantly lower normalization. However, this is a result of varying infall times for the satellites rather than a change in satellite disruption stemming from the modifications to the merger history. Thus we find no impact of merger history on satellite survival in our simulations for both galaxy formation models.

\subsubsection{Does halo formation time impact satellite survival?}
As well as mergers, overall halo formation time and assembly history can also impact the satellite populations of the host, with younger hosts expected to contain more galaxies than older ones \citep[e.g.][]{Bose2019,Fielder2019,Wu2022}. We therefore consider whether halo formation time impacts the preceding results. In Fig.~\ref{fig:nsatCollapseTime}, we again show the numbers of satellites, now for individual simulations \LF{}, \EF{} and \XC{}, and again with low- and high-mass satellites combined. As a reminder, the \XC{} ICs were selected to have a merger history similar to that of the MW, with a significant merger at $z \approx 2$ and a quiet merger history with no major mergers after this time. It has a halo formation time of 4.0~Gyr and 4.3~Gyr in the \tng{} and \vgn{} runs respectively (defined as $t_{50}$, the time at which the halo attains half of its present day halo mass). The \EF{} simulations have an earlier formation time of 3.2~Gyr and 3.1~Gyr respectively, and no significant mergers (of mass ratios 1:10 or greater) after $z\approx3$ in both runs. On the other hand, \LF{} forms later at 5.2~Gyr and 5.3~Gyr respectively and has several significant mergers until $z=0.7$. 

Fig.~\ref{fig:nsatCollapseTime} shows that the satellite abundances in the \tng{} simulations have a clear dependence on halo formation time. The numbers of satellites grow more gradually and are higher at later times in the late-forming \LF{} simulation; in the early-forming \EF{} simulation, the number of satellites rises rapidly at early times and then decreases to present-day. The numbers for the \XC{} simulations are intermediate between the two. To confirm that this is a causal connection, we performed a further simulation \XCEC{}, which evolved a modified version of the \XC{} ICs, altered to obtain an earlier collapse of the host halo similar to the \EF{} simulations. The modifications result in it not having a $z \approx 2$ merger while retaining the large-scale structure and environment of the \XC{} simulation, thereby mimicking the \EF{} merger and assembly history (with a halo formation time of 3.2~Gyr). The results for this simulation are also shown in the bottom panels of Fig.~\ref{fig:nsatCollapseTime} with open markers, which show an early increase in satellite numbers followed by a gradual decline, similar to what we find for the \EF{} simulation. Such a dependence of satellite abundance on halo formation time is also evident in the \vgn{} simulations, albeit with some scatter.

We omit plots considering the disruption fractions and disruption timescales for brevity. We found no significant trends in the satellite disruption fractions with halo formation time. The disruption timescales do show a mild dependence, with the \LF{} satellites having longer disruption timescales compared to the \EF{} ones by $\sim1-2$~Gyr at $z\gtrsim0.5$ (and \XC{} having intermediate values). However, the \XCEC{} satellites have the longest disruption timescales at these times, and so this does not appear to be a causal link. These results suggest that the dependence of satellite abundance on halo formation time is mainly due to early-forming haloes accreting proportionally more satellites at early times.


\begin{figure*}
    \centering
    \includegraphics[width=\linewidth]{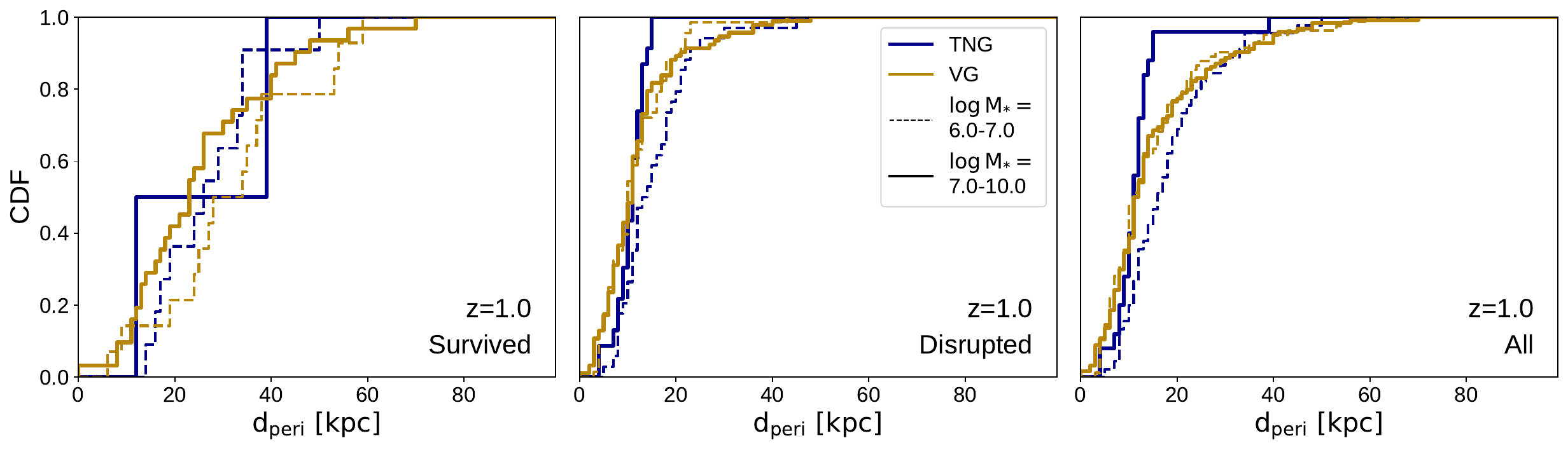}
    \includegraphics[width=\linewidth]{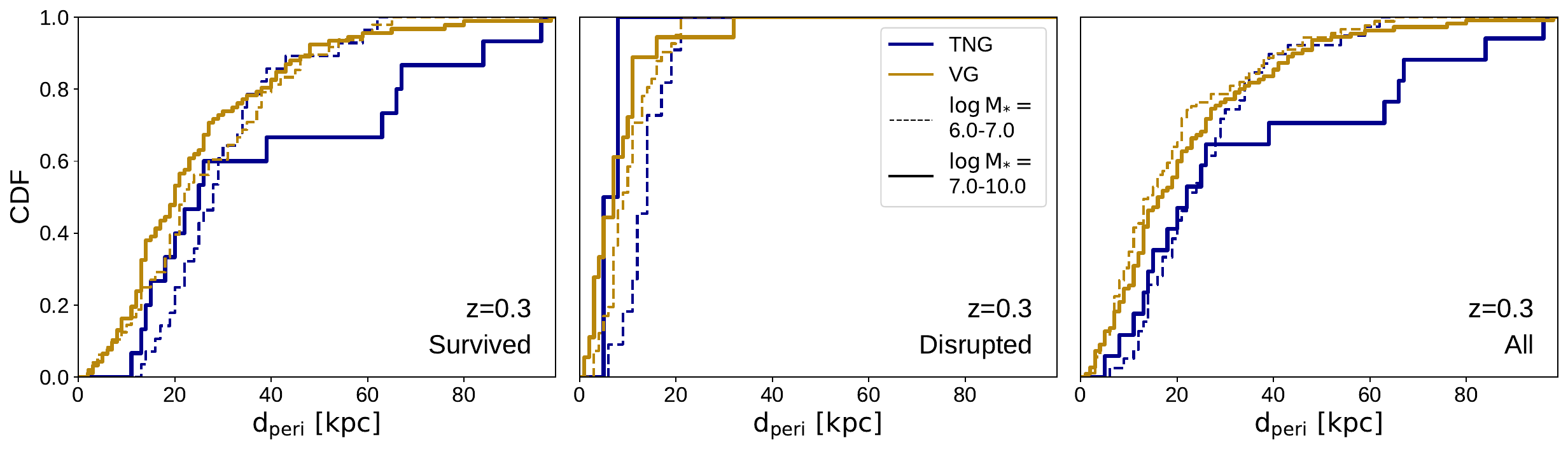}
    \caption{Distribution of pericentric distances for satellites selected at $z=1.0$ (top) and $z=0.3$ (bottom). Left and middle panels show surviving and disrupted satellites separately, while the right panels show all satellites. Surviving satellites have significantly larger pericentric distances compared to disrupted ones, indicating $d_{\rm peri}$ is an important factor in determining whether a satellite is disrupted. When considering all satellites together, low-mass \vgn{} satellites tend towards smaller pericentric distances at all times compared to \tng{}. The same is true for high-mass satellites at $z\leq0.4$, but at earlier epochs, \tng{} high-mass satellites have smaller pericentric distances than \vgn{}.}
    \label{fig:SurvivalVsPericentre}
\end{figure*}

\begin{figure*}
    \centering
    \includegraphics[width=\linewidth]{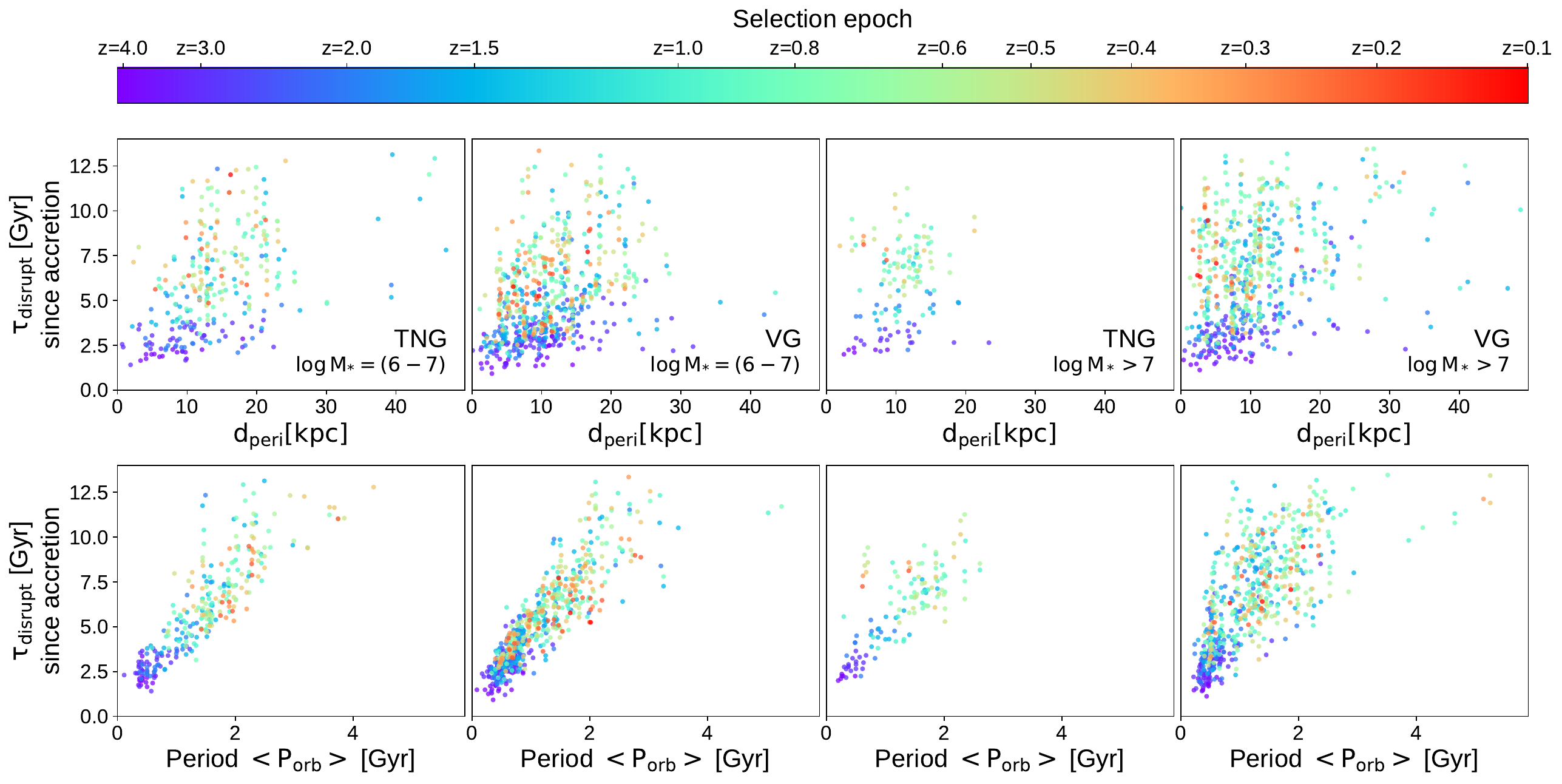}
    \caption{Disruption timescale since accretion as a function of pericentric distance (top) and orbital period (bottom) for disrupted satellites. Satellites selected at each epoch are shown with different colours (note that several satellites appear multiple times here as they can be included at more than one selection epoch). There is no significant correlation seen between disruption timescale and pericentric distance, apart from smaller $d_{\rm peri}$ at earlier epochs as expected. By contrast, orbital period and disruption timescale show a strong positive correlation. The correlation is marginally steeper, i.e. longer disruption timescales for a given orbital period, for \vgn{} satellites compared to \tng{} ones.}
    \label{fig:tauDisruptVsPericentre}
\end{figure*}

\section{Factors affecting satellite disruption} \label{sec:discContributingFactors}
In the previous sections, we find good qualitative agreement between the \tng{} and \vgn{} models in terms of disruption fractions and disruption timescales as a function of cosmic time, but there are noticeable quantitative differences. Here we explore possible origins for these differences by considering various factors that would boost or impede satellite disruption. These can be broadly subdivided into a few categories: 1) the orbits of the satellites that determine the environment they encounter, 2) their structure and other intrinsic properties that make them more or less resilient to disruption, 3) the structural properties of their hosts, which shape their interactions, and 4) the population of satellites available to be accreted and subsequently disrupted. Of these, (2) and (3) were found to be inconclusive, i.e.
although there are significant differences in the structural properties of the \tng{} and \vgn{} satellites and hosts, these properties do not play a major role in determining either the likelihood of satellites being disrupted or their disruption timescales. These results are shown in Appendix~\ref{sec:appDisrupt} for the interested reader. We discuss (1) and (4) in detail below.

\subsection{Orbital properties} \label{sec:discOrbitalProps}
The orbits of satellites can be an important factor in determining whether and how quickly they are disrupted in the host halo. They can be characterized in terms of their peri- and apo-centric distances, orbital eccentricities and periods, and the number of orbits undergone by the satellite between the time of accretion and disruption; we explore the impact of each on both disruption fractions and timescales. To determine when peri- and apo-centric passages occur, we use the positions of tracked particles as described in Sections~\ref{sec:methodsTrackForward} and \ref{sec:methodsTrackBackward} in 125~Myr intervals. The peri- and apo-centric distances are measured at the times when the motion of the satellite reverses direction. In the case of multiple such events, we use the minimum peri-centric and maximum apo-centric distance. For satellites on first infall, we measure the pericentric distance as the minimum distance it has reached. Eccentricities are calculated as 
\begin{equation}
    \epsilon = \frac{r_{\rm a}-r_{\rm p}}{r_{\rm a}+r_{\rm p}}
\end{equation}
where $r_{\rm a}$ and $r_{\rm p}$ are the distances for pairs of consecutive apo- and peri-centric passages, starting from the first peri-centric passage. For multiple orbits, we take the average value of eccentricity over all such pairs. Similarly, we measure the orbital period as the average of $2\times$ the time interval between consecutive peri- and apo-centric passages. Finally we calculate the number of orbits a satellite has completed as 
\begin{equation}
    n_{\rm orbits} = \frac{1}{2}\left(n_{\rm peri}+n_{\rm apo}-1\right)
\end{equation}
such that e.g. an orbit with a single pericentric passage is considered to be half complete. Note that we consider the entire time interval between accretion and disruption (for disrupted satellites) or present-day (for surviving satellites) in these measurements. 

\paragraph*{Impact on disruption fractions:}
In Fig.~\ref{fig:SurvivalVsPericentre}, we show the distribution of pericentric distances for satellites at two epochs, $z=1.0$ and $z=0.3$. The left and middle panels show the distribution for surviving and disrupted satellites (separately for low- and high-mass ones), while the right panel shows all satellites. The left and middle panels show that the surviving and disrupted satellites have markedly different distributions of pericentric distance. (KS tests indicate the distributions are different with at least 95\% confidence in nearly all cases except for the \tng{} high-mass satellites where the sample sizes are too small.) This indicates that pericentric distance is an important contributing factor in determining whether satellites are disrupted. We find similar differences in the distributions of orbital periods and apocentric distance (with $>99$ and $>95$\% confidence respectively in most cases), but no effects from orbital eccentricity and number of orbits (only the results for pericentric distance are shown for brevity). Disrupted satellites have substantially smaller pericentric distances, shorter orbital periods and are on marginally more eccentric (radial) orbits than surviving satellites. It should be noted that that several of these properties are correlated with each other, e.g. $d_{\rm apo}$ and $P_{\rm orb}$ are strongly correlated with each other, while there is a strong anti-correlation between $n_{\rm orbits}$ and both $d_{\rm apo}$ and $P_{\rm orb}$ -- i.e. longer orbital periods require larger apocentric distances and satellites with long orbital periods undergo fewer orbits, as expected. Similarly, $\epsilon$ is correlated with $P_{\rm orb}$ and anti-correlated with $d_{\rm peri}$ -- i.e. satellites on radial orbits have longer orbital periods and smaller pericentric distances, although with considerable scatter. It should also be noted that the pericentric distances shown here are relatively small; this is because we use the minimum pericentric distance (note that this still requires a true pericentric passage and is not simply the minimum distance ever reached). We have examined the same results when considering the distance of the first pericentric passage instead and find that while the distances are larger, the trends discussed above remain unchanged.

When we now consider \emph{all} satellites at a given epoch, comparing the distributions of e.g. pericentric distance (right panels of Fig.~\ref{fig:SurvivalVsPericentre}) between the two models and between low- and high- mass satellites can provide some explanation for the quantitative differences in disruption fractions seen in Fig.~\ref{fig:FDisrupt1Rvir}. At both times, low-mass satellites tend towards smaller pericentric distances in the \vgn{} simulations compared to \tng{}. In fact this is the case at all times after $z=2$. The same is true for high-mass satellites at $z=0.3$, but the opposite is true at $z=1.0$ i.e. the \tng{} satellites have shorter pericentric distances. We have considered the same results at other epochs and find the trend reverses for high-mass \tng{} satellites at $z=0.5$. These results are then consistent with the disruption fractions seen in Fig.~\ref{fig:FDisrupt1Rvir} -- low-mass \vgn{} satellites have higher disruption fractions than \tng{} at all times after $z=2$, correlated with smaller pericentric distances. High-mass \vgn{} satellites have lower disruption fractions than \tng{} at $2>z>0.5$, along with larger pericentric distances; at later epochs, both trends are reversed. Additionally, we find that \tng{} satellites of all masses undergo significantly fewer orbits than their \vgn{} counterparts at all epochs. While we see some differences in the distributions of $P_{\rm orb}$, $\epsilon$ and $d_{\rm apo}$ for all satellites at early epochs, there is little difference seen at later times and hence, we do not expect these properties to contribute significantly to the different disruption fractions between \tng{} and \vgn{}. Thus, satellites are more likely to be disrupted primarily if their orbits have small peri-centric distances, and secondarily if they go through several, shorter, less eccentric orbits rather than fewer, longer, more radial ones. The \tng{} satellites have, on average, smaller pericentric distances and undergo fewer orbits compared to the \vgn{} satellites, with the exception of high-mass satellites at $z\geq0.5$ which show the opposite trend. Preliminary analysis suggests that the different distributions of orbital properties between the two models are largely due to them populating a different subset of (sub)haloes with galaxies, as discussed in Section \ref{sec:smhmRelations}.

\paragraph*{Impact on disruption timescales:}
In Fig.~\ref{fig:tauDisruptVsPericentre}, we show the disruption timescale since accretion as a function of peri-centric distance (top panels) and orbital period (bottom panels), with colour indicating the epoch at which a satellite is selected. We find no significant correlation between disruption timescale and peri-centric distance. The trend of shorter disruption timescales at larger peri-centric distances is entirely due to satellites at $z\leq3$, when the relevant physical scales are smaller. On the other hand, there is a strong correlation between orbital period and disruption timescales, which is perhaps to be expected -- satellites with long orbital periods spend substantial periods of time in the host halo while minimizing close encounters with the central galaxy during which the disruptive tidal forces are strongest. We find no significant correlation with other properties i.e. $d_{\rm apo}$, $\epsilon$ or $n_{\rm orbits}$.

At early epochs ($z\geq2$), the disruption timescales are short and nearly identical between \tng{} and \vgn{} as seen in Fig.~\ref{fig:TDisruptMeanAcc1Rvir}. The differences seen at later epochs have different origins for low- and high-mass satellites. The correlation between $P_{\rm orb}$ and $\tau_{\rm disrupt}$ is similar between \tng{} and \vgn{} for low-mass satellites. The difference in the disruption timescales between the two sets of simulations in Fig.~\ref{fig:TDisruptMeanAcc1Rvir} are primarily due to the \vgn{} low-mass satellites having shorter orbital periods on average than \tng{} ones. In the case of high-mass satellites, the \tng{} correlation is shallower than the \vgn{} one, i.e. \tng{} satellites have shorter disruption timescales for a given orbital period compared to \vgn{}. At the same time, the \vgn{} satellites again have shorter orbital periods on average. The former trend would drive the median \vgn{} disruption timescales to be longer, while the latter would drive them to be shorter. Thus we see at $1.5>z>0.2$, the two competing factors, coupled with low-number statistics, can result in higher or lower disruption timescales for \vgn{} versus \tng{} at any given epoch, as is seen in Fig.~\ref{fig:FDisrupt1Rvir}.

\subsection{Normalization of satellite LFs} \label{sec:smhmRelations}

\begin{figure}
    \centering
    \includegraphics[width=\linewidth]{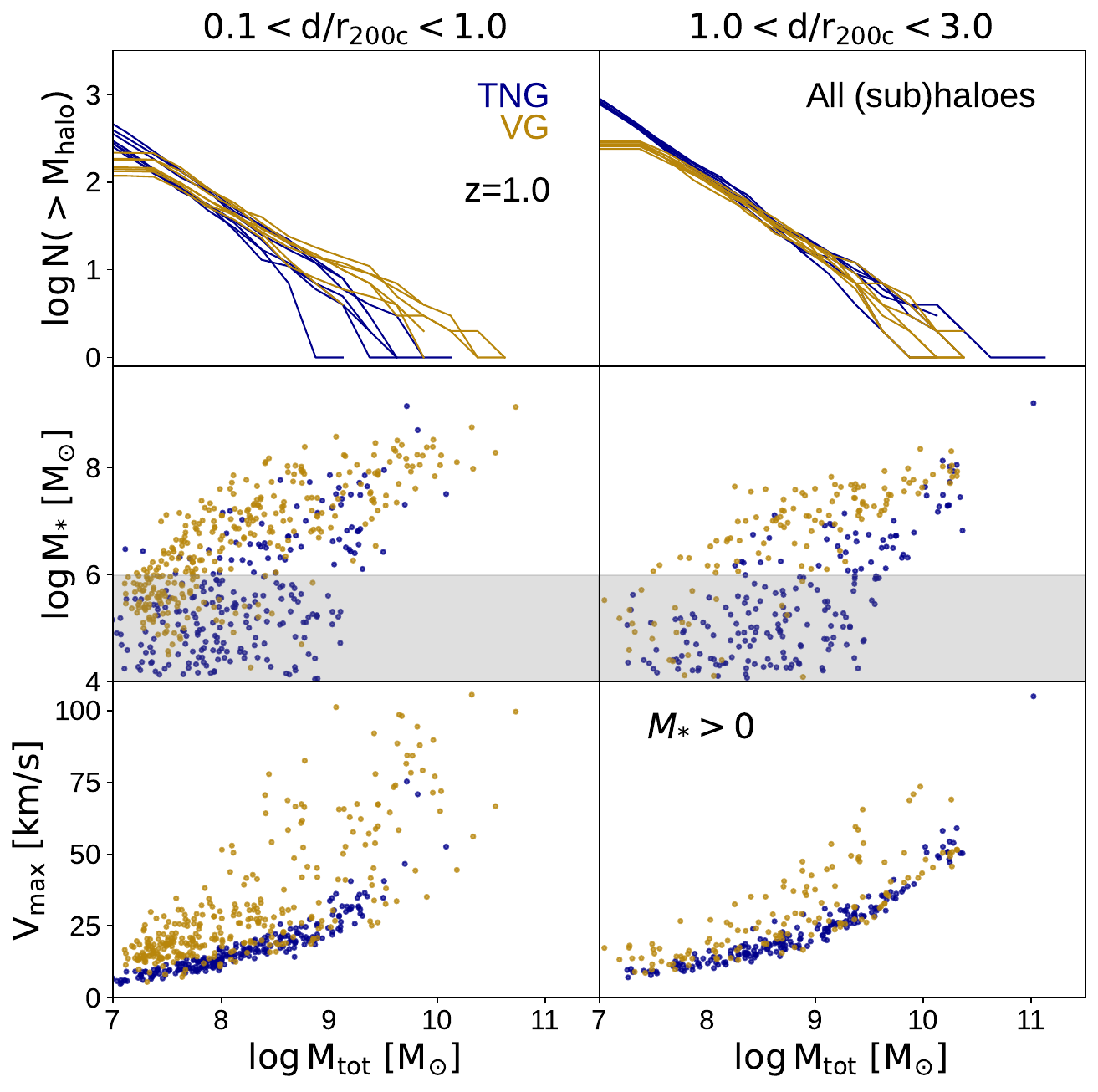}
    \includegraphics[width=\linewidth]{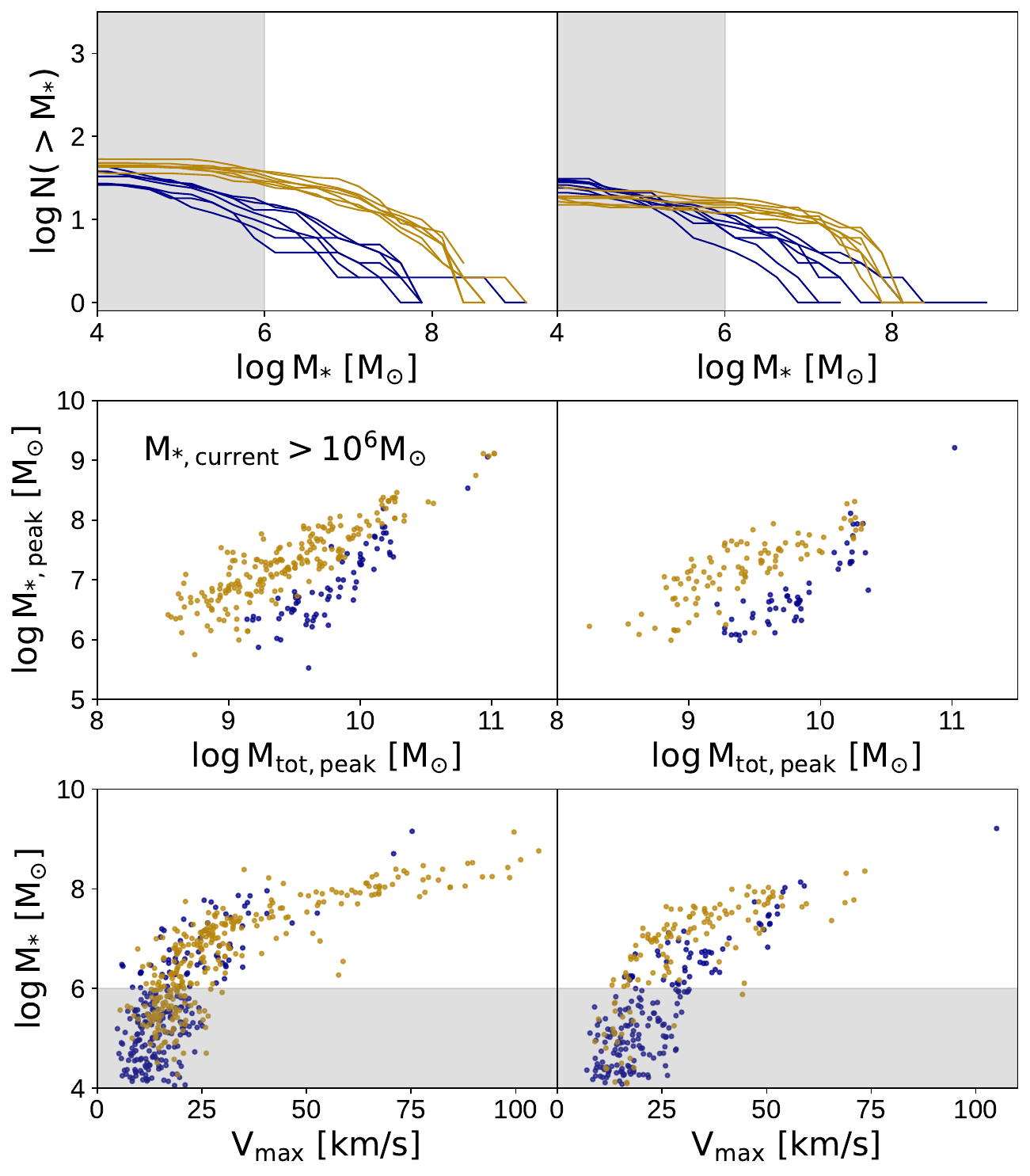}
    \caption{Satellite mass functions and scaling relations for satellites within $\RHOST$ (left) and infalling satellites within $(1-3)\,\RHOST$ at $z=1$. \emph{Row 1}: Halo MFs for all (sub)haloes in the simulations. Each curve represents an individual simulation. \emph{Row 2}: SMHM relation. \emph{Row 3}: Maximum circular velocity as a function of halo mass. \emph{Row 4}: Stellar MFs. \emph{Row 5}: SMHM relation considering peak halo mass and stellar mass at the time of peak halo mass, with the selection criteria in Section~\ref{sec:methodsSample}. \emph{Row 6}: Stellar mass as a function of maximum circular velocity. Grey shaded regions indicate stellar masses below our lower mass limit. Both \tng{} and \vgn{} predict similar HMFs but the \vgn{} model populates these subhaloes with significantly more stellar mass at a given halo mass. This is also true for $\MSTAR$ at peak halo mass, although this relation converges at $M_{\rm tot,peak}>10^{10.5}\MSUN$. The \vgn{} model also predicts higher circular velocities by 30-60~km/s for a given halo mass, indicating more concentrated mass profiles compared to the \tng{} satellites.}
    \label{fig:smhmRelations}
\end{figure}

The abundance of satellites around MW hosts at later epochs may also be impacted by the availability of galaxies to become satellites at earlier epochs, which in turn is dictated by the SMFs, HMFs and SMHM relations in the simulations at early times. If the two galaxy formation models populate DM haloes differently, they will naturally lead to different galaxy/satellite SMFs, despite having similar DM HMFs. We have therefore compared the HMFs, SMFs, SMHM relations between the two \XC{} simulations in detail. We also consider $V_{\rm max}$, the maximum of the circular velocity curve, which can be influenced by the overall structure of the satellites. We have confirmed that our results are not significantly impacted by the use of two different halo finders by comparing the \tng{} \XC{} satellite population using both halo finders (not shown). We found the HMFs, SMFs and SMHM relations from both to be virtually identical. Given this agreement between the two halo finders on a single simulation, we can continue to compare results from the two halo finders directly.

In the top row of Fig.~\ref{fig:smhmRelations}, we first show the HMFs of all (sub)haloes, including those with no stellar mass, within $\RHOST$ (left column) for each individual simulation at $z=1$. Since MFs and certain scaling relations can be substantially different between satellites and field galaxies, we also extend this analysis to infalling galaxies found at $(1-3)\,\RHOST$ (right column). For these results, we impose the same selection criteria as listed in Section~\ref{sec:methodsSample} with regards to contamination by low-res DM particles, excising the inner $0.1\,\RHOST$ region, minimum DM fraction and minimum halo mass, but do not require a minimum stellar mass. Instead, we indicate this stellar mass limit with grey shaded regions in the relevant panels of Fig.~\ref{fig:smhmRelations}. The HMFs are nearly identical, especially outside the virial radius, albeit with some scatter at halo masses $>10^{10}\MSUN$ (note that the large difference at the low-mass end is simply due to the two halo finders being configured to have different lower limits on the number of particles in the (sub)haloes). Row~4 of Fig.~\ref{fig:smhmRelations} shows the corresponding SMF, where there is a marked difference between the two sets of simulations -- the \vgn{} simulations have many more luminous galaxies with masses $M_{*}>10^{6}\MSUN$, but at lower masses, \tng{} is seen to `catch-up' by producing many more small luminous satellites (albeit that are not well-resolved). In fact, the number of satellites with $M_{*}>10^{4}\MSUN$ is nearly identical between the two sets of simulations, and even higher in \tng{} when considering infalling satellites. This suggests that both models are populating the same subhaloes, but with markedly different amounts of stellar mass. 
 
More explicitly, Row~2 of the figure shows the SMHM relations for these satellites, showing that the \vgn{} relation is indeed higher by $\sim1$~dex than the \tng{} one for both current satellites (which have been subject to environmental effects within the halo) and infalling satellites (which are more similar to field galaxies). We have confirmed that this is the case at all epochs. The effect is most prominent at $M_{\rm tot}<10^{10}\MSUN$, and marginally larger within the virial radius than beyond. This higher SMHM relation is partially due to the population of early-forming satellites that produce most of their stars in a single starburst as discussed in Section \ref{sec:resultsZ0LFs}. To confirm that these results are not simply due to tidal mass loss, Row~5 of Fig.~\ref{fig:smhmRelations} shows the SMHM relation by considering the peak halo mass of the satellites and the stellar mass at the time this peak halo mass is reached. Note that for these panels, we only consider the satellites that meet all our selection criteria. Here again, the \vgn{} satellites have stellar masses approximately 1~dex higher for a given halo mass compared to the \tng{} satellites for $M_{\rm tot,peak}<10^{10}\MSUN$. For more massive haloes however, this discrepancy is reduced such that there is no significant difference at $M_{\rm tot,peak}>10^{10.5}\MSUN$ ($M_{\rm *,peak}\gtrsim10^{8.5}\MSUN$). Thus, these differences in the SMHM relations, along with our imposed minimum stellar mass limit of $10^{6}\MSUN$ is responsible for the stark difference in satellite abundances in the two models.

We find that the difference in SMHM relation is the combined result of the $V_{\rm max}-M_{\rm tot}$ correlation and the $M_{*}-V_{\rm max}$ correlation, as shown in Rows 3 and 6 of Fig.~\ref{fig:smhmRelations}. For the infalling satellites, the \vgn{} satellites have $V_{\rm max}$ values that are up to 30~km/s higher than the \tng{} ones for a given halo mass, and $M_{*}$ up to 1~dex higher at constant $V_{\rm max}$. The maximum $V_{\rm max}$ values are similar between the two simulations. In the case of the current satellites, $V_{\rm max}$ for a given halo mass can be up to 60~km/s higher in \vgn{} than in \tng{}. However, unlike the infalling satellites, there is no significant difference in $M_{*}$ at constant $V_{\rm max}$ and the \vgn{} $V_{\rm max}$ distributions extend to higher velocities than the \tng{} ones. The higher values of $V_{\rm max}$ at fixed $M_{\rm tot}$ for the \vgn{} satellites are consistent with their DM density profiles being more concentrated, as seen in Fig.~\ref{fig:satDensityProfiles}. Note that the \vgn{} $M_{*}-V_{\rm max}$ correlation remains nearly the same between the current and infalling populations, and it is in fact the \tng{} correlation for infalling satellites moving to higher $V_{\rm max}$ values that reduces the difference between the two simulations. Hence the agreement in the $M_{*}-V_{\rm max}$ correlation between the two models for current satellites appears to be coincidental.

\subsection{What determines satellite abundances?} \label{sec:calcSatAbundances}
The sections above highlight the differences in the way the two galaxy formation models populate subhaloes with galaxies, which provides important context for the degree to which the satellites are disrupted. It is clear that the SMHM relations are vastly different between the \tng{} and \vgn{} models and that this dominates the difference in satellite abundances at late times. However, it is also important to understand any differences in how effectively they disrupt satellites. To do so, we predict the number of satellites that should exist in each model if no satellites were ever disrupted, and compare these to the true numbers seen in Fig.~\ref{fig:Nsats1Rvir}. If the true and predicted numbers -- and more importantly the ratio of the numbers from the two models -- differs, this would indicate that the two models disrupt satellites differently. 

We predict the numbers of predicted satellites as follows: a halo-occupation distribution (HOD) is first obtained by measuring the average number of low- ($10^{6}<M_{*}/\MSUN<10^{7}$) and high-mass ($M_{*}>10^{7}$) satellites associated with a halo of a given mass (in bins of 0.25 dex in halo mass), using the \textit{infalling} satellites at $z=3$. This HOD is then convolved with HMFs at later epochs to predict the number of low- and high-mass satellites there should be at these times, had there been no satellite disruption. Using infalling satellites at $z=3$ provides a statistically large enough population of galaxies while simultaneously ensuring that the galaxies have not been significantly influenced by environmental effects, thus making the HOD representative of the model.

In both \tng{} and \vgn{}, the predicted number of low-mass satellites are higher than the true number by a factor of 1.51 and 1.81 respectively, while the predicted number of high-mass satellites is lower than the true number by a factor of 0.96 and 0.80. This would suggest that within both simulations, low-mass satellites are more affected by disruptive processes. When comparing between the two simulations, without satellite disruption these calculations predict $N_{\rm sat,\vgn{}}/N_{\rm sat,\tng{}}=2.15$ for low-mass satellites, while the true proportion is 1.83, i.e. there are relatively fewer low-mass \vgn{} satellites than expected directly from the HMFs and SMFs. This implies that the \vgn{} low-mass satellites are indeed disrupted more, consistent with the higher \vgn{} disruption fractions seen in Fig.~\ref{fig:FDisrupt1Rvir}. This partially brings the \vgn{} and \tng{} satellite numbers into closer agreement than the SMHM relation would suggest on its own. 

In the high mass case, we predict $N_{\rm sat,\vgn{}}/N_{\rm sat,\tng{}}=6.14$ without disruption, while the true value is 6.62, which indicates the opposite trend i.e. there are relatively more \vgn{} satellites than expected, implying that they are less susceptible to disruption. In fact, if we consider the time between $z>2>0.5$ and $0.4>z>0.1$ separately, the true proportions measured are 7.62 and 6.05, reflecting the lower \vgn{} disruption fractions at $z\geq0.5$ and higher at later epochs. 

These simple calculations highlight that while the difference in satellite abundances are driven primarily by different SMHM relations in the two models, the degree and rate at which these satellites are eventually disrupted also plays an important role, reducing the differences in satellite abundances by up to 15 and 25 per cent. 


\section{Star-formation status of present-day satellites} \label{sec:resultsQuenching}

\begin{figure}
    \centering
    \includegraphics[width=\linewidth]{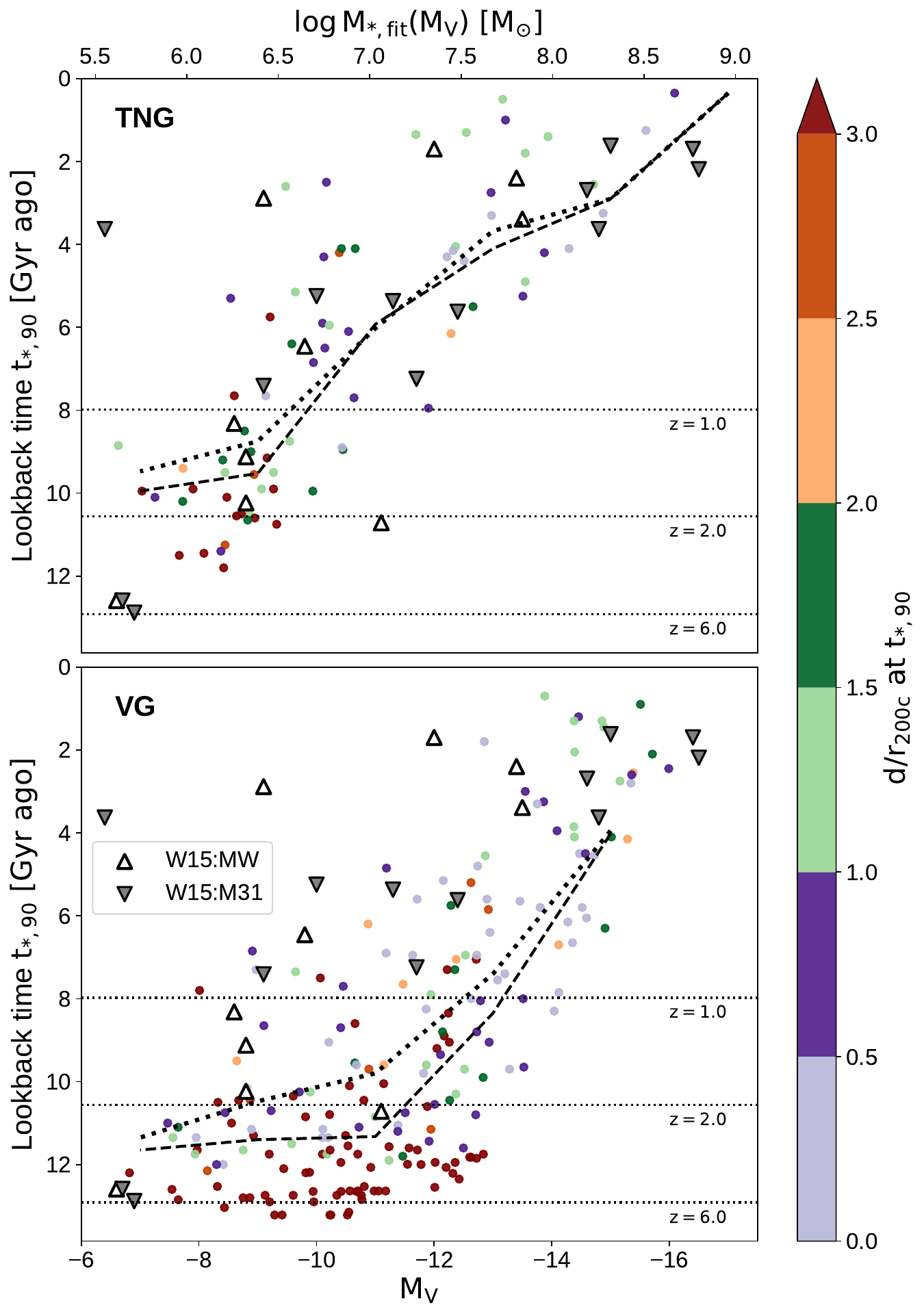}
    \caption{Quenching time as a function of absolute \emph{V}-band magnitude for satellites within 300~kpc at $z=0$. We use $t_{*,90}$, the time at which the satellite has assembled 90 per cent of its present-day stellar mass (based on stellar ages), as a proxy for quenching time. Colours indicate the distance of the satellite from the central MW at the time of quenching, normalized by the MW's virial radius at that time. The dashed curve indicates the median quenching time in bins of 2~mags, while the dotted curve is the median if we exclude satellites quenched beyond $1.5\,\RHOST$. We stack all simulations in each panel, but results for \EF{} are excluded since the \vgn{} run ends at $z=0.11$. For comparison, we show results from \citet{Weisz2015} for MW (open) and M31 (filled) satellites as triangle markers in both panels. In both sets of simulations, fainter satellites quench early, while brighter satellites quench later, qualitatively matching the observed trend for MW and M31 satellites. For satellites that quenched after $z=1$, the vast majority did so within $1.5\,\RHOST$ of the host. At earlier times however, the majority quenched beyond this radius. The \tng{} quenched more recently compared to the \vgn{} ones, making them quantitatively more consistent with the observations.} \label{fig:quenchTimes}
\end{figure}

Finally, we examine the star-formation status of the present-day surviving satellites in Fig.~\ref{fig:quenchTimes}, where we show the quenching times of all satellites in our sample within 300~kpc at $z=0$ from their host centre as a function of absolute \emph{V}-band magnitude. We measure SFHs as the normalized cumulative histogram of the stellar ages of the satellite, only considering particles within $R_{\rm SB=30}$. As is common practice, we use $t_{*,90}$, the time at which the satellite has assembled 90 per cent of its present-day stellar mass, as a proxy for its quenching time. The colours of the datapoints show the distance of the satellites to the central MW at the time of quenching, normalized by the MW's virial radius at that time. For comparison, we show the quenching times for observed satellites of the MW and M31 from \citet{Weisz2015} (with magnitudes and distances from \citealt{McConnachie2012}), who reconstruct the SFHs of each satellite from colour magnitude diagrams of resolved stellar populations. The authors note that while their sample is not complete due to the availability of archival data, it is reasonably representative in terms of stellar mass and distance from host centre. From the observational dataset, we include all satellites within $D_{\rm host}<300$~kpc from their respective hosts, consistent with common definitions for the extent of the MW halo.

Fig.~\ref{fig:quenchTimes} shows that while satellites from both sets of simulations follow the same trend of more recent quenching times for brighter satellites, qualitatively consistent with the observations, there are significant differences between the two galaxy formation models. On average, \tng{} satellites of all magnitudes (and therefore masses) quench more recently compared to the \vgn{} ones, by $\sim2$~Gyr for the brightest and faintest satellites ($M_{V}<-14$ and $M_{V}>-10$) and up to 5~Gyr at intermediate magnitudes, as shown by the median quenching times (dashed curves). The \tng{} quenching times are quantitatively consistent with observations for MW and M31 satellites. Furthermore, while most massive satellites were quenched within $1.5\,\RHOST$ in both sets of simulations, low-mass satellites were mostly quenched outside the host. In the \vgn{} case, however, we find a small but significant portion of low- and intermediate-mass satellites that were quenched early, but within the host halo (purple and light green markers). Additionally, \tng{} median quenching times are not significantly changed when we exclude satellites that were quenched beyond $1.5\,\RHOST$ (dotted curves), while the \vgn{} median times are changed to be up to 1~Gyr later (i.e. more recent).

Another key difference is the presence of a significant population of satellites in the \vgn{} simulations with $M_{V}>-13.5$ (corresponding to $\log{M_{*}}<7.8$) that were quenched at very early times i.e. $z=2-10$ and well outside ($>3\,r_{200\rm c}$) the MW progenitor's host halo. While their quenching time suggests they could be quenched due to reionization, they are too massive to have their SF significantly suppressed by reionization \citep[e.g. see][]{RodriguezWimberly2019}. Instead, these are the early-forming galaxies that undergo rapid starbursts at high redshifts as discussed in Section~\ref{sec:resultsZ0LFs}. The absence of such a population in the \tng{} simulations on the other hand is because the model does not form satellites in these haloes due to its higher temperature floor of $10^{4}$~K. These results show that the two models have significant differences in their modelling of the ISM and SF. Combined with differing resolutions in our simulations, this results in the \vgn{} satellites being quenched earlier than observed for satellites of the MW and M31. 

Understanding the differences in quenching times between the two simulations requires an exploration of the various processes, both secular and environmental, that lead to the end of SF in these satellites. Several simulation studies, using a range of galaxy formation models, highlight the importance of ram-pressure stripping in quenching satellites around MW-mass systems, particularly for those of masses $\MSTAR\sim 10^{6-8}\MSUN$ \citep[e.g.][]{Simpson2018,Akins2021,Engler2021}. Furthermore, \citet{Samuel2023} show that impulsive ram pressure is more efficient at quenching satellites than smoothly varying ram pressure. Such an exploration of the particular quenching pathways of these satellites is beyond the scope of the current paper, but will be examined in detail in an upcoming paper (Rodriguez-Cardoso et al., in prep.), which will also include a thorough investigation of the impact of the early bursts of SF in the \vgn{} model.


\section{Comparison to previous results} \label{sec:discussion}

When comparing this work with previous studies, it is important to bear in mind that definitions for the disruption timescale will differ. Our definition of satellite disruption assumes that as satellites are either destroyed by tidal forces or merge with the central MW, their stellar particles are dispersed to large relative distances. We chose this definition to minimize the uncertainty in tracking the satellites stemming from the halo finders. The absolute value of the disruption timescales presented in this study are naturally sensitive to the precise definition used i.e. $R_{\rm 50}/R_{\rm 50,initial}>5$. Using a different critical radius e.g. $R_{\rm 20}$ or $R_{\rm 80}$ changes the average disruption timescales by up to $+2$~Gyr and $-1$~Gyr respectively, as the more bound material in $R_{\rm 20}$ takes longer to be dispersed and the more unbound material within $R_{\rm 80}$ is easier to disperse. Similarly, changing the critical factor $R_{\rm 50}/R_{\rm 50,initial}$ from 5 to 10 or 2 corresponds to an increase/decrease in the disruption timescale by $\sim1$~Gyr respectively. These differences are more prominent for the more massive satellites in our sample. However, we have verified that the trends we have shown are independent of the precise definition we use. 

Methods relying solely on halo finders usually define satellite disruption as the time at which a (sub)halo no longer meets their identification criteria. However, such satellites may still exist for a short time after this as a small remnant, whether it be in the process of merging with the central or while being destroyed by tidal forces. Our requirement of particle dispersion aims to mimic the end result of these processes and thus account for this additional time. Therefore, the disruption timescales we report are expected to be longer than those presented in other studies that define disruption based solely on the halo finders/merger trees.

Another factor to consider when examining the disruption fractions and timescales of the satellites is the role of artificial disruption. Studies such as \citet{vanDenBosch2018} and \citet{Benson2022} have shown that artificial disruption of subhaloes, as a result of insufficient resolution or inadequate force softening, can account for a significant proportion or even most of the `disrupted' subhaloes in N-body simulations; analytical modelling instead indicates that most subhaloes can retain a bound remnant even after undergoing $>90$\% mass loss due to tidal stripping. This is likely a contributing factor to the disruption of satellites in our sample, particularly at the low-mass end. However, the analysis of \citet{vanDenBosch2018} indicates that haloes resolved by at least $30,000-100,000$ particles do not suffer from significant artificial disruption at the softening lengths used in this work. This translates to a halo mass of $\sim(5-10)\times 10^{9}\MSUN$ in our simulations. Furthermore, most previous studies on artificial disruption rely on N-body simulations, and it is as yet unclear to what degree this affects satellites in cosmological simulations with baryonic physics.

Previous studies based on N-body simulations and semi-analytical modelling such as \citet{Taffoni2003} and \citet{Taylor2004} have reported a mass dependence for their satellite disruption timescales with less massive subhaloes being more long-lived and undergoing more orbits compared to more massive ones (by a few Gyr for a 1~dex difference in satellite mass). Note that \citet{Taffoni2003} generate an idealized simulation with rigid satellites orbiting in a static potential described by an NFW density profile, while \citet{Taylor2004} model merger trees based on analytic expressions for the growth of a host halo from primordial density fluctuations. Furthermore, both studies compare satellites based on the ratio $M_{\rm tot,satellite}/M_{\rm tot,host}$, for a range of masses between $M_{\rm tot,satellite}/M_{\rm tot,host}\sim0.02-0.1$ and $0.0001-0.5$ respectively, for a MW-mass host halo. Studies such as \citet{Engler2021} with the IllustrisTNG-50 simulations show that $\sim80$ per cent of satellites with $M_{*}>10^{7}$ that were accreted since $z=2$ have been disrupted by $z=0$, while a negligible proportion of those accreted after $z=0.7$ have been disrupted, suggesting a maximum survival time of $\sim 7$~Gyr. Additionally, efforts such as \citet{Brooks2013,Nadler2018,Engler2021} also emphasize that the presence of baryons leads to significantly more efficient satellite disruption, by factors of up to 2 for all satellites ever accreted by MW-mass haloes, due to feedback processes and disk shocking.

Our results are broadly in line with those of \citet{Shipp2024} using the AURIGA cosmological zoom-in hydrodynamical simulations. They adopt a definition of satellite disruption whereby a satellite is considered intact as long as at least 97 per cent of the stellar particles it had at infall remain bound to it according to the halo finder \textsc{subfind}. They find that the majority of satellites accreted by MW-mass haloes have been either phase-mixed by $z=0$, or tidally disrupted into stellar streams. According to the definition of disruption we employ, both these categories are considered `disrupted' in our analysis. \citet{Shipp2024} report that most phase-mixed satellites were accreted $>8$~Gyr ago and/or were on highly eccentric orbits with $d_{\rm peri}<10$~kpc, while surviving satellites were accreted $<6$~Gyr ago on more radial orbits and with $d_{\rm peri}>50$~kpc. While the disrupted satellites in our simulations may have $d_{\rm peri}$ out to larger distances and on average lower orbital eccentricities, this is likely due to the inclusion of `streams' in our disrupted population.

On the other hand, with the ARTEMIS simulations, \citet{Grimozzi2024} report longer disruption timescales for more massive satellites, by now comparing by maximum stellar mass of the satellites (for $M_{\rm *,max}=10^{6}-10^{10}\MSUN$). Note that they define satellite disruption based on when a satellite can no longer be identified by the \textsc{subfind} halo finder. This trend is mainly due to their lower mass satellites ($M_{\rm *,max}<10^{8}\MSUN$) having been accreted very early ($z>2$), when the host halo is small and has short dynamical times, while the more massive satellites are accreted later, when the hosts are also larger and therefore disruption takes longer. Our results, especially for the \tng{} simulations, appear to be in tension with these previous results in several ways. Firstly, we find no mass dependence for the \tng{} satellites' average disruption timescales at any epoch, while for the \vgn{} satellites, there is a mild trend of longer disruption timescales for more massive satellites (up to 1~Gyr per dex in stellar mass) at $z<2$, matching the trend seen by \citet{Grimozzi2024} (see fig. 14 of their paper). Secondly, we find a minimum disruption timescale of 2~Gyr, whereas \citet{Grimozzi2024} find disruption timescales $<0.5$~Gyr for their lower mass satellites. Furthermore, we find a larger range in disruption timescales i.e. $\tau_{\rm disrupt}=2-13$~Gyr for our low mass satellites from both sets of simulations, in contrast to the \citet{Grimozzi2024} results (at higher masses, our results are more in line with theirs). The difference in absolute values of $\tau_{\rm disrupt}$ is expected due to the different definitions of disruption used in the two studies as discussed above. The larger range of disruption timescales in our simulations reflects a larger range of accretion times. It remains to be seen whether these differences arise from differing distributions of host properties and/or assembly histories.


\section{Summary \& Conclusions} \label{sec:summary}
Using the PARADIGM suite of zoom-in hydrodynamical simulations, we explore the fate of satellites around MW-like haloes over several epochs. We examine the fraction of satellites at different epochs that are disrupted by present-day and the timescales over which they disrupt. For the satellites that survive, we also study when and where they quench and whether this is consistent with satellite disruption. 

The PARADIGM suite consists of two sets of simulations, evolving the same ICs with two vastly different galaxy formation models, VINTERGATAN (\vgn{}) and IllustrisTNG (\tng{}). The ICs consist of a fiducial simulation, \XC{}, mimicking the expected assembly history for the MW, an early-former (\EF{}) in which the central halo collapses earlier, and a late-former (\LF{}) that undergoes several significant mergers until $z\sim0.7$. Additionally, it includes four `genetically modified' versions of the \XC{} ICs that vary the strength of a significant \targetMerger{} merger, making the mass ratio smaller (\XXC{}, \XVC{}) or larger (\XCX{}, \XCXX{}). Finally, we also include another modification of the \XC{} IC, \XCEC{}, requiring an earlier halo formation similar to \EF{}. These simulations allow us to simultaneously explore the impact of varying halo assembly histories in a controlled manner and varying input physics in the galaxy formation models on various properties of the central MW, its haloes and its satellite populations. 

We find several common trends between the \tng{} and \vgn{} sets of simulations:
\begin{itemize}
    \item The number of satellites within $\RHOST$ remain approximately constant after $z=1$ following an initial increase at early times (Fig.~\ref{fig:Nsats1Rvir}).
    \item Nearly all satellites at $z\geq2$, but almost none at $z\leq0.1$, have been disrupted by $z=0$, with the disruption fraction steadily decreasing from early epochs to late ones (Fig.~\ref{fig:FDisrupt1Rvir}).
    \item The timescales over which the satellites disrupt after being accreted gradually rise from $1-2$~Gyr at $z=4$ to $\sim 6-8$~Gyr by $z=1$ and then remain approximately constant at late times, regardless of stellar mass (Fig.~\ref{fig:TDisruptMeanAcc1Rvir}).
    \item We find no significant dependence on halo merger history based on the four GM ICs that modify the \targetMerger{} merger of the \XC{} simulation (Fig.~\ref{fig:fDisruptMergerHistory}). However, satellite abundances show a clear response to the varying halo growth histories of the \EF{}, \XC{},  \LF{} and \XCEC{} simulations (Fig.~\ref{fig:nsatCollapseTime}). An earlier halo formation time correlates with a smaller number of satellites at late times. This is because early accreted satellites are disrupted and not replenished. On the other hand, a later halo formation time results in a slow but sustained assembly of the satellite population, leaving more survivors at $z=0$.
    \item Whether or not a satellite is disrupted is primarily determined by several orbital factors, most important of which is the pericentric distance, followed by orbital period and number of orbits. Disrupted satellites have substantially smaller pericentric and apocentric distances with shorter orbital periods. The impact of orbital eccentricities and number of orbits is mild at best (Fig.~\ref{fig:SurvivalVsPericentre}).
    \item Disruption timescales are strongly correlated with orbital period, but only show weak or no correlation with other orbital properties (Fig.~\ref{fig:tauDisruptVsPericentre}).
    \item Satellites that survive to present-day are found to have been quenched up to 12~Gyr ago. Brighter satellites were quenched more recently ($\lesssim 8$~Gyr ago) and within the MW host halo, while fainter satellites were quenched at earlier epochs and usually outside the MW host halo, consistent with the $6-8$~Gyr timescale over which satellites can survive in the host. These trends match observed results from \citet{Weisz2015}, qualitatively in the \vgn{} case and quantitatively in \tng{} case (Fig.~\ref{fig:quenchTimes}).
\end{itemize}

These common trends show that both galaxy formation models predict similar behaviours for satellites of MW-mass hosts despite their vastly different implementations of key physical processes. There are however quantitative differences: 
\begin{itemize}
    \item The most important of these is that the \vgn{} simulations overpredict the number of satellites at all masses/luminosities by a factor of 2 to 3. The large differences in satellite numbers between the two sets of simulations are primarily due to significantly different SMHM relations especially at low-halo masses, resulting in higher stellar masses for a given halo mass in the \vgn{} simulations (by up to 1.5~dex). This in turn is most probably due to the formation of relatively massive galaxies at very early times $z>5$ in the \vgn{} model, which form most of their stellar mass through a single starburst. This issue has a smaller impact on more massive galaxies which continue building stellar mass through more regulated SF at later times, but appears to be unphysical. It will be investigated in detail by Rodriguez-Cardoso et al. (in prep).
    \item Satellite disruption fractions are higher for low-mass \vgn{} satellites than \tng{} by $10-20$ percentage points. For high-mass \vgn{} satellites, these fractions are lower than \tng{} by up to 40 percentage points until $z\sim0.5$, after which the trend is reversed. Thus, the \tng{} satellites are more resistant to disruption, except in the case of high-mass satellites at $z>0.5$.
    \item The disruption timescales for low-mass \tng{} satellites are longer than for \vgn{} by $\sim 1$~Gyr; for high-mass satellites the difference can be $\pm 1$~Gyr for earlier/later epochs.
    \item The \vgn{} satellites were quenched earlier at all masses compared to the \tng{} ones. The \vgn{} simulations also contain a subpopulation of surviving satellites that quenched over 10~Gyr ago, at large distances from the host ($>3\,\RHOST$), corresponding to the early-forming galaxies mentioned above. 
\end{itemize}
The quantitative differences in the satellite disruption fractions and timescales between the \tng{} and \vgn{} simulations are primarily the result of their satellites having different distributions of orbital properties, particularly of pericentric distance. We emphasize that this is not due to the two models producing different orbital properties, but rather due to them populating different subsets of DM haloes with galaxies due to their differing SMHM relations Overall, \vgn{} satellites (both disrupted and surviving to $z=0$) undergo several more orbits compared to the \tng{} ones. While there are considerable differences in the stellar and DM concentrations of the satellites and the host haloes themselves, these properties are subdominant to orbital properties in accounting for the differences in satellite disruption fractions. Finally, while the substantial differences in the SMHM relations of the two models dominate the differences in their satellite abundance, the manner in which their satellites are disrupted also plays an important role (see Section~\ref{sec:calcSatAbundances}).

Our work highlights the importance of halo assembly history and satellite orbits in the disruption of satellites, showing that it is a crucial but complex contributing factor in driving diversity in satellite populations around MW-mass hosts. On the other hand, the \tng{} and \vgn{} galaxy formation models represent two different strategies for cosmological simulations emphasizing different physical ingredients and differing levels of agreement to observational constraints, and yet we find good agreement in their resultant behaviour in terms of satellite disruption. This suggests that these and other different approaches to simulating galaxy formation models are now able to reach a coherent account of emergent interactions between satellites and their hosts, despite the considerable challenges to a fully converged account of galaxy formation.

These results serve as a case study for MW-mass haloes, but provide important insights into the process of satellite disruption. The near constant disruption timescales we find, regardless of the hosts' formation histories and with at best weak dependence on satellite mass, indicate that we may be able to reliably constrain such disruption timescales as a function of halo mass using large-volume simulations. Crucially, while previous studies have shown that the inclusion of baryonic physics can substantially increase the efficiency of satellite disruption, the agreement we find between two vastly different implementations of baryonic physics suggest that the disruption process may be less sensitive to the details of the implementation. Determining the history of satellite disruption in MW-like hosts in the Universe requires us to observe their stellar haloes as well as to distinguish between in-situ and ex-situ stellar populations in the central galaxy. Currently this is only possible within the MW and nearby galaxies, but new and future observing facilities will allow us to explore a much larger population of host galaxies sampling a wider range of formation histories. By contrast, satellite quenching is highly dependent on the model, particularly of the ISM and feedback processes, as expected. Upcoming data from surveys such as EUCLID will greatly expand our understanding of when and where satellites are quenched, and will be an important benchmark for galaxy formation models.

\section*{Acknowledgements}

We thank the anonymous referee for their thoughtful comments which helped improve the manuscript. This project has received funding from the European Union’s Horizon 2020 research and innovation programme under grant agreement No. 818085 GMGalaxies. This study used computing equipment funded by the Research Capital Investment Fund (RCIF) provided by UKRI, and partially funded by the University College London Cosmoparticle Initiative. OA acknowledges support from the Knut and Alice Wallenberg Foundation, the Swedish Research Council (grant 2019-04659), and the Swedish National Space Agency (SNSA Dnr 2023-00164). We acknowledge PRACE for awarding us access to Joliot-Curie at GENCI/CEA, France to perform the simulations presented in this work. Computations presented in this work were in part performed on resources provided by the Swedish National Infrastructure for Computing (SNIC) at the Tetralith supercomputer, part of the National Supercomputer Centre, Linköping University. JR would like to thank the STFC for support from grants ST/Y002865/1 and ST/Y002857/1.

Author contributions are as follows.
\newline
\textbf{GJ}: Conceptualization, data curation, formal analysis, investigation, writing -- original draft (lead). \textbf{AP}: Conceptualization, funding acquisition, methodology, resources, writing -- original draft (support). \textbf{OA}: Conceptualization, data curation, methodology, resources, writing -- review \& editing. \textbf{JR}: Conceptualization, writing -- review \& editing. \textbf{MR}: Conceptualization, data curation, writing -- review \& editing.

\section*{Data Availability}

The data underlying this article will be shared upon reasonable request to the corresponding author.



\bibliographystyle{mnras}
\bibliography{PARADIGM_satellite_destruction_and_quenching} 




\appendix

\section{Mitigating the effect of early bursty SF in the \vgn{} model} \label{sec:appVGFiltering}

\begin{figure*}
    \includegraphics[width=\linewidth]{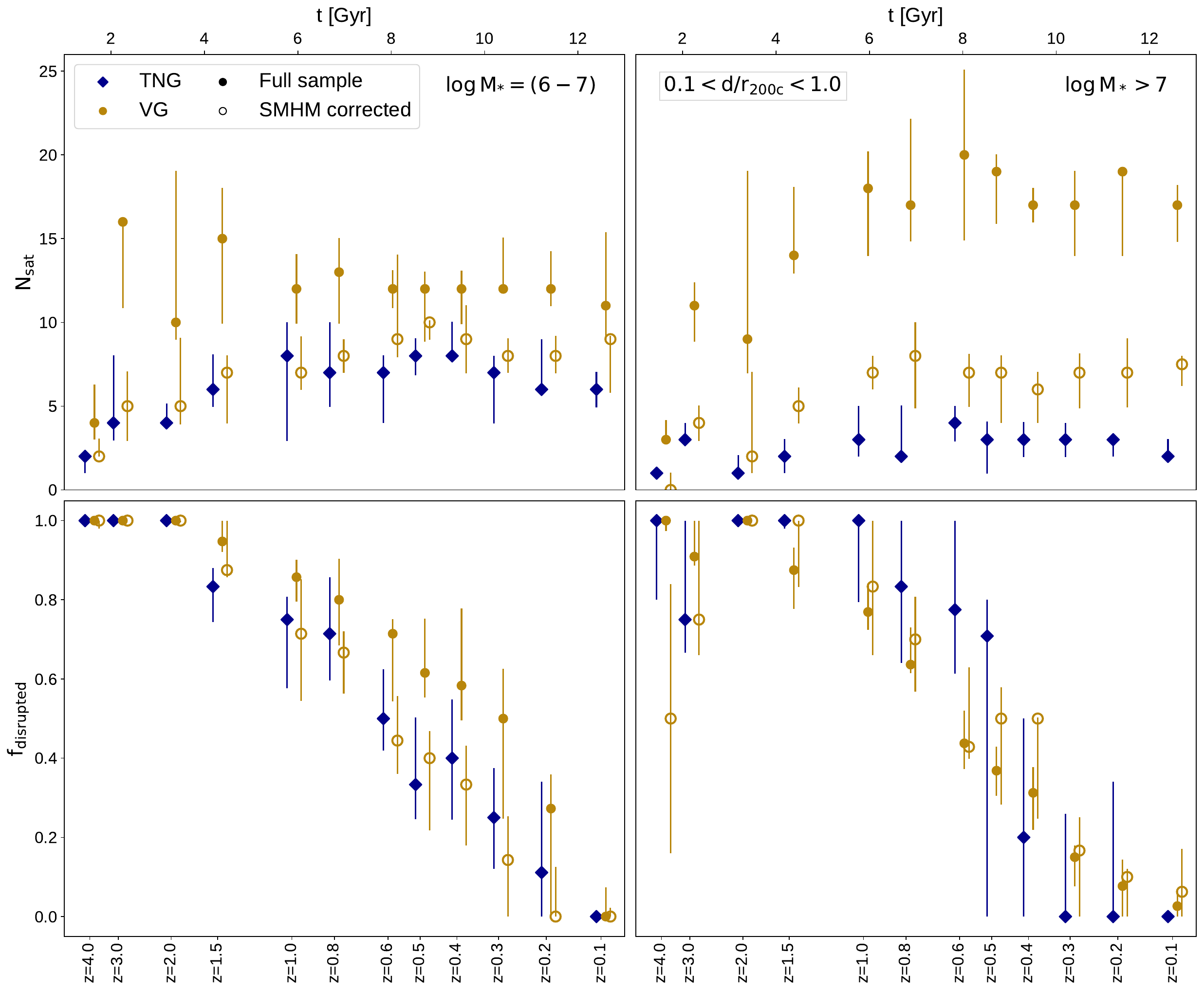}
    \caption{Numbers of satellites (top panels) and disruption fractions (bottom panels) as a function of selection epoch as in Fig. \ref{fig:Nsats1Rvir} and \ref{fig:FDisrupt1Rvir} respectively. Filled markers replicate the results from the mentioned figures. Open markers show the results for \vgn{} after correcting the satellite stellar masses to match the average SMHM relation of the \tng{} field satellites at $z=0$; these have been slightly displaced in the x-direction for clarity. The correction brings the numbers of \vgn{} satellites into better agreement with the \tng{} results, as expected. Additionally, the disruption fractions are also in better agreement with the \tng{} results, particularly in the low-mass case.} \label{fig:smhmCorrections}
\end{figure*}

As discussed in Section \ref{sec:resultsZ0LFs}, the \vgn{} model overpredicts the number of satellites in the MW hosts compared to both the \tng{} model and observations of the MW and M31. This is largely due to an early phase of bursty, poorly regulated SF which results in a significant mass of stars produced often in a single starburst at $z>8$. This issue impacts low-mass galaxies ($M_{\rm tot}<10^{9}\MSUN$) significantly, resulting in a likely unphysical population of satellites that form most of their stellar mass in this starburst phase. For more massive galaxies, continued SF that is regulated by stellar feedback at later times mitigates the impact of this early phase; nonetheless, intermediate mass ($M_{\rm tot}=10^{9-10.5}\MSUN$) galaxies also contain too much stellar mass for their halo mass to varying degrees, resulting in an overall higher SMHM relation as discussed in Section \ref{sec:smhmRelations}. Simply filtering the low-mass galaxies that form all their stellar mass in a single starburst will therefore introduce significant biases in the \vgn{} sample. Instead, we attempt to correct the stellar masses of the \vgn{} satellites here and re-evaluate the results of Sections \ref{sec:resultsNSatEvol}, \ref{sec:resultsDisFracs} and \ref{sec:resultsDisTimes}.

To correct the \vgn{} stellar masses, we measure the SMHM relations of the \tng{} and \vgn{} field galaxies (i.e. at $1<d/\RHOST<3$) at $z=0$ as a linear fit between \emph{peak} halo mass, $\log{(M_{\rm tot,peak})}$, and stellar mass at peak halo mass, $\log{(M_{\rm *,peak})}$. We then obtain corrected stellar masses for the \vgn{} satellites as:

\begin{equation}
    M_{\rm *,corrected} = M_{\rm *,original} \times \frac{M_{\rm *,predicted,TNG}(M_{\rm tot,peak})}{M_{\rm *,predicted,VG}(M_{\rm tot,peak})}.
\end{equation}
This allows us to correct the stellar masses at a population level, while retaining the scatter in the SMHM relation from the \vgn{} model. We caution here that this is by no means a rigorous method to account for the early SF in the \vgn{} model, which requires careful consideration of the precise conditions under which such starbursts occur as well as any secondary effects of the early-forming stars on the subsequent evolution of the galaxy. Rather, we use this simple correction to indicate the magnitude of any biases introduced by the early SF to some of our key results.

We now re-evaluate the satellites abundances, disruption fractions and disruption timescales of satellites over the selection epochs with the corrected \vgn{} stellar masses. In Fig. \ref{fig:smhmCorrections}, we show the average number of satellites (top panels) and disruption fractions (bottom panels) as a function of selection epoch, as in Figs. \ref{fig:Nsats1Rvir} and \ref{fig:FDisrupt1Rvir} respectively. Filled markers recreate the results from the previous figures, while open markers show the stellar mass-corrected \vgn{} satellites. The top panels show that the stellar mass correction brings the satellite abundances in the two sets of simulations into much better agreement, particularly in the low-mass case; in the high-mass case, although the numbers of \vgn{} satellites are significantly lowered, they remain higher than the \tng{} values at all times. The agreement in satellite numbers is largely to be expected as the \vgn{} stellar masses were corrected down for haloes of $M_{\rm tot,peak}<10^{10.5}\MSUN$. Our main conclusions in the paper show that the rates of disruption are broadly comparable between the two models. Therefore, once the mean field SMHM is corrected, the agreement in satellite numbers is to be expected. The remaining differences are largely due to different amounts of scatter in the SMHM relation between the two models.

The bottom panels of Fig. \ref{fig:smhmCorrections} then show that this change in the satellite population brings the \vgn{} disruption fractions into even better agreement with the \tng{} ones. In the low-mass case, the disruption fractions are nearly identical within errorbars. In the high mass case, the results are in better agreement at earlier epochs $3>z>1$, while the impact is minimal at later times, albeit with increased scatter. Note that the large deviation at $z=4$ is due to low-number statistics.

We also considered the results of Fig. \ref{fig:TDisruptMeanAcc1Rvir}, i.e. disruption timescales rather than disruption fractions, with the corrected stellar masses and found the results to be nearly identical (a corresponding figure is omitted for brevity). These results show that although the early bursty SF phase in the \vgn{} model results in significantly higher satellite abundances than observed, it is does not bias our results on satellite disruption to a significant degree.

\section{Additional factors affecting satellite disruption} \label{sec:appDisrupt}

\begin{figure*}
    \centering
    \includegraphics[width=\linewidth]{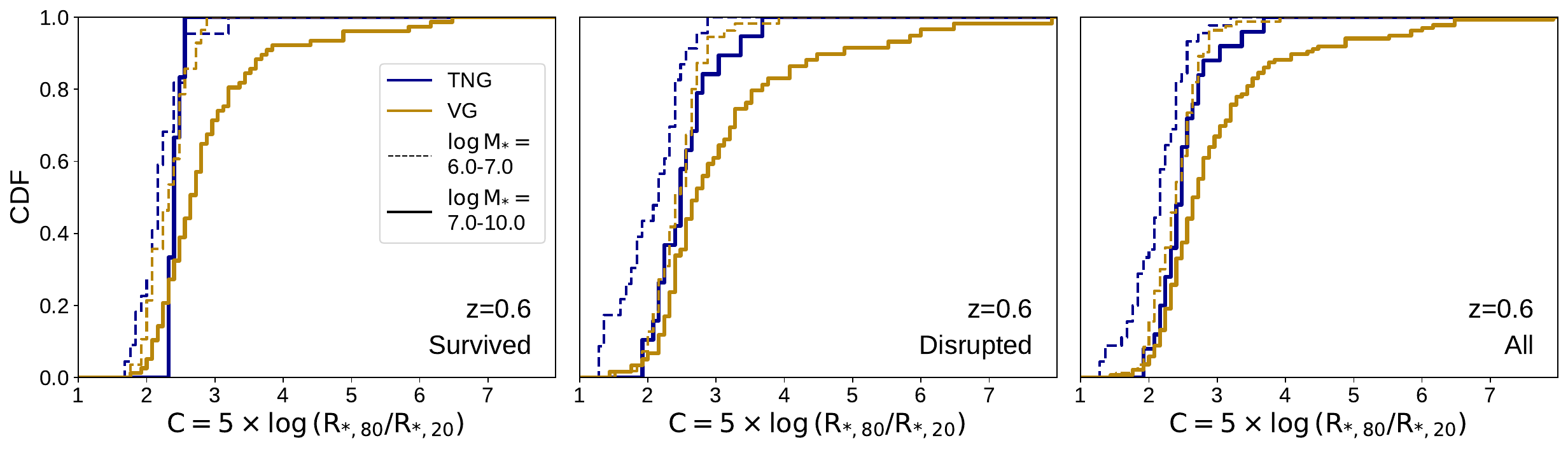}
    \caption{Distribution of stellar concentration $C$ for satellites selected at $z=0.6$; left and middle panels show surviving and disrupted satellites separately, while the right panels show all satellites. There are significant differences in the concentrations of \tng{} and \vgn{} satellites, particularly at higher masses. However, disruption fractions are found not to depend on concentration.}
    \label{fig:SurvivalVsConcentration}
\end{figure*}

In Section~\ref{sec:discContributingFactors}, we discussed several factors found to impact the proportion of satellites that are disrupted and the timescales over which they disrupt in our simulations. Here we discuss additional factors that were also considered, but found to have little impact on satellite disruption.

\subsection{Intrinsic galaxy properties} \label{sec:discIntrinsicProps}
As in Section~\ref{sec:discOrbitalProps} with orbital characteristics, we also considered different intrinsic galaxy properties that could make a satellite more or less resilient to disruption by tidal forces, namely stellar and gas mass fractions at peak halo mass, $f_{\rm *,peak}$ and $f_{\rm gas,peak}$, and stellar concentration $C$. Following common definitions, we measure concentration as
\begin{equation}
    C = 5\times\log{\frac{R_{*,80}}{R_{*,20}}}
\end{equation}
where $R_{*,20}$ and $R_{*,80}$ are the 3D radii containing 20 and 80 per cent of the stellar mass of the satellite respectively, at the selection epoch. As with the satellite tracking, we exclude the outermost 20 per cent of stellar particles so as to minimize the impact of this diffuse outer material. Note that, as we discussed in detail in Section~\ref{sec:smhmRelations}, the \tng{} and \vgn{} satellites have significantly different SMHM relations and therefore the distributions of $f_{\rm *, peak}$ for the two models are significantly different at all times and for all masses. However, we can still compare the distributions between disrupted and surviving satellites to determine whether this factor is important in determining satellite disruption. 

In Fig.~\ref{fig:SurvivalVsConcentration}, we show the distributions of stellar concentration $C$ for surviving and disrupted satellites separately (left and middle panels) and all satellites (right panels) at selection epoch $z=0.6$, when there are sufficient numbers of both disrupted and surviving satellites in all subsamples. It is clear that the \vgn{} satellites, particularly the high-mass ones, are significantly more concentrated than the \tng{} ones, showing a substantial tail in the distribution towards high values of $C$. This is seen at all epochs, although the differences are most evident at early times. However, we find no clear connection between stellar concentration and the probability of satellites to be disrupted. In most cases, KS tests indicate there is less than 80\% confidence that the surviving and disrupted distributions are different, except for the \vgn{} low-mass case at some epochs. This suggests that concentration is not an important factor in determining whether satellites disrupt. We find a similar lack of correlation between disruption fraction and stellar and gas mass fractions at peak halo mass. Similarly, we find no correlation between disruption timescale and any of these properties. Although by no means an exhaustive list of all intrinsic properties that could impact satellite disruption, our analysis suggests that the probability of satellite disruption and disruption timescales are independent of any intrinsic properties of the satellites, according to both galaxy formation models.

\begin{figure*}
    \includegraphics[width=\linewidth]{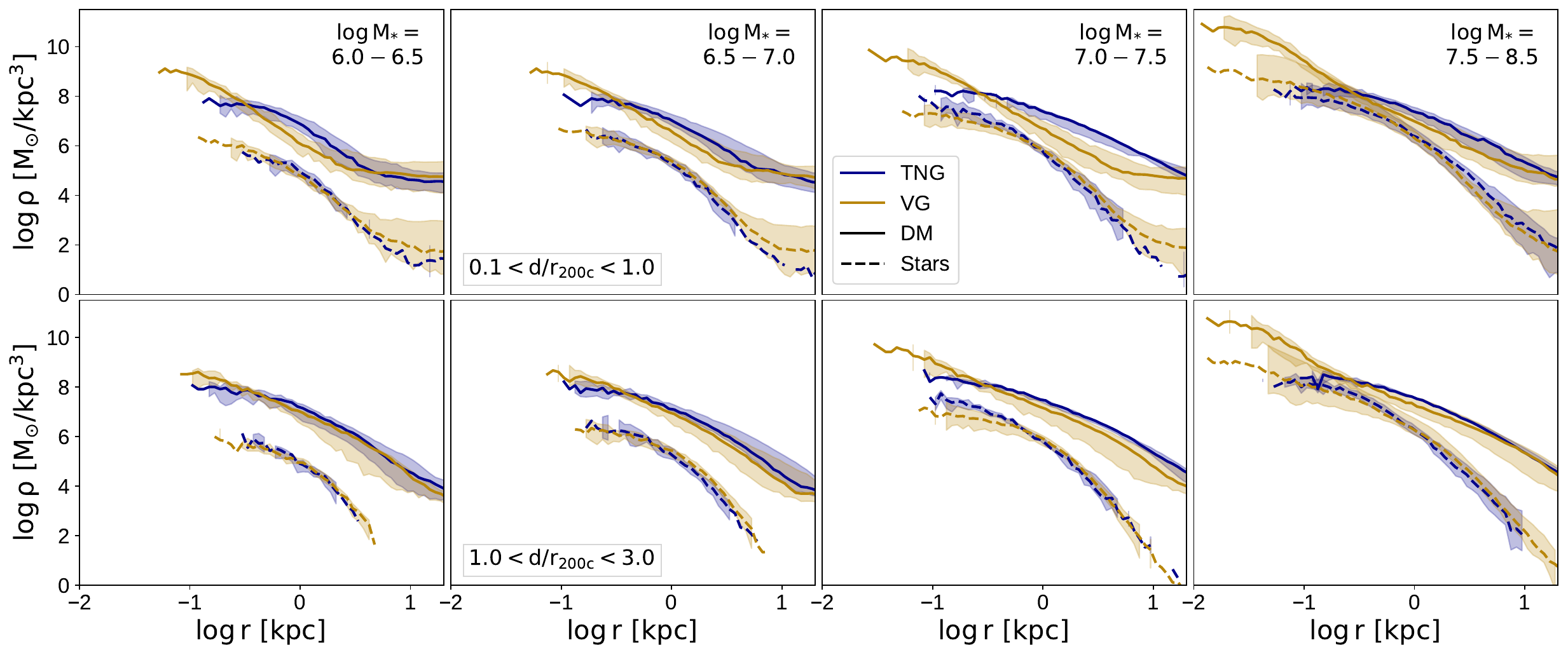}
    \caption{Density profiles of $z=0$ satellites in bins of stellar mass, showing the density of DM (solid curves) and stars (dashed curves). The profiles of all satellites in the mass bin within the host's virial radius (top row) and between $(1-3)\,\RHOST$ (bottom row) are stacked from all simulations; the solid/dashed curves show the median while the shaded regions indicate the $16^{\rm th}-84^{\rm th}$ percentile ranges. The stellar density profiles are nearly identical at the low masses ($M_{*}<10^{7}\MSUN$), and for more massive satellites at radii $>1$~kpc. The DM density profiles on the other hand are more concentrated for the \vgn{} satellites, especially those found within the hosts' virial radii.} \label{fig:satDensityProfiles}
\end{figure*}

\begin{figure*}
    \centering
    \includegraphics[width=\linewidth]{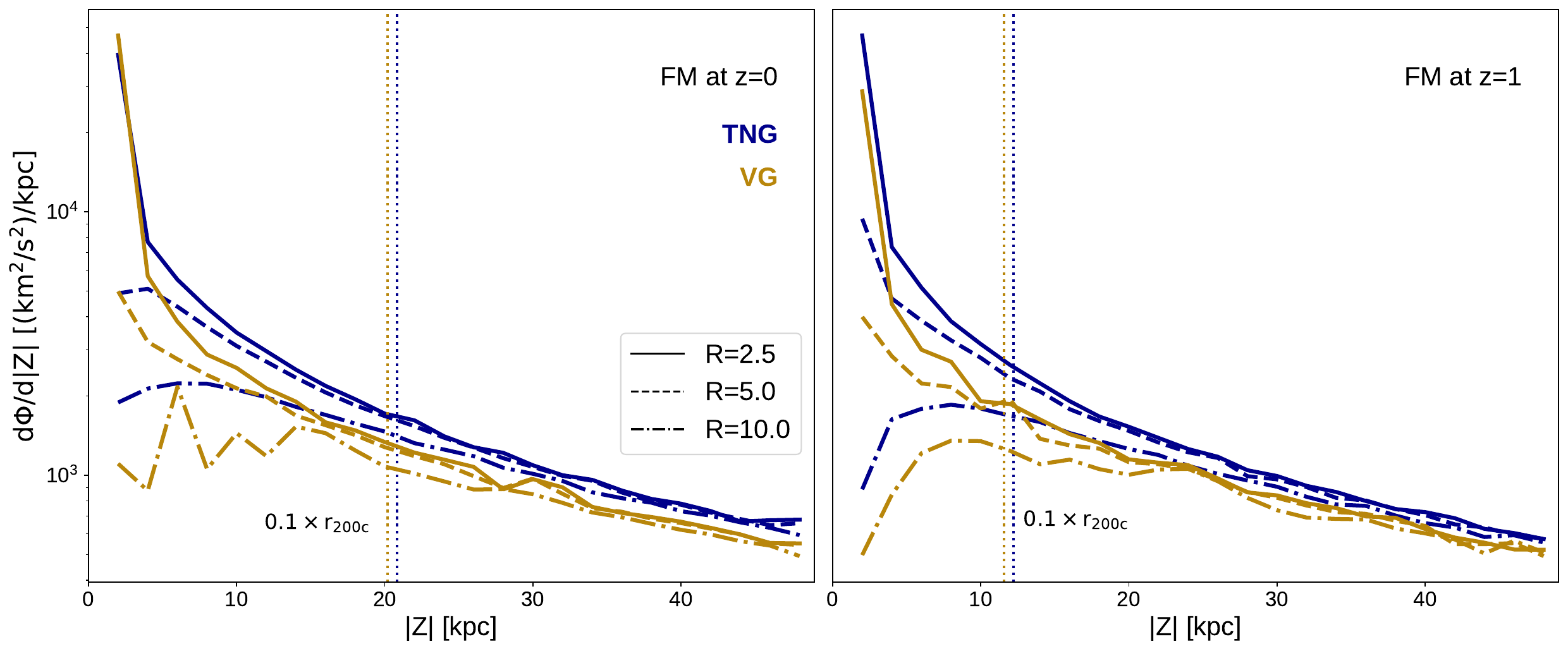}
    \caption{Vertical gravitational acceleration profiles, measured as $d\Phi/d|Z|$, for the \XC{} simulations at $z=0$ and $z=1$. Different line-styles indicate vertical profiles taken at different face-on projected distances from the centre in kpc. Vertical dotted lines indicate $0.1\times$ the virial radius of the host halo. At all (vertical) distances beyond $\sim$5~kpc, the \vgn{} central galaxy has lower vertical acceleration compared to the \tng{} galaxy at both times.}
    \label{fig:cenGravForceProfile}
\end{figure*}

\subsection{Comparison of satellite and host structure} \label{sec:discStructure}
Another factor that may have bearing on whether satellites are disrupted or not is the structure of the satellites themselves, as well as the that of the hosts. In Fig.~\ref{fig:satDensityProfiles}, we first show the median DM and stellar density profiles for the $z=0$ satellites in bins of stellar mass within the MW halo (top row). Since satellite density profiles may be significantly impacted by their host environment, we include similar profiles for infalling galaxies outside the virial radius of the host found between $(1-3)\,\RHOST$, to isolate the impact of the host. Satellites from all simulations in our sample are stacked to produce the median curves and $16^{\rm th}-84^{\rm th}$ percentile ranges indicated by the shaded regions. Within the MW halo, the DM density profiles for the \vgn{} satellites are more concentrated with higher central densities and steeper slopes compared to the \tng{} satellites at all masses, but particularly for $M_{*}>10^{7}\MSUN$. Furthermore, the higher spatial resolution (and smaller softening length) of the \vgn{} simulations allows the DM (and stellar) material to be found at radii as small as $10$ pc. These differences are also seen in satellites outside the MW halo, although to a lesser degree. 

The differences between the stellar mass profiles are less stark. At lower masses ($M_{*}<10^{7.5}\MSUN$), the stellar mass profiles appear nearly identical both within and outside the host halo, with the caveat that the \vgn{} profiles for satellites within the host extend to slightly smaller radii. In the highest mass bin however, the \vgn{} profiles extent to significantly smaller radii as with the DM profiles. These results suggest that there is little different in the stellar density profiles between the \tng{} and \vgn{} satellites, except for the innermost regions. We caution however that these profiles probe a different spatial regime than the concentrations $C$ measured in Section~\ref{sec:discIntrinsicProps}, which rely on the cumulative mass profiles and effectively linear radial binning and therefore, are not necessarily inconsistent with the findings of Section~\ref{sec:discIntrinsicProps}. The $r_{20}$ and $r_{80}$ values used to calculate $C$ are of the order $0.5$ and $1.5$~kpc respectively which spans a narrow region in these figures. 

The difference in satellite disruption may also stem from encountering different gravitational potentials from their hosts. Satellites with different pericentric distances may interact with the plane of the central galaxy's disc at various projected distances in the plane, and travel through a range of vertical distances. The gravitational forces (and tidal fields) they encounter will play a significant role in their disruption. In Fig.~\ref{fig:cenGravForceProfile}, we show the profiles of vertical acceleration (perpendicular to the central disc) i.e. $d\Phi/d|Z|$, for the \XC{} simulations at $z=0$ and $z=1$ at different projected distances from the host centre. The satellites were accreted at various times in cosmic history and therefore will have encountered different host potentials; however, comparing the potential profiles at these two times provides an indication of how much this environment changes with time. The \vgn{} host exerts lower gravitational forces everywhere compared to the \tng{} host at both epochs, except for the central most regions where the forces are comparable. We have also examined the profiles of tidal force exerted by the hosts, $F_{\rm tidal} \propto d^{2}\Phi/d|Z|^{2}$ and find these to be similar between the two models (not shown). Given that the \vgn{} low-mass satellites have higher disruption fractions than \tng{}, as do high-mass satellites at $z<0.4$, the difference in host potentials does not appear to be an important contributing factor in determining the satellite disruption fractions.


\bsp	
\label{lastpage}
\end{document}

%% file: figures/halo_props_all.tex
\XXC{}	& 211.4 & 99.3 & 6.9 & 4.5 & 4.8 & 7.1 & 13.6 & 18.0 & 1.28 & 4.07 \\
\XVC{}	& 210.9 & 98.6 & 6.6 & 4.7 & 4.8 & 6.7 & 12.6 & 16.7 & 0.83 & 4.29 \\
\XC{}	& 208.2 & 95.0 & 4.7 & 5.2 & 4.0 & 4.6 & 10.0 & 12.4 & 0.30 & 4.46 \\
\XCX{}	& 207.5 & 94.0 & 5.3 & 3.9 & 3.7 & 4.6 & 12.9 & 10.7 & 0.34 & 3.27 \\
\XCXX{}	& 215.5 & 105.3 & 4.6 & 3.8 & 4.3 & 1.3 & 5.3 & 4.2 & 0.01 & 2.55 \\
\XCEC{}	& 204.2 & 89.6 & 4.0 & 4.1 & 3.2 & 4.0 & 7.4 & 11.4 & 0.14 & 3.45 \\
\EF{}	& 222.0 & 115.1 & 6.2 & 6.4 & 3.2 & 6.7 & 13.7 & 21.5 & 0.84 & 5.75 \\
\LF{}	& 216.3 & 106.5 & 7.7 & 5.3 & 5.2 & 9.6 & 14.4 & 25.2 & 2.66 & 4.98 \\
\hline
\multicolumn{11}{c}{VG} \\
\hline
\XXC{}	& 198.6 & 82.4 & 1.9 & 1.8 & 4.9 & 1.8 & 2.9 & 3.3 & 0.17 & 1.10 \\
\XVC{}	& 197.7 & 81.3 & 2.1 & 1.6 & 5.0 & 1.6 & 1.8 & 2.8 & 0.18 & 1.05 \\
\XC{}	& 201.8 & 86.5 & 2.6 & 1.8 & 4.3 & 1.0 & 2.2 & 2.2 & 0.12 & 1.14 \\
\XCX{}	& 200.2 & 84.4 & 1.9 & 1.6 & 4.2 & 1.2 & 1.7 & 2.3 & 0.14 & 1.03 \\
\XCXX{}	& 202.6 & 87.5 & 2.7 & 1.8 & 4.7 & 0.5 & 0.7 & 1.7 & 0.03 & 1.20 \\
\EF{}	& 203.7 & 99.3 & 1.6 & 2.8 & 3.1 & 2.6 & 6.5 & 6.4 & 0.18 & 1.77 \\
\LF{}	& 204.5 & 89.9 & 2.3 & 1.3 & 5.3 & 2.1 & 1.6 & 3.4 & 0.23 & 0.88 \\
\hline

%% file: PARADIGM_satellite_destruction_and_quenching.bbl
\begin{thebibliography}{}
\makeatletter
\relax
\def\mn@urlcharsother{\let\do\@makeother \do\$\do\&\do\#\do\^\do\_\do\%\do\~}
\def\mn@doi{\begingroup\mn@urlcharsother \@ifnextchar [ {\mn@doi@} {\mn@doi@[]}}
\def\mn@doi@[#1]#2{\def\@tempa{#1}\ifx\@tempa\@empty \href {http://dx.doi.org/#2} {doi:#2}\else \href {http://dx.doi.org/#2} {#1}\fi \endgroup}
\def\mn@eprint#1#2{\mn@eprint@#1:#2::\@nil}
\def\mn@eprint@arXiv#1{\href {http://arxiv.org/abs/#1} {{\tt arXiv:#1}}}
\def\mn@eprint@dblp#1{\href {http://dblp.uni-trier.de/rec/bibtex/#1.xml} {dblp:#1}}
\def\mn@eprint@#1:#2:#3:#4\@nil{\def\@tempa {#1}\def\@tempb {#2}\def\@tempc {#3}\ifx \@tempc \@empty \let \@tempc \@tempb \let \@tempb \@tempa \fi \ifx \@tempb \@empty \def\@tempb {arXiv}\fi \@ifundefined {mn@eprint@\@tempb}{\@tempb:\@tempc}{\expandafter \expandafter \csname mn@eprint@\@tempb\endcsname \expandafter{\@tempc}}}

\bibitem[\protect\citeauthoryear{{Agertz} et~al.,}{{Agertz} et~al.}{2021}]{Agertz2021}
{Agertz} O.,  et~al., 2021, \mn@doi [\mnras] {10.1093/mnras/stab322}, \href {https://ui.adsabs.harvard.edu/abs/2021MNRAS.503.5826A} {503, 5826}

\bibitem[\protect\citeauthoryear{{Akins}, {Christensen}, {Brooks}, {Munshi}, {Applebaum}, {Engelhardt}  \& {Chamberland}}{{Akins} et~al.}{2021}]{Akins2021}
{Akins} H.~B.,  {Christensen} C.~R.,  {Brooks} A.~M.,  {Munshi} F.,  {Applebaum} E.,  {Engelhardt} A.,   {Chamberland} L.,  2021, \mn@doi [\apj] {10.3847/1538-4357/abe2ab}, \href {https://ui.adsabs.harvard.edu/abs/2021ApJ...909..139A} {909, 139}

\bibitem[\protect\citeauthoryear{{Applebaum}, {Brooks}, {Christensen}, {Munshi}, {Quinn}, {Shen}  \& {Tremmel}}{{Applebaum} et~al.}{2021}]{Applebaum2021JusticeLeague}
{Applebaum} E.,  {Brooks} A.~M.,  {Christensen} C.~R.,  {Munshi} F.,  {Quinn} T.~R.,  {Shen} S.,   {Tremmel} M.,  2021, \mn@doi [\apj] {10.3847/1538-4357/abcafa}, \href {https://ui.adsabs.harvard.edu/abs/2021ApJ...906...96A} {906, 96}

\bibitem[\protect\citeauthoryear{{Bennet}, {Sand}, {Crnojevi{\'c}}, {Spekkens}, {Karunakaran}, {Zaritsky}  \& {Mutlu-Pakdil}}{{Bennet} et~al.}{2020}]{Bennet2020}
{Bennet} P.,  {Sand} D.~J.,  {Crnojevi{\'c}} D.,  {Spekkens} K.,  {Karunakaran} A.,  {Zaritsky} D.,   {Mutlu-Pakdil} B.,  2020, \mn@doi [\apjl] {10.3847/2041-8213/ab80c5}, \href {https://ui.adsabs.harvard.edu/abs/2020ApJ...893L...9B} {893, L9}

\bibitem[\protect\citeauthoryear{{Benson} \& {Du}}{{Benson} \& {Du}}{2022}]{Benson2022}
{Benson} A.~J.,  {Du} X.,  2022, \mn@doi [\mnras] {10.1093/mnras/stac2750}, \href {https://ui.adsabs.harvard.edu/abs/2022MNRAS.517.1398B} {517, 1398}

\bibitem[\protect\citeauthoryear{{Bose}, {Eisenstein}, {Hernquist}, {Pillepich}, {Nelson}, {Marinacci}, {Springel}  \& {Vogelsberger}}{{Bose} et~al.}{2019}]{Bose2019}
{Bose} S.,  {Eisenstein} D.~J.,  {Hernquist} L.,  {Pillepich} A.,  {Nelson} D.,  {Marinacci} F.,  {Springel} V.,   {Vogelsberger} M.,  2019, \mn@doi [\mnras] {10.1093/mnras/stz2546}, \href {https://ui.adsabs.harvard.edu/abs/2019MNRAS.490.5693B} {490, 5693}

\bibitem[\protect\citeauthoryear{{Boselli} et~al.,}{{Boselli} et~al.}{2021}]{Boselli2021}
{Boselli} A.,  et~al., 2021, \mn@doi [\aap] {10.1051/0004-6361/202039046}, \href {https://ui.adsabs.harvard.edu/abs/2021A&A...646A.139B} {646, A139}

\bibitem[\protect\citeauthoryear{{Boylan-Kolchin} \& {Ma}}{{Boylan-Kolchin} \& {Ma}}{2007}]{BoylanKolchin2007}
{Boylan-Kolchin} M.,  {Ma} C.-P.,  2007, \mn@doi [\mnras] {10.1111/j.1365-2966.2006.11276.x}, \href {https://ui.adsabs.harvard.edu/abs/2007MNRAS.374.1227B} {374, 1227}

\bibitem[\protect\citeauthoryear{{Brooks}, {Kuhlen}, {Zolotov}  \& {Hooper}}{{Brooks} et~al.}{2013}]{Brooks2013}
{Brooks} A.~M.,  {Kuhlen} M.,  {Zolotov} A.,   {Hooper} D.,  2013, \mn@doi [\apj] {10.1088/0004-637X/765/1/22}, \href {https://ui.adsabs.harvard.edu/abs/2013ApJ...765...22B} {765, 22}

\bibitem[\protect\citeauthoryear{{Buck}, {Macci{\`o}}, {Dutton}, {Obreja}  \& {Frings}}{{Buck} et~al.}{2019}]{Buck2019}
{Buck} T.,  {Macci{\`o}} A.~V.,  {Dutton} A.~A.,  {Obreja} A.,   {Frings} J.,  2019, \mn@doi [\mnras] {10.1093/mnras/sty2913}, \href {https://ui.adsabs.harvard.edu/abs/2019MNRAS.483.1314B} {483, 1314}

\bibitem[\protect\citeauthoryear{{Bullock} \& {Johnston}}{{Bullock} \& {Johnston}}{2005}]{Bullock2005}
{Bullock} J.~S.,  {Johnston} K.~V.,  2005, \mn@doi [\apj] {10.1086/497422}, \href {https://ui.adsabs.harvard.edu/abs/2005ApJ...635..931B} {635, 931}

\bibitem[\protect\citeauthoryear{{Bullock}, {Kravtsov}  \& {Weinberg}}{{Bullock} et~al.}{2001}]{Bullock2001}
{Bullock} J.~S.,  {Kravtsov} A.~V.,   {Weinberg} D.~H.,  2001, \mn@doi [\apj] {10.1086/318681}, \href {https://ui.adsabs.harvard.edu/abs/2001ApJ...548...33B} {548, 33}

\bibitem[\protect\citeauthoryear{{Calzetti}, {Armus}, {Bohlin}, {Kinney}, {Koornneef}  \& {Storchi-Bergmann}}{{Calzetti} et~al.}{2000}]{Calzetti2000}
{Calzetti} D.,  {Armus} L.,  {Bohlin} R.~C.,  {Kinney} A.~L.,  {Koornneef} J.,   {Storchi-Bergmann} T.,  2000, \mn@doi [\apj] {10.1086/308692}, \href {https://ui.adsabs.harvard.edu/abs/2000ApJ...533..682C} {533, 682}

\bibitem[\protect\citeauthoryear{{Carlin} et~al.,}{{Carlin} et~al.}{2016}]{Carlin2016}
{Carlin} J.~L.,  et~al., 2016, \mn@doi [\apjl] {10.3847/2041-8205/828/1/L5}, \href {https://ui.adsabs.harvard.edu/abs/2016ApJ...828L...5C} {828, L5}

\bibitem[\protect\citeauthoryear{{Carlsten}, {Greene}, {Beaton}, {Danieli}  \& {Greco}}{{Carlsten} et~al.}{2022}]{Carlsten2022}
{Carlsten} S.~G.,  {Greene} J.~E.,  {Beaton} R.~L.,  {Danieli} S.,   {Greco} J.~P.,  2022, \mn@doi [\apj] {10.3847/1538-4357/ac6fd7}, \href {https://ui.adsabs.harvard.edu/abs/2022ApJ...933...47C} {933, 47}

\bibitem[\protect\citeauthoryear{{Chabrier}}{{Chabrier}}{2003}]{Chabrier2003}
{Chabrier} G.,  2003, \mn@doi [\pasp] {10.1086/376392}, \href {https://ui.adsabs.harvard.edu/abs/2003PASP..115..763C} {115, 763}

\bibitem[\protect\citeauthoryear{{Choi}, {Weinberg}  \& {Katz}}{{Choi} et~al.}{2009}]{Choi2009}
{Choi} J.-H.,  {Weinberg} M.~D.,   {Katz} N.,  2009, \mn@doi [\mnras] {10.1111/j.1365-2966.2009.15556.x}, \href {https://ui.adsabs.harvard.edu/abs/2009MNRAS.400.1247C} {400, 1247}

\bibitem[\protect\citeauthoryear{{Collins} \& {Read}}{{Collins} \& {Read}}{2022}]{Collins2022}
{Collins} M. L.~M.,  {Read} J.~I.,  2022, \mn@doi [Nature Astronomy] {10.1038/s41550-022-01657-4}, \href {https://ui.adsabs.harvard.edu/abs/2022NatAs...6..647C} {6, 647}

\bibitem[\protect\citeauthoryear{{Crnojevi{\'c}} et~al.,}{{Crnojevi{\'c}} et~al.}{2019}]{Crnojevic2019}
{Crnojevi{\'c}} D.,  et~al., 2019, \mn@doi [\apj] {10.3847/1538-4357/aafbe7}, \href {https://ui.adsabs.harvard.edu/abs/2019ApJ...872...80C} {872, 80}

\bibitem[\protect\citeauthoryear{{D'Onghia}, {Springel}, {Hernquist}  \& {Keres}}{{D'Onghia} et~al.}{2010}]{DOnghia2010}
{D'Onghia} E.,  {Springel} V.,  {Hernquist} L.,   {Keres} D.,  2010, \mn@doi [\apj] {10.1088/0004-637X/709/2/1138}, \href {https://ui.adsabs.harvard.edu/abs/2010ApJ...709.1138D} {709, 1138}

\bibitem[\protect\citeauthoryear{{Davis}, {Efstathiou}, {Frenk}  \& {White}}{{Davis} et~al.}{1985}]{Davis1985}
{Davis} M.,  {Efstathiou} G.,  {Frenk} C.~S.,   {White} S.~D.~M.,  1985, \mn@doi [\apj] {10.1086/163168}, \href {http://adsabs.harvard.edu/abs/1985ApJ...292..371D} {292, 371}

\bibitem[\protect\citeauthoryear{{Dolag}, {Borgani}, {Murante}  \& {Springel}}{{Dolag} et~al.}{2009}]{Dolag2009}
{Dolag} K.,  {Borgani} S.,  {Murante} G.,   {Springel} V.,  2009, \mn@doi [\mnras] {10.1111/j.1365-2966.2009.15034.x}, \href {http://adsabs.harvard.edu/abs/2009MNRAS.399..497D} {399, 497}

\bibitem[\protect\citeauthoryear{{Engler} et~al.,}{{Engler} et~al.}{2021}]{Engler2021}
{Engler} C.,  et~al., 2021, \mn@doi [\mnras] {10.1093/mnras/stab2437}, \href {https://ui.adsabs.harvard.edu/abs/2021MNRAS.507.4211E} {507, 4211}

\bibitem[\protect\citeauthoryear{{Faltenbacher} \& {Mathews}}{{Faltenbacher} \& {Mathews}}{2005}]{Faltenbacher2005}
{Faltenbacher} A.,  {Mathews} W.~G.,  2005, \mn@doi [\mnras] {10.1111/j.1365-2966.2005.09334.x}, \href {https://ui.adsabs.harvard.edu/abs/2005MNRAS.362..498F} {362, 498}

\bibitem[\protect\citeauthoryear{{Fattahi}, {Navarro}, {Frenk}, {Oman}, {Sawala}  \& {Schaller}}{{Fattahi} et~al.}{2018}]{Fattahi2018}
{Fattahi} A.,  {Navarro} J.~F.,  {Frenk} C.~S.,  {Oman} K.~A.,  {Sawala} T.,   {Schaller} M.,  2018, \mn@doi [\mnras] {10.1093/mnras/sty408}, \href {https://ui.adsabs.harvard.edu/abs/2018MNRAS.476.3816F} {476, 3816}

\bibitem[\protect\citeauthoryear{{Faucher-Gigu{\`e}re}, {Lidz}, {Zaldarriaga}  \& {Hernquist}}{{Faucher-Gigu{\`e}re} et~al.}{2009}]{FaucherGiguere2009}
{Faucher-Gigu{\`e}re} C.-A.,  {Lidz} A.,  {Zaldarriaga} M.,   {Hernquist} L.,  2009, \mn@doi [\apj] {10.1088/0004-637X/703/2/1416}, \href {https://ui.adsabs.harvard.edu/abs/2009ApJ...703.1416F} {703, 1416}

\bibitem[\protect\citeauthoryear{{Feldmann} et~al.,}{{Feldmann} et~al.}{2023}]{Feldmann2023Firebox}
{Feldmann} R.,  et~al., 2023, \mn@doi [\mnras] {10.1093/mnras/stad1205}, \href {https://ui.adsabs.harvard.edu/abs/2023MNRAS.522.3831F} {522, 3831}

\bibitem[\protect\citeauthoryear{{Fielder}, {Mao}, {Newman}, {Zentner}  \& {Licquia}}{{Fielder} et~al.}{2019}]{Fielder2019}
{Fielder} C.~E.,  {Mao} Y.-Y.,  {Newman} J.~A.,  {Zentner} A.~R.,   {Licquia} T.~C.,  2019, \mn@doi [\mnras] {10.1093/mnras/stz1098}, \href {https://ui.adsabs.harvard.edu/abs/2019MNRAS.486.4545F} {486, 4545}

\bibitem[\protect\citeauthoryear{{Font} et~al.,}{{Font} et~al.}{2020}]{Font2020Artemis}
{Font} A.~S.,  et~al., 2020, \mn@doi [\mnras] {10.1093/mnras/staa2463}, \href {https://ui.adsabs.harvard.edu/abs/2020MNRAS.498.1765F} {498, 1765}

\bibitem[\protect\citeauthoryear{{Font}, {McCarthy}  \& {Belokurov}}{{Font} et~al.}{2021}]{Font2021}
{Font} A.~S.,  {McCarthy} I.~G.,   {Belokurov} V.,  2021, \mn@doi [\mnras] {10.1093/mnras/stab1332}, \href {https://ui.adsabs.harvard.edu/abs/2021MNRAS.505..783F} {505, 783}

\bibitem[\protect\citeauthoryear{{Garrison-Kimmel} et~al.,}{{Garrison-Kimmel} et~al.}{2017}]{GarrisonKimmel2017}
{Garrison-Kimmel} S.,  et~al., 2017, \mn@doi [\mnras] {10.1093/mnras/stx1710}, \href {https://ui.adsabs.harvard.edu/abs/2017MNRAS.471.1709G} {471, 1709}

\bibitem[\protect\citeauthoryear{{Garrison-Kimmel} et~al.,}{{Garrison-Kimmel} et~al.}{2019}]{GarrisonKimmel2019}
{Garrison-Kimmel} S.,  et~al., 2019, \mn@doi [\mnras] {10.1093/mnras/stz1317}, \href {https://ui.adsabs.harvard.edu/abs/2019MNRAS.487.1380G} {487, 1380}

\bibitem[\protect\citeauthoryear{{Geha} et~al.,}{{Geha} et~al.}{2017}]{Geha2017}
{Geha} M.,  et~al., 2017, \mn@doi [\apj] {10.3847/1538-4357/aa8626}, \href {https://ui.adsabs.harvard.edu/abs/2017ApJ...847....4G} {847, 4}

\bibitem[\protect\citeauthoryear{{Geha} et~al.,}{{Geha} et~al.}{2024}]{Geha2024}
{Geha} M.,  et~al., 2024, \mn@doi [\apj] {10.3847/1538-4357/ad61e7}, \href {https://ui.adsabs.harvard.edu/abs/2024ApJ...976..118G} {976, 118}

\bibitem[\protect\citeauthoryear{{Gill}, {Knebe}  \& {Gibson}}{{Gill} et~al.}{2004}]{Gill2004}
{Gill} S. P.~D.,  {Knebe} A.,   {Gibson} B.~K.,  2004, \mn@doi [\mnras] {10.1111/j.1365-2966.2004.07786.x}, \href {https://ui.adsabs.harvard.edu/abs/2004MNRAS.351..399G} {351, 399}

\bibitem[\protect\citeauthoryear{{Gozman} et~al.,}{{Gozman} et~al.}{2024}]{Gozman2024}
{Gozman} K.,  et~al., 2024, \mn@doi [\apj] {10.3847/1538-4357/ad8c3a}, \href {https://ui.adsabs.harvard.edu/abs/2024ApJ...977..179G} {977, 179}

\bibitem[\protect\citeauthoryear{{Grand} et~al.,}{{Grand} et~al.}{2017}]{Grand2017Auriga}
{Grand} R. J.~J.,  et~al., 2017, \mn@doi [\mnras] {10.1093/mnras/stx071}, \href {https://ui.adsabs.harvard.edu/abs/2017MNRAS.467..179G} {467, 179}

\bibitem[\protect\citeauthoryear{{Greene}, {Danieli}, {Carlsten}, {Beaton}, {Jiang}  \& {Li}}{{Greene} et~al.}{2023}]{Greene2023}
{Greene} J.~E.,  {Danieli} S.,  {Carlsten} S.,  {Beaton} R.,  {Jiang} F.,   {Li} J.,  2023, \mn@doi [\apj] {10.3847/1538-4357/acc58c}, \href {https://ui.adsabs.harvard.edu/abs/2023ApJ...949...94G} {949, 94}

\bibitem[\protect\citeauthoryear{{Grimozzi}, {Font}  \& {De Rossi}}{{Grimozzi} et~al.}{2024}]{Grimozzi2024}
{Grimozzi} S.~E.,  {Font} A.~S.,   {De Rossi} M.~E.,  2024, \mn@doi [\mnras] {10.1093/mnras/stae878}, \href {https://ui.adsabs.harvard.edu/abs/2024MNRAS.530...95G} {530, 95}

\bibitem[\protect\citeauthoryear{{Haardt} \& {Madau}}{{Haardt} \& {Madau}}{1996}]{Haardt1996}
{Haardt} F.,  {Madau} P.,  1996, \mn@doi [\apj] {10.1086/177035}, \href {https://ui.adsabs.harvard.edu/abs/1996ApJ...461...20H} {461, 20}

\bibitem[\protect\citeauthoryear{{Hayashi}, {Navarro}, {Taylor}, {Stadel}  \& {Quinn}}{{Hayashi} et~al.}{2003}]{Hayashi2003}
{Hayashi} E.,  {Navarro} J.~F.,  {Taylor} J.~E.,  {Stadel} J.,   {Quinn} T.,  2003, \mn@doi [\apj] {10.1086/345788}, \href {https://ui.adsabs.harvard.edu/abs/2003ApJ...584..541H} {584, 541}

\bibitem[\protect\citeauthoryear{{Heesters} et~al.,}{{Heesters} et~al.}{2025}]{Heesters2025}
{Heesters} N.,  et~al., 2025, \mn@doi [arXiv e-prints] {10.48550/arXiv.2505.18307}, \href {https://ui.adsabs.harvard.edu/abs/2025arXiv250518307H} {p. arXiv:2505.18307}

\bibitem[\protect\citeauthoryear{{Henriques} \& {Thomas}}{{Henriques} \& {Thomas}}{2010}]{Henriques2010}
{Henriques} B. M.~B.,  {Thomas} P.~A.,  2010, \mn@doi [\mnras] {10.1111/j.1365-2966.2009.16151.x}, \href {https://ui.adsabs.harvard.edu/abs/2010MNRAS.403..768H} {403, 768}

\bibitem[\protect\citeauthoryear{{Hopkins}}{{Hopkins}}{2017}]{Hopkins2017}
{Hopkins} P.~F.,  2017, \mn@doi [\mnras] {10.1093/mnras/stw3306}, \href {https://ui.adsabs.harvard.edu/abs/2017MNRAS.466.3387H} {466, 3387}

\bibitem[\protect\citeauthoryear{{Hopkins}, {Kere{\v{s}}}, {O{\~n}orbe}, {Faucher-Gigu{\`e}re}, {Quataert}, {Murray}  \& {Bullock}}{{Hopkins} et~al.}{2014}]{Hopkins2014}
{Hopkins} P.~F.,  {Kere{\v{s}}} D.,  {O{\~n}orbe} J.,  {Faucher-Gigu{\`e}re} C.-A.,  {Quataert} E.,  {Murray} N.,   {Bullock} J.~S.,  2014, \mn@doi [\mnras] {10.1093/mnras/stu1738}, \href {https://ui.adsabs.harvard.edu/abs/2014MNRAS.445..581H} {445, 581}

\bibitem[\protect\citeauthoryear{{Jester} et~al.,}{{Jester} et~al.}{2005}]{Jesters2005}
{Jester} S.,  et~al., 2005, \mn@doi [\aj] {10.1086/432466}, \href {https://ui.adsabs.harvard.edu/abs/2005AJ....130..873J} {130, 873}

\bibitem[\protect\citeauthoryear{{Joshi}, {Pontzen}, {Agertz}, {Rey}, {Read}  \& {Renaud}}{{Joshi} et~al.}{2024}]{Joshi2024}
{Joshi} G.~D.,  {Pontzen} A.,  {Agertz} O.,  {Rey} M.~P.,  {Read} J.,   {Renaud} F.,  2024, \mn@doi [\mnras] {10.1093/mnras/stae129}, \href {https://ui.adsabs.harvard.edu/abs/2024MNRAS.528.2346J} {528, 2346}

\bibitem[\protect\citeauthoryear{{Joshi}, {Pontzen}, {Agertz}, {Rey}, {Read}  \& {Pillepich}}{{Joshi} et~al.}{2025}]{Joshi2025}
{Joshi} G.~D.,  {Pontzen} A.,  {Agertz} O.,  {Rey} M.~P.,  {Read} J.,   {Pillepich} A.,  2025, \mn@doi [\mnras] {10.1093/mnras/staf276}, \href {https://ui.adsabs.harvard.edu/abs/2025MNRAS.537.3792J} {537, 3792}

\bibitem[\protect\citeauthoryear{{Kazantzidis}, {Mayer}, {Callegari}, {Dotti}  \& {Moustakas}}{{Kazantzidis} et~al.}{2017}]{Kazantzidis2017}
{Kazantzidis} S.,  {Mayer} L.,  {Callegari} S.,  {Dotti} M.,   {Moustakas} L.~A.,  2017, \mn@doi [\apjl] {10.3847/2041-8213/aa5b8f}, \href {https://ui.adsabs.harvard.edu/abs/2017ApJ...836L..13K} {836, L13}

\bibitem[\protect\citeauthoryear{{Kelley}, {Bullock}, {Garrison-Kimmel}, {Boylan-Kolchin}, {Pawlowski}  \& {Graus}}{{Kelley} et~al.}{2019}]{Kelley2019}
{Kelley} T.,  {Bullock} J.~S.,  {Garrison-Kimmel} S.,  {Boylan-Kolchin} M.,  {Pawlowski} M.~S.,   {Graus} A.~S.,  2019, \mn@doi [\mnras] {10.1093/mnras/stz1553}, \href {https://ui.adsabs.harvard.edu/abs/2019MNRAS.487.4409K} {487, 4409}

\bibitem[\protect\citeauthoryear{{Kenney}, {Geha}, {J{\'a}chym}, {Crowl}, {Dague}, {Chung}, {van Gorkom}  \& {Vollmer}}{{Kenney} et~al.}{2014}]{Kenney2014}
{Kenney} J. D.~P.,  {Geha} M.,  {J{\'a}chym} P.,  {Crowl} H.~H.,  {Dague} W.,  {Chung} A.,  {van Gorkom} J.,   {Vollmer} B.,  2014, \mn@doi [\apj] {10.1088/0004-637X/780/2/119}, \href {https://ui.adsabs.harvard.edu/abs/2014ApJ...780..119K} {780, 119}

\bibitem[\protect\citeauthoryear{{Kennicutt}}{{Kennicutt}}{1998}]{Kennicutt1998}
{Kennicutt} Robert~C. J.,  1998, \mn@doi [\apj] {10.1086/305588}, \href {https://ui.adsabs.harvard.edu/abs/1998ApJ...498..541K} {498, 541}

\bibitem[\protect\citeauthoryear{{Kim} \& {Ostriker}}{{Kim} \& {Ostriker}}{2015}]{Kim2015}
{Kim} C.-G.,  {Ostriker} E.~C.,  2015, \mn@doi [\apj] {10.1088/0004-637X/802/2/99}, \href {https://ui.adsabs.harvard.edu/abs/2015ApJ...802...99K} {802, 99}

\bibitem[\protect\citeauthoryear{{Kim} et~al.,}{{Kim} et~al.}{2014}]{Kim2014Agora}
{Kim} J.-h.,  et~al., 2014, \mn@doi [\apjs] {10.1088/0067-0049/210/1/14}, \href {https://ui.adsabs.harvard.edu/abs/2014ApJS..210...14K} {210, 14}

\bibitem[\protect\citeauthoryear{{Kim} et~al.,}{{Kim} et~al.}{2024}]{Kim2024}
{Kim} S.~Y.,  et~al., 2024, \mn@doi [arXiv e-prints] {10.48550/arXiv.2408.15214}, \href {https://ui.adsabs.harvard.edu/abs/2024arXiv240815214K} {p. arXiv:2408.15214}

\bibitem[\protect\citeauthoryear{{Knollmann} \& {Knebe}}{{Knollmann} \& {Knebe}}{2009}]{Knollmann2009}
{Knollmann} S.~R.,  {Knebe} A.,  2009, \mn@doi [\apjs] {10.1088/0067-0049/182/2/608}, \href {https://ui.adsabs.harvard.edu/abs/2009ApJS..182..608K} {182, 608}

\bibitem[\protect\citeauthoryear{{Libeskind} et~al.,}{{Libeskind} et~al.}{2020}]{Libeskind2020Hestia}
{Libeskind} N.~I.,  et~al., 2020, \mn@doi [\mnras] {10.1093/mnras/staa2541}, \href {https://ui.adsabs.harvard.edu/abs/2020MNRAS.498.2968L} {498, 2968}

\bibitem[\protect\citeauthoryear{{Mao}, {Geha}, {Wechsler}, {Weiner}, {Tollerud}, {Nadler}  \& {Kallivayalil}}{{Mao} et~al.}{2021}]{Mao2021}
{Mao} Y.-Y.,  {Geha} M.,  {Wechsler} R.~H.,  {Weiner} B.,  {Tollerud} E.~J.,  {Nadler} E.~O.,   {Kallivayalil} N.,  2021, \mn@doi [\apj] {10.3847/1538-4357/abce58}, \href {https://ui.adsabs.harvard.edu/abs/2021ApJ...907...85M} {907, 85}

\bibitem[\protect\citeauthoryear{{Mao} et~al.,}{{Mao} et~al.}{2024}]{Mao2024}
{Mao} Y.-Y.,  et~al., 2024, \mn@doi [\apj] {10.3847/1538-4357/ad64c4}, \href {https://ui.adsabs.harvard.edu/abs/2024ApJ...976..117M} {976, 117}

\bibitem[\protect\citeauthoryear{{Marleau} et~al.,}{{Marleau} et~al.}{2025}]{Marleau2025}
{Marleau} F.~R.,  et~al., 2025, \mn@doi [arXiv e-prints] {10.48550/arXiv.2503.15335}, \href {https://ui.adsabs.harvard.edu/abs/2025arXiv250315335M} {p. arXiv:2503.15335}

\bibitem[\protect\citeauthoryear{{Martin}, {Ibata}, {McConnachie}, {Mackey}, {Ferguson}, {Irwin}, {Lewis}  \& {Fardal}}{{Martin} et~al.}{2013}]{Martin2013}
{Martin} N.~F.,  {Ibata} R.~A.,  {McConnachie} A.~W.,  {Mackey} A.~D.,  {Ferguson} A. M.~N.,  {Irwin} M.~J.,  {Lewis} G.~F.,   {Fardal} M.~A.,  2013, \mn@doi [\apj] {10.1088/0004-637X/776/2/80}, \href {https://ui.adsabs.harvard.edu/abs/2013ApJ...776...80M} {776, 80}

\bibitem[\protect\citeauthoryear{{Martin} et~al.,}{{Martin} et~al.}{2016}]{Martin2016}
{Martin} N.~F.,  et~al., 2016, \mn@doi [\apj] {10.3847/1538-4357/833/2/167}, \href {https://ui.adsabs.harvard.edu/abs/2016ApJ...833..167M} {833, 167}

\bibitem[\protect\citeauthoryear{{Mayer}, {Mastropietro}, {Wadsley}, {Stadel}  \& {Moore}}{{Mayer} et~al.}{2006}]{Mayer2006}
{Mayer} L.,  {Mastropietro} C.,  {Wadsley} J.,  {Stadel} J.,   {Moore} B.,  2006, \mn@doi [\mnras] {10.1111/j.1365-2966.2006.10403.x}, \href {https://ui.adsabs.harvard.edu/abs/2006MNRAS.369.1021M} {369, 1021}

\bibitem[\protect\citeauthoryear{{Mazzarini}, {Just}, {Macci{\`o}}  \& {Moetazedian}}{{Mazzarini} et~al.}{2020}]{Mazzarini2020}
{Mazzarini} M.,  {Just} A.,  {Macci{\`o}} A.~V.,   {Moetazedian} R.,  2020, \mn@doi [\aap] {10.1051/0004-6361/202037558}, \href {https://ui.adsabs.harvard.edu/abs/2020A&A...636A.106M} {636, A106}

\bibitem[\protect\citeauthoryear{{McConnachie}}{{McConnachie}}{2012}]{McConnachie2012}
{McConnachie} A.~W.,  2012, \mn@doi [\aj] {10.1088/0004-6256/144/1/4}, \href {https://ui.adsabs.harvard.edu/abs/2012AJ....144....4M} {144, 4}

\bibitem[\protect\citeauthoryear{{McConnachie}, {Venn}, {Irwin}, {Young}  \& {Geehan}}{{McConnachie} et~al.}{2007}]{McConnachie2007}
{McConnachie} A.~W.,  {Venn} K.~A.,  {Irwin} M.~J.,  {Young} L.~M.,   {Geehan} J.~J.,  2007, \mn@doi [\apjl] {10.1086/524887}, \href {https://ui.adsabs.harvard.edu/abs/2007ApJ...671L..33M} {671, L33}

\bibitem[\protect\citeauthoryear{{McConnachie} et~al.,}{{McConnachie} et~al.}{2018}]{McConnachie2018}
{McConnachie} A.~W.,  et~al., 2018, \mn@doi [\apj] {10.3847/1538-4357/aae8e7}, \href {https://ui.adsabs.harvard.edu/abs/2018ApJ...868...55M} {868, 55}

\bibitem[\protect\citeauthoryear{{Merritt}, {van Dokkum}  \& {Abraham}}{{Merritt} et~al.}{2014}]{Merritt2014}
{Merritt} A.,  {van Dokkum} P.,   {Abraham} R.,  2014, \mn@doi [\apjl] {10.1088/2041-8205/787/2/L37}, \href {https://ui.adsabs.harvard.edu/abs/2014ApJ...787L..37M} {787, L37}

\bibitem[\protect\citeauthoryear{{M{\"u}ller}, {Pawlowski}, {Revaz}, {Venhola}, {Rejkuba}, {Hilker}  \& {Lutz}}{{M{\"u}ller} et~al.}{2024}]{Muller2024}
{M{\"u}ller} O.,  {Pawlowski} M.~S.,  {Revaz} Y.,  {Venhola} A.,  {Rejkuba} M.,  {Hilker} M.,   {Lutz} K.,  2024, \mn@doi [\aap] {10.1051/0004-6361/202348969}, \href {https://ui.adsabs.harvard.edu/abs/2024A&A...684L...6M} {684, L6}

\bibitem[\protect\citeauthoryear{{M{\"u}ller}, {Jerjen}, {Taibi}, {Heesters}, {Crosby}  \& {Pawlowski}}{{M{\"u}ller} et~al.}{2025a}]{Muller2025b}
{M{\"u}ller} O.,  {Jerjen} H.,  {Taibi} S.,  {Heesters} N.,  {Crosby} E.,   {Pawlowski} M.~S.,  2025a, \mn@doi [arXiv e-prints] {10.48550/arXiv.2504.11608}, \href {https://ui.adsabs.harvard.edu/abs/2025arXiv250411608M} {p. arXiv:2504.11608}

\bibitem[\protect\citeauthoryear{{M{\"u}ller} et~al.,}{{M{\"u}ller} et~al.}{2025b}]{Muller2025a}
{M{\"u}ller} O.,  et~al., 2025b, \mn@doi [\aap] {10.1051/0004-6361/202450143}, \href {https://ui.adsabs.harvard.edu/abs/2025A&A...693A..44M} {693, A44}

\bibitem[\protect\citeauthoryear{{Mutlu-Pakdil} et~al.,}{{Mutlu-Pakdil} et~al.}{2024}]{MutluPakdil2024}
{Mutlu-Pakdil} B.,  et~al., 2024, \mn@doi [\apj] {10.3847/1538-4357/ad36c4}, \href {https://ui.adsabs.harvard.edu/abs/2024ApJ...966..188M} {966, 188}

\bibitem[\protect\citeauthoryear{{Nadler}, {Mao}, {Wechsler}, {Garrison-Kimmel}  \& {Wetzel}}{{Nadler} et~al.}{2018}]{Nadler2018}
{Nadler} E.~O.,  {Mao} Y.-Y.,  {Wechsler} R.~H.,  {Garrison-Kimmel} S.,   {Wetzel} A.,  2018, \mn@doi [\apj] {10.3847/1538-4357/aac266}, \href {https://ui.adsabs.harvard.edu/abs/2018ApJ...859..129N} {859, 129}

\bibitem[\protect\citeauthoryear{{Nelson} et~al.,}{{Nelson} et~al.}{2019}]{TNG50Nelson2019}
{Nelson} D.,  et~al., 2019, \mn@doi [\mnras] {10.1093/mnras/stz2306}, \href {https://ui.adsabs.harvard.edu/abs/2019MNRAS.490.3234N} {490, 3234}

\bibitem[\protect\citeauthoryear{{Nichols} \& {Bland-Hawthorn}}{{Nichols} \& {Bland-Hawthorn}}{2011}]{Nichols2011}
{Nichols} M.,  {Bland-Hawthorn} J.,  2011, \mn@doi [\apj] {10.1088/0004-637X/732/1/17}, \href {https://ui.adsabs.harvard.edu/abs/2011ApJ...732...17N} {732, 17}

\bibitem[\protect\citeauthoryear{{Pathak}, {Christensen}, {Brooks}, {Munshi}, {Wright}  \& {Carter}}{{Pathak} et~al.}{2025}]{Pathak2025}
{Pathak} D.,  {Christensen} C.~R.,  {Brooks} A.~M.,  {Munshi} F.,  {Wright} A.~C.,   {Carter} C.,  2025, \mn@doi [arXiv e-prints] {10.48550/arXiv.2505.22742}, \href {https://ui.adsabs.harvard.edu/abs/2025arXiv250522742P} {p. arXiv:2505.22742}

\bibitem[\protect\citeauthoryear{{Pe{\~n}arrubia}, {Benson}, {Walker}, {Gilmore}, {McConnachie}  \& {Mayer}}{{Pe{\~n}arrubia} et~al.}{2010}]{Penarrubia2010}
{Pe{\~n}arrubia} J.,  {Benson} A.~J.,  {Walker} M.~G.,  {Gilmore} G.,  {McConnachie} A.~W.,   {Mayer} L.,  2010, \mn@doi [\mnras] {10.1111/j.1365-2966.2010.16762.x}, \href {https://ui.adsabs.harvard.edu/abs/2010MNRAS.406.1290P} {406, 1290}

\bibitem[\protect\citeauthoryear{{Pillepich} et~al.,}{{Pillepich} et~al.}{2018}]{TNGMethodsPillepich2018}
{Pillepich} A.,  et~al., 2018, \mn@doi [\mnras] {10.1093/mnras/stx2656}, \href {https://ui.adsabs.harvard.edu/abs/2018MNRAS.473.4077P} {473, 4077}

\bibitem[\protect\citeauthoryear{{Pillepich} et~al.,}{{Pillepich} et~al.}{2019}]{TNG50Pillepich2019}
{Pillepich} A.,  et~al., 2019, \mn@doi [\mnras] {10.1093/mnras/stz2338}, \href {https://ui.adsabs.harvard.edu/abs/2019MNRAS.490.3196P} {490, 3196}

\bibitem[\protect\citeauthoryear{{Planck Collaboration} et~al.,}{{Planck Collaboration} et~al.}{2016}]{Planck2016}
{Planck Collaboration} et~al., 2016, \mn@doi [\aap] {10.1051/0004-6361/201525830}, \href {https://ui.adsabs.harvard.edu/abs/2016A&A...594A..13P} {594, A13}

\bibitem[\protect\citeauthoryear{{Pontzen} \& {Tremmel}}{{Pontzen} \& {Tremmel}}{2018}]{Pontzen2018}
{Pontzen} A.,  {Tremmel} M.,  2018, \mn@doi [\apjs] {10.3847/1538-4365/aac832}, \href {https://ui.adsabs.harvard.edu/abs/2018ApJS..237...23P} {237, 23}

\bibitem[\protect\citeauthoryear{{Pontzen}, {Ro{\v{s}}kar}, {Stinson}  \& {Woods}}{{Pontzen} et~al.}{2013}]{Pontzen2013}
{Pontzen} A.,  {Ro{\v{s}}kar} R.,  {Stinson} G.,   {Woods} R.,  2013, {pynbody: N-Body/SPH analysis for python}, Astrophysics Source Code Library, record ascl:1305.002 (\mn@eprint {ascl} {1305.002})

\bibitem[\protect\citeauthoryear{{Power}, {Navarro}, {Jenkins}, {Frenk}, {White}, {Springel}, {Stadel}  \& {Quinn}}{{Power} et~al.}{2003}]{Power2003}
{Power} C.,  {Navarro} J.~F.,  {Jenkins} A.,  {Frenk} C.~S.,  {White} S.~D.~M.,  {Springel} V.,  {Stadel} J.,   {Quinn} T.,  2003, \mn@doi [\mnras] {10.1046/j.1365-8711.2003.05925.x}, \href {https://ui.adsabs.harvard.edu/abs/2003MNRAS.338...14P} {338, 14}

\bibitem[\protect\citeauthoryear{{Purcell}, {Bullock}  \& {Zentner}}{{Purcell} et~al.}{2007}]{Purcell2007}
{Purcell} C.~W.,  {Bullock} J.~S.,   {Zentner} A.~R.,  2007, \mn@doi [\apj] {10.1086/519787}, \href {https://ui.adsabs.harvard.edu/abs/2007ApJ...666...20P} {666, 20}

\bibitem[\protect\citeauthoryear{{Read} \& {Erkal}}{{Read} \& {Erkal}}{2019}]{Read2019}
{Read} J.~I.,  {Erkal} D.,  2019, \mn@doi [\mnras] {10.1093/mnras/stz1320}, \href {https://ui.adsabs.harvard.edu/abs/2019MNRAS.487.5799R} {487, 5799}

\bibitem[\protect\citeauthoryear{{Read}, {Wilkinson}, {Evans}, {Gilmore}  \& {Kleyna}}{{Read} et~al.}{2006}]{Read2006}
{Read} J.~I.,  {Wilkinson} M.~I.,  {Evans} N.~W.,  {Gilmore} G.,   {Kleyna} J.~T.,  2006, \mn@doi [\mnras] {10.1111/j.1365-2966.2005.09959.x}, \href {https://ui.adsabs.harvard.edu/abs/2006MNRAS.367..387R} {367, 387}

\bibitem[\protect\citeauthoryear{{Rey} \& {Pontzen}}{{Rey} \& {Pontzen}}{2018}]{Rey2018}
{Rey} M.~P.,  {Pontzen} A.,  2018, \mn@doi [\mnras] {10.1093/mnras/stx2744}, \href {https://ui.adsabs.harvard.edu/abs/2018MNRAS.474...45R} {474, 45}

\bibitem[\protect\citeauthoryear{{Rey} \& {Starkenburg}}{{Rey} \& {Starkenburg}}{2022}]{Rey2022}
{Rey} M.~P.,  {Starkenburg} T.~K.,  2022, \mn@doi [\mnras] {10.1093/mnras/stab3709}, \href {https://ui.adsabs.harvard.edu/abs/2022MNRAS.510.4208R} {510, 4208}

\bibitem[\protect\citeauthoryear{{Rey}, {Pontzen}, {Agertz}, {Orkney}, {Read}, {Saintonge}  \& {Pedersen}}{{Rey} et~al.}{2019}]{Rey2019}
{Rey} M.~P.,  {Pontzen} A.,  {Agertz} O.,  {Orkney} M. D.~A.,  {Read} J.~I.,  {Saintonge} A.,   {Pedersen} C.,  2019, \mn@doi [\apjl] {10.3847/2041-8213/ab53dd}, \href {https://ui.adsabs.harvard.edu/abs/2019ApJ...886L...3R} {886, L3}

\bibitem[\protect\citeauthoryear{{Rey} et~al.,}{{Rey} et~al.}{2023}]{Rey2023}
{Rey} M.~P.,  et~al., 2023, \mn@doi [\mnras] {10.1093/mnras/stad513}, \href {https://ui.adsabs.harvard.edu/abs/2023MNRAS.521..995R} {521, 995}

\bibitem[\protect\citeauthoryear{{Rey} et~al.,}{{Rey} et~al.}{2025}]{Rey2025}
{Rey} M.~P.,  et~al., 2025, \mn@doi [\mnras] {10.1093/mnras/staf1058}, \href {https://ui.adsabs.harvard.edu/abs/2025MNRAS.541.1195R} {541, 1195}

\bibitem[\protect\citeauthoryear{{Riley} et~al.,}{{Riley} et~al.}{2024}]{Riley2024}
{Riley} A.~H.,  et~al., 2024, \mn@doi [arXiv e-prints] {10.48550/arXiv.2410.09144}, \href {https://ui.adsabs.harvard.edu/abs/2024arXiv241009144R} {p. arXiv:2410.09144}

\bibitem[\protect\citeauthoryear{{Roca-F{\`a}brega} et~al.,}{{Roca-F{\`a}brega} et~al.}{2021}]{RocaFabrega2021Agora}
{Roca-F{\`a}brega} S.,  et~al., 2021, \mn@doi [\apj] {10.3847/1538-4357/ac088a}, \href {https://ui.adsabs.harvard.edu/abs/2021ApJ...917...64R} {917, 64}

\bibitem[\protect\citeauthoryear{{Rodr{\'\i}guez-Cardoso} et~al.,}{{Rodr{\'\i}guez-Cardoso} et~al.}{2025}]{RodriguezCardoso2025}
{Rodr{\'\i}guez-Cardoso} R.,  et~al., 2025, \mn@doi [\aap] {10.1051/0004-6361/202453639}, \href {https://ui.adsabs.harvard.edu/abs/2025A&A...698A.303R} {698, A303}

\bibitem[\protect\citeauthoryear{{Rodriguez Wimberly}, {Cooper}, {Fillingham}, {Boylan-Kolchin}, {Bullock}  \& {Garrison-Kimmel}}{{Rodriguez Wimberly} et~al.}{2019}]{RodriguezWimberly2019}
{Rodriguez Wimberly} M.~K.,  {Cooper} M.~C.,  {Fillingham} S.~P.,  {Boylan-Kolchin} M.,  {Bullock} J.~S.,   {Garrison-Kimmel} S.,  2019, \mn@doi [\mnras] {10.1093/mnras/sty3357}, \href {https://ui.adsabs.harvard.edu/abs/2019MNRAS.483.4031R} {483, 4031}

\bibitem[\protect\citeauthoryear{{Roth}, {Pontzen}  \& {Peiris}}{{Roth} et~al.}{2016}]{Roth2016}
{Roth} N.,  {Pontzen} A.,   {Peiris} H.~V.,  2016, \mn@doi [\mnras] {10.1093/mnras/stv2375}, \href {https://ui.adsabs.harvard.edu/abs/2016MNRAS.455..974R} {455, 974}

\bibitem[\protect\citeauthoryear{{Sales}, {Helmi}  \& {Battaglia}}{{Sales} et~al.}{2010}]{Sales2010}
{Sales} L.~V.,  {Helmi} A.,   {Battaglia} G.,  2010, \mn@doi [Advances in Astronomy] {10.1155/2010/194345}, \href {https://ui.adsabs.harvard.edu/abs/2010AdAst2010E..18S} {2010, 194345}

\bibitem[\protect\citeauthoryear{{Samuel}, {Pardasani}, {Wetzel}, {Santistevan}, {Boylan-Kolchin}, {Moreno}  \& {Faucher-Gigu{\`e}re}}{{Samuel} et~al.}{2023}]{Samuel2023}
{Samuel} J.,  {Pardasani} B.,  {Wetzel} A.,  {Santistevan} I.,  {Boylan-Kolchin} M.,  {Moreno} J.,   {Faucher-Gigu{\`e}re} C.-A.,  2023, \mn@doi [\mnras] {10.1093/mnras/stad2576}, \href {https://ui.adsabs.harvard.edu/abs/2023MNRAS.525.3849S} {525, 3849}

\bibitem[\protect\citeauthoryear{{Sawala} et~al.,}{{Sawala} et~al.}{2016}]{Sawala2016Apostle}
{Sawala} T.,  et~al., 2016, \mn@doi [\mnras] {10.1093/mnras/stw145}, \href {https://ui.adsabs.harvard.edu/abs/2016MNRAS.457.1931S} {457, 1931}

\bibitem[\protect\citeauthoryear{{Schmidt}}{{Schmidt}}{1959}]{Schmidt1959}
{Schmidt} M.,  1959, \mn@doi [\apj] {10.1086/146614}, \href {https://ui.adsabs.harvard.edu/abs/1959ApJ...129..243S} {129, 243}

\bibitem[\protect\citeauthoryear{{Shipp} et~al.,}{{Shipp} et~al.}{2024}]{Shipp2024}
{Shipp} N.,  et~al., 2024, \mn@doi [arXiv e-prints] {10.48550/arXiv.2410.09143}, \href {https://ui.adsabs.harvard.edu/abs/2024arXiv241009143S} {p. arXiv:2410.09143}

\bibitem[\protect\citeauthoryear{{Simpson}, {Grand}, {G{\'o}mez}, {Marinacci}, {Pakmor}, {Springel}, {Campbell}  \& {Frenk}}{{Simpson} et~al.}{2018}]{Simpson2018}
{Simpson} C.~M.,  {Grand} R. J.~J.,  {G{\'o}mez} F.~A.,  {Marinacci} F.,  {Pakmor} R.,  {Springel} V.,  {Campbell} D. J.~R.,   {Frenk} C.~S.,  2018, \mn@doi [\mnras] {10.1093/mnras/sty774}, \href {https://ui.adsabs.harvard.edu/abs/2018MNRAS.478..548S} {478, 548}

\bibitem[\protect\citeauthoryear{{Smercina}, {Bell}, {Price}, {D'Souza}, {Slater}, {Bailin}, {Monachesi}  \& {Nidever}}{{Smercina} et~al.}{2018}]{Smercina2018}
{Smercina} A.,  {Bell} E.~F.,  {Price} P.~A.,  {D'Souza} R.,  {Slater} C.~T.,  {Bailin} J.,  {Monachesi} A.,   {Nidever} D.,  2018, \mn@doi [\apj] {10.3847/1538-4357/aad2d6}, \href {https://ui.adsabs.harvard.edu/abs/2018ApJ...863..152S} {863, 152}

\bibitem[\protect\citeauthoryear{{Springel}}{{Springel}}{2010}]{Springel2010}
{Springel} V.,  2010, \mn@doi [\mnras] {10.1111/j.1365-2966.2009.15715.x}, \href {http://adsabs.harvard.edu/abs/2010MNRAS.401..791S} {401, 791}

\bibitem[\protect\citeauthoryear{{Springel}, {White}, {Tormen}  \& {Kauffmann}}{{Springel} et~al.}{2001}]{Springel2001}
{Springel} V.,  {White} S.~D.~M.,  {Tormen} G.,   {Kauffmann} G.,  2001, \mn@doi [\mnras] {10.1046/j.1365-8711.2001.04912.x}, \href {http://adsabs.harvard.edu/abs/2001MNRAS.328..726S} {328, 726}

\bibitem[\protect\citeauthoryear{{Stopyra}, {Pontzen}, {Peiris}, {Roth}  \& {Rey}}{{Stopyra} et~al.}{2021}]{Stopyra2021}
{Stopyra} S.,  {Pontzen} A.,  {Peiris} H.,  {Roth} N.,   {Rey} M.~P.,  2021, \mn@doi [\apjs] {10.3847/1538-4365/abcd94}, \href {https://ui.adsabs.harvard.edu/abs/2021ApJS..252...28S} {252, 28}

\bibitem[\protect\citeauthoryear{{Taffoni}, {Mayer}, {Colpi}  \& {Governato}}{{Taffoni} et~al.}{2003}]{Taffoni2003}
{Taffoni} G.,  {Mayer} L.,  {Colpi} M.,   {Governato} F.,  2003, \mn@doi [\mnras] {10.1046/j.1365-8711.2003.06395.x}, \href {https://ui.adsabs.harvard.edu/abs/2003MNRAS.341..434T} {341, 434}

\bibitem[\protect\citeauthoryear{{Taylor} \& {Babul}}{{Taylor} \& {Babul}}{2004}]{Taylor2004}
{Taylor} J.~E.,  {Babul} A.,  2004, \mn@doi [\mnras] {10.1111/j.1365-2966.2004.07395.x}, \href {https://ui.adsabs.harvard.edu/abs/2004MNRAS.348..811T} {348, 811}

\bibitem[\protect\citeauthoryear{{Teyssier}}{{Teyssier}}{2002}]{Teyssier2002}
{Teyssier} R.,  2002, \mn@doi [\aap] {10.1051/0004-6361:20011817}, \href {https://ui.adsabs.harvard.edu/abs/2002A&A...385..337T} {385, 337}

\bibitem[\protect\citeauthoryear{{Toro}, {Spruce}  \& {Speares}}{{Toro} et~al.}{1994}]{Toro1994}
{Toro} E.~F.,  {Spruce} M.,   {Speares} W.,  1994, \mn@doi [Shock Waves] {10.1007/BF01414629}, \href {https://ui.adsabs.harvard.edu/abs/1994ShWav...4...25T} {4, 25}

\bibitem[\protect\citeauthoryear{{Tremmel}, {Karcher}, {Governato}, {Volonteri}, {Quinn}, {Pontzen}, {Anderson}  \& {Bellovary}}{{Tremmel} et~al.}{2017}]{Tremmel2017Romulus25}
{Tremmel} M.,  {Karcher} M.,  {Governato} F.,  {Volonteri} M.,  {Quinn} T.~R.,  {Pontzen} A.,  {Anderson} L.,   {Bellovary} J.,  2017, \mn@doi [\mnras] {10.1093/mnras/stx1160}, \href {https://ui.adsabs.harvard.edu/abs/2017MNRAS.470.1121T} {470, 1121}

\bibitem[\protect\citeauthoryear{{Wang}, {Dutton}, {Stinson}, {Macci{\`o}}, {Penzo}, {Kang}, {Keller}  \& {Wadsley}}{{Wang} et~al.}{2015}]{Wang2015Nihao}
{Wang} L.,  {Dutton} A.~A.,  {Stinson} G.~S.,  {Macci{\`o}} A.~V.,  {Penzo} C.,  {Kang} X.,  {Keller} B.~W.,   {Wadsley} J.,  2015, \mn@doi [\mnras] {10.1093/mnras/stv1937}, \href {https://ui.adsabs.harvard.edu/abs/2015MNRAS.454...83W} {454, 83}

\bibitem[\protect\citeauthoryear{{Weinberger} et~al.,}{{Weinberger} et~al.}{2017}]{TNGMethodsWeinberger2017}
{Weinberger} R.,  et~al., 2017, \mn@doi [\mnras] {10.1093/mnras/stw2944}, \href {https://ui.adsabs.harvard.edu/abs/2017MNRAS.465.3291W} {465, 3291}

\bibitem[\protect\citeauthoryear{{Weisz}, {Dolphin}, {Skillman}, {Holtzman}, {Gilbert}, {Dalcanton}  \& {Williams}}{{Weisz} et~al.}{2015}]{Weisz2015}
{Weisz} D.~R.,  {Dolphin} A.~E.,  {Skillman} E.~D.,  {Holtzman} J.,  {Gilbert} K.~M.,  {Dalcanton} J.~J.,   {Williams} B.~F.,  2015, \mn@doi [\apj] {10.1088/0004-637X/804/2/136}, \href {https://ui.adsabs.harvard.edu/abs/2015ApJ...804..136W} {804, 136}

\bibitem[\protect\citeauthoryear{{Wetzel}, {Hopkins}, {Kim}, {Faucher-Gigu{\`e}re}, {Kere{\v{s}}}  \& {Quataert}}{{Wetzel} et~al.}{2016}]{Wetzel2016Latte}
{Wetzel} A.~R.,  {Hopkins} P.~F.,  {Kim} J.-h.,  {Faucher-Gigu{\`e}re} C.-A.,  {Kere{\v{s}}} D.,   {Quataert} E.,  2016, \mn@doi [\apjl] {10.3847/2041-8205/827/2/L23}, \href {https://ui.adsabs.harvard.edu/abs/2016ApJ...827L..23W} {827, L23}

\bibitem[\protect\citeauthoryear{{Wu} et~al.,}{{Wu} et~al.}{2022}]{Wu2022}
{Wu} J.~F.,  et~al., 2022, \mn@doi [\apj] {10.3847/1538-4357/ac4eea}, \href {https://ui.adsabs.harvard.edu/abs/2022ApJ...927..121W} {927, 121}

\bibitem[\protect\citeauthoryear{{Yang}, {Ianjamasimanana}, {Hammer}, {Higgs}, {Namumba}, {Carignan}, {J{\'o}zsa}  \& {McConnachie}}{{Yang} et~al.}{2022}]{Yang2022}
{Yang} Y.,  {Ianjamasimanana} R.,  {Hammer} F.,  {Higgs} C.,  {Namumba} B.,  {Carignan} C.,  {J{\'o}zsa} G. I.~G.,   {McConnachie} A.~W.,  2022, \mn@doi [\aap] {10.1051/0004-6361/202243307}, \href {https://ui.adsabs.harvard.edu/abs/2022A&A...660L..11Y} {660, L11}

\bibitem[\protect\citeauthoryear{{van den Bosch} \& {Ogiya}}{{van den Bosch} \& {Ogiya}}{2018}]{vanDenBosch2018}
{van den Bosch} F.~C.,  {Ogiya} G.,  2018, \mn@doi [\mnras] {10.1093/mnras/sty084}, \href {https://ui.adsabs.harvard.edu/abs/2018MNRAS.475.4066V} {475, 4066}

\makeatother
\end{thebibliography}
